\documentclass[final,5p,times,twocolumn]{elsarticle}

\makeatletter
\def\ps@pprintTitle{%
  \let\@oddhead\@empty
  \let\@evenhead\@empty
  \def\@oddfoot{}%
  \let\@evenfoot\@oddfoot
}
\makeatother

\usepackage{graphicx, epsfig} % Required for images

\usepackage[utf8]{inputenc}
\usepackage[english]{babel}
\usepackage[T1]{fontenc}
\usepackage{csquotes}
\usepackage{color}

\usepackage{amssymb,amsmath,amsthm,stmaryrd}
\usepackage{mathrsfs,mathtools}
\biboptions{sort&compress}

\usepackage{subcaption}{}
\usepackage{bm,color}
\usepackage{algorithm}
\usepackage{algpseudocode}
\usepackage{siunitx}
\usepackage{booktabs}
\usepackage{cuted}

\newcommand{\R}{\mathbb{R}}
\newcommand{\de}{\partial}
\newcommand{\vc}[1]{\boldsymbol{#1}}  %for vectors
\newcommand{\vt}[1]{\mathsf{#1}}  %for tensors
\newcommand{\tsp}{\mathsf{T}}  %for transpose

\newcommand{\tr}{\operatorname{\mathrm{tr}}}

\newcommand{\DD}{\nabla^S\vc{u}}

\newcommand{\I}{{\vt{I}}}

\newcommand{\Adv}[1]{\mathcal{D}_{#1}}

\newcommand{\taur}{\tau_\mathrm{r}}
\newcommand{\T}{{\vt{T}}}
\newcommand{\Tel}{\T_\mathrm{el}}

\newcommand{\gd}{\dot{\gamma}}

\newcommand{\etaeff}{\eta_\mathrm{eff}}

\newcommand{\Tvi}{\T_\mathrm{vi}}

\newcommand{\Bel}{\vt{B}_{\mathrm{el}}}

\newcommand{\Wi}{\mathit{Wi}}
\newcommand{\El}{\xi}
\newcommand{\Rey}{\mathit{Re}}

\newcommand{\f}{\frac}
\newcommand{\del}{\partial}
\newcommand{\nab}{\nabla}

\newcommand{\nn}{\nonumber}

\newcommand{\bs}{\boldsymbol}

\newcommand{\Belhat}{\hat{\vt B}_{\mathrm{el}}}

\newcommand{\uhat}{\hat{\vc{u}}}
\newcommand{\un}{\vc{u}^n}
\newcommand{\td}[1]{\tilde{#1}}

\newcommand{\Belhath}{\hat{\vt B}_{\mathrm{el}_h}}
\newcommand{\Belh}{{\vt B}_{\mathrm{el}_h}}

\usepackage[colorlinks=true,breaklinks=true,linkcolor=blue]{hyperref}
\usepackage{cleveref}
\crefname{figure}{Fig.}{Figs.}
\Crefname{figure}{Figure}{Figures}
\crefname{equation}{equation}{equations}

\usepackage{textgreek}
\makeatletter
\renewcommand*\env@matrix[1][\arraystretch]{%
  \edef\arraystretch{#1}%
  \hskip -\arraycolsep
  \let\@ifnextchar\new@ifnextchar
  \array{*\c@MaxMatrixCols c}}
\makeatother

\begin{document}
% \onehalfspacing

\begin{frontmatter}
\title{A numerical study to analyze the interplay of Weissenberg number and viscosity ratio in a log-strain tensorial model for viscoelastic fluids}

%% use optional labels to link authors explicitly to addresses:
\author[label1]{Nehal Dash}
\ead{nehaldash.dash@gmail.com}
\affiliation[label1]
        {organization={Department of Mathematics ``Tullio Levi-Civita'', University of Padova},
        % addressline={Via Trieste, 63},
        city={Padova},
        postcode={35121},
        % state={},
        country={Italy}}
        
\author[label3,label4]{Ramon Codina}
\ead{ramon.codina@upc.edu}
\affiliation[label3]
        {organization={Department of Civil and Environmental Engineering, Polytechnic University of Catalonia},
        % addressline={Jordi Girona 1-3, Edifici C1},
        city={Barcelona},
        postcode={08034},
        % state={},
        country={Spain}}
        
\affiliation[label4]
        {organization={International Centre for Numerical Methods in Engineering},
        % addressline={Gran Capità S/N},
        city={Barcelona},
        postcode={08034},
        % state={},
        country={Spain}}
        
\author[label1,label2]{Giulio G. Giusteri\corref{cor1}}
    \ead{giulio.giusteri@unipd.it}
    \cortext[cor1]{Corresponding author.}
\affiliation[label2]{
            organization={Gruppo Nazionale per la Fisica Matematica, Istituto Nazionale di Alta Matematica ``Francesco Severi''},
            city={Sezione di Padova},
            country={Italy}}

%% ======== Abstract ===============
\begin{abstract}

We present a computational study aimed at exploring the different and independent roles of the Weissenberg number and of the ratio between the polymeric and solvent viscosity contributions in a viscoelastic fluid model.
The tensorial model under investigation, recently proposed, is based on a logarithmic relation between the elastic (or recoverable) strain and the elastic stress. 
In this model, the elastic strain plays the role of a conformation tensor and its evolution equation inherently preserves its determinant and positive definiteness. 
These properties are also enforced in the computational method employed in the study.
A finite-difference discretization in time is combined with a stabilized mixed finite element formulation based on the Variational Multiscale method for the spatial discretization and with a generalized Lie derivative approach for the advection terms.
The behavior of the model is analyzed in paradigmatic pressure-driven flows and we find that the value of the viscosity ratio is crucial in determining to which extent non-Newtonian flow profiles are observed upon increasing the Weissenberg number.
By comparing the solutions of the log-strain tensorial model with those of a suitable Generalized Newtonian Fluid model, we show that flow-type dependence and memory play a significant role even in the simple planar flow past a cylinder.

\end{abstract}

%% ======== End of Abstract ===============

% =======   Graphical abstract ===============
% \begin{graphicalabstract}
% %\includegraphics{grabs}
% \end{graphicalabstract}

% =========   Research highlights ============
% \begin{highlights}

% \item The independent roles of the Weissenberg number and viscosity ratio are investigated in a log-strain viscoelastic model.

% \item The computational method combines stabilized mixed finite elements with a generalized Lie derivative formulation.

% \item The model behavior is explored in three benchmark planar flows. 

% \item Comparisons with a Generalized Newtonian Fluid model highlight viscoelastic effects beyond rate-dependent viscosity.

% \end{highlights}

%% Keywords

\end{frontmatter}

%% Add \usepackage{lineno} before \begin{document} and uncomment 
%% following line to enable line numbers
%% \linenumbers

%% main text
%%

%% Use \section commands to start a section

\section{Introduction}
\label{sec:intro}

The mechanical response of matter traditionally presents two limiting cases, Newtonian viscous fluids and elastic solids. The former deform irreversibly under applied stress, dissipating energy continuously, whereas the latter stores elastic energy and recovers its original configuration upon unloading. Viscoelastic fluids lie between these two extremes. Their response is governed by the instantaneous rate of deformation, the history of applied stresses, and the relaxation dynamics of the material microstructure. Therefore, they exhibit different responses depending on the deformation time scale. Such complex behavior is central to a wide range of engineering, medical, and natural science applications, where microstructural evolution governs the macroscopic flow dynamics. Viscoelastic fluids exhibit distinctive phenomena, including turbulent drag reduction \cite{graham2004drag}, elastic turbulence \citep{groisman2000elastic} at low Reynolds numbers, and viscoelastic flow instabilities \citep{datta2022perspectives, sanchez2022understanding}, all arising from the interplay between elastic stress storage and viscous dissipation in evolving flows. The development of viscoelastic constitutive models is essential to provide descriptive and predictive computational tools for a variety of relevant applications.

Viscoelastic models in continuum mechanics have their roots in the classical works of Maxwell, and were later placed on a more rigorous mathematical foundation by Oldroyd through the introduction of an invariant constitutive framework \citep{oldroyd1950formulation, oldroyd1958non}. This development has since served as a basis for a wide class of viscoelastic modeling approaches. Many of these models, such as those by \citet{giesekus1966elastizitat, giesekus1982simple} and \citet{thien1977new}, are based on the tensorial structure of the governing equations and have proven useful for fitting experimental results. Differential models couple the flow equations with evolution equations for the elastic stress or conformation tensor to describe memory effects and time-dependent material response under deformation. These approaches include extra-stress models, such as the upper-convected Maxwell, Oldroyd-B, and corotational Maxwell models, as well as conformation-tensor formulations, including the log-conformation \citep{fattal2004constitutive}, square-root conformation \citep{balci2011symmetric}, and kernel-conformation methods \citep{afonso2012kernel}. 
In spite of its long history and the presence of several well-established methods \citep{bird1986dynamics, phan2003understanding, alves2021numerical,shitikova2022models}, viscoelastic fluid modeling is still a very active research field \citep{renardy1988recent, zhou2020viscoelastic,  renardy2021mathematician, beris2021continuum,stephanou2014continuum,ferras2019generalised}.

Recently, \citet{alrashdi2024evolution} introduced a class of tensorial continuum models for viscoelastic fluids that interpolate between purely viscous and solid-like behavior. The formulation, built upon concepts of solid plasticity, incorporates logarithmic stress-strain relations in which a central role is played by the Hencky strain. This class of models, even in its simplest form with constant material parameters, was shown to qualitatively reproduce several experimental observations in simple shear and extensional flows of viscoelastic fluids, including dependence on the deformation rate and on the flow type.
Despite having the same number of parameters as the upper-convected Maxwell model, the log-strain model presents features comparable to those of the Giesekus model \citep{alrashdi2024evolution}.

The present work focuses on further exploring the predictions of the simplest log-strain model by applying a stabilized mixed finite element formulation to the simulation of benchmark problems in viscoelastic fluid mechanics.
A key feature of the model is the presence of three independent dimensionless parameters: the Reynolds number $\Rey$, the Weissenberg number $\Wi$, and the viscosity ratio $\El$. 
Both the $\Rey$ and $\Wi$ are defined in the standard way, while $\El$ is the ratio between the polymeric contribution to the viscosity, product of elastic modulus times the relaxation time, and the solvent viscosity. Precise definitions of these dimensionless quantities are given in Section~\ref{sec:model}. The presence of multiple dimensionless parameters motivates a systematic study of the interplay between inertia, elasticity, and dissipation in viscoelastic flows. 
We explore regimes with low $\Rey$ to emphasize the effects of $\El$ and $\Wi$. We stress the importance of analyzing the effect $\El$, often neglected in studies focused only on the role of $\Wi$.
Furthermore, to clearly identify the role of viscoelastic memory, as opposed to simpler inelastic approximations, the log-strain model is compared with a Generalized Newtonian Fluid (GNF) model in which the rate-dependent viscosity is deduced from a fitting of the effective shear viscosity of the log-strain model.

It is worth noticing that numerical simulations at high $\Wi$, where elastic effects become dominant, can encounter substantial difficulties. This phenomenon is referred to as the \textit{High Weissenberg Number Problem} (HWNP) \citep{owens2002computational}, which has been a significant challenge in computational rheology since its origin. The difficulties are further exacerbated by the strong coupling of velocity and stress evolutions, the presence of nonlinear advective terms, and the need to enforce the incompressibility constraint. We can identify two main sources of the HWNP. These are the loss of positive definiteness of the conformation tensor (an internal variable that must remain symmetric and positive definite to ensure physical admissibility) and numerical breakdown in regions of very high deformation rates \citep{hulsen1997simulation,fattal2004constitutive, fattal2005time}. 
To overcome this, \citet{fattal2004constitutive} introduced the log-conformation formulation. {This reformulation of the standard viscoelastic equations allows to represent large values of the conformation tensor and of the associated elastic stress by means of a quantity that grows much more slowly and enables simulations over a broader range of $\Wi$.} Apart from the log-conformation reformulation, several alternative numerical approaches have been proposed to enhance stability and accuracy, including square-root conformation representations \citep{balci2011symmetric}, matrix kernel-transformation methods \citep{afonso2012kernel}, Discontinuous Galerkin elements implementations with log-conformation and upwind stabilization \citep{hulsen2005flow}, and formulations based on linear interpolation of the convective term \citep{coronado2007simple}, reflecting the continued efforts to treat high $\Wi$ flows.

To address these computational challenges, stabilized techniques have proven very effective as they eliminate the necessity of satisfying the technical inf–sup condition and allow for the use of simpler finite element spaces. One such approach is the Variational Multiscale (VMS) framework, which provides a systematic stabilization strategy and enables the use of equal-order, piecewise linear interpolation for all primary variables. The foundational ideas were first introduced for scalar convection-diffusion-reaction problems by \citet{HUGHES19983}. This was later extended to the Navier–Stokes equations through a series of works \citep{codina2000stabilization, codina2002stabilized, codina2008analysis} where the sub-grid scale space is assumed to be orthogonal to the finite element space. Subsequently, this framework has been applied to mixed and multiphysics formulations, including three-field Navier--Stokes systems \citep{castillo2014stabilized}, viscoelastic flow models \citep{castillo2014variational, castillo2017numerical, barrenechea2019time}, and Oldroyd-type fluids \citep{kwack2010three, kwack2017stabilized}. Motivated by these developments, the present study adopts a VMS-based stabilized finite element formulation to ensure numerical stability and robustness across a wide range of parameter regimes. With this approach we perform a systematic investigation of the flow behavior reaching relatively high values of $\Wi$ in different pressure-driven flows. {The maximum value of $\Wi$ that we simulated ranges from $7$, for the flow past a cylinder, to $35$, in a straight-channel flow.}

The article is organized as follows: \Cref{sec:model} introduces the log-strain tensorial model for viscoelastic fluids.
In \Cref{sec:computational framework}, we present the computational method with our discretization and solution strategies.
The numerical results are presented and discussed in \Cref{sec:numerical results}. Specifically, we consider planar pressure-driven flows in a straight channel (\Cref{subsec:channel}), past a cylinder (\Cref{subsec:fpc}), and through a 4:1 contraction (\Cref{subsec:contraction}).
The comparison of the full model with the GNF model is presented in \Cref{subsec:compare_GNF}. Conclusions are drawn in \Cref{sec:conclusions}, along with directions for future work.

\section{A logarithmic strain-based viscoelastic model}
\label{sec:model}

Let us begin by presenting the equations governing the flow of a viscoelastic fluid moving in a domain $\Omega$ of $\R^d$ ($d$ = $2,\,3$) with a boundary $\de \Omega$ and in the time interval $[0,\,t_{\rm f}]$.
Considering an incompressible and isothermal flow, the governing equations are the conservation of momentum and mass, expressed as
\begin{align}
    \rho \f{\del \vc{u}}{\del t}+\rho (\vc{u} \cdot \nab) \vc{u} - \nab \cdot \T +\nab p &= \vc{f} \:\ {\rm in} \:\ \Omega\times (0,\,t_{\rm f}),\label{e:consermom}\\
    \nab \cdot \vc{u} &=0 \:\ {\rm in} \:\ \Omega\times (0,\,t_{\rm f}).\label{e:consermass}
\end{align}
Here, $\rho$ represents the constant density, $p:\Omega \times (0,\,t_{\rm f}) \rightarrow \R$ is the pressure field, $\vc{u}:\Omega \times (0,\,t_{\rm f}) \rightarrow \R^d$ is the velocity field, $\vc{f}:\Omega \times (0,\,t_{\rm f}) \rightarrow \R^d$ is the force field and $\T:\Omega \times (0,\,t_{\rm f}) \rightarrow \mathrm{Mat}_d(\R)$ denotes the deviatoric extra stress tensor. In general, the extra stress tensor $\T$ can be decomposed into the sum of viscous $\left(\Tvi\right)$ and elastic $\left(\Tel\right)$ contributions as
\begin{align}
\T=\Tvi + \Tel.
\end{align}
We assume a linear viscous contribution given by $\Tvi=2\eta \nabla^S\vc u$, where $\eta \geq 0$ denotes the solvent viscosity of the fluid, and $\nabla^S\vc u := \left(\nab \vc{u} +\nab {\vc{u}}^{\tsp}\right)/2$ is the rate-of-strain tensor represented by the symmetric part of the velocity gradient.

To complete the viscoelastic flow problem, we must give the constitutive equation for the elastic stress {and in this we follow the proposal by \citet{alrashdi2024evolution}. The constitutive prescription for the elastic stress $\Tel$ is
\begin{equation}\label{e:Teldef}
    \Tel=\kappa \log \Bel,
\end{equation}
with $\kappa> 0$ an elastic modulus and $\Bel$ the elastic (recoverable) left Cauchy--Green strain tensor.} 

{Equation \eqref{e:Teldef} corresponds to the classical Hencky model for an incompressible material in nonlinear elasticity. Compared to most of the other nonlinearly elastic models, and besides providing a good fit for several experimental tests, the Hencky stress in the compressible case has the distinct advantage of preventing the unphysical state of total compression. It is then reasonable to employ this model also in the incompressible case, where it gives internal forces that are tangential to the manifold of incompressible deformations without the need for additional Lagrange multipliers.}

{\citet{alrashdi2024evolution} have shown that, by assuming an evolution of the relaxed material state with a rate that is proportional to the Hencky stress, the evolution of the elastic strain tensor $\Bel$ is governed by}
\begin{align}
    \f{\de\Bel}{\de t}+(\vc{u}\cdot \nab)\Bel - \nab \vc{u} \cdot \Bel - \Bel \cdot \nabla \vc{u}^\tsp=-\f{1}{\taur}\Bel \cdot \log \Bel.
\label{e:goveqBel}
\end{align}
In our conventions, the components of the velocity gradient are given by
\[
(\nabla\vc{u})_{ik}=\frac{\de u_i}{\de x_k}.
\]
In this way, we see that the left-hand side of \cref{e:goveqBel} is the upper-convected derivative of $\Bel$, while the right-hand side is the term responsible for the relaxation of $\Bel$ towards the identity matrix.
Such a nonlinear relaxation term has three important features:
\begin{itemize} 
\item it vanishes in the absence of elastic strain, that is when $\Bel$ equals the identity matrix $\I$;
\item it entails an exponential stress relaxation in step-strain experiments with relaxation time $\taur$;
\item it preserves the symmetry and the determinant of $\Bel$ during the evolution, so we can guarantee that $\det\Bel=1$, $\tr(\log\Bel)=0$, and $\Bel$ is symmetric positive definite at all times.
\end{itemize}

Substituting the definition of the extra stress tensor $\T$ into the momentum equation results in the set of the governing equations for the current log-strain viscoelastic model:
\begin{align}
    &\rho \f{\del \vc{u}}{\del t}+\rho (\vc{u} \cdot \nab) \vc{u} - 2\eta \nab\cdot(\nab^S\vc{u}) - \kappa\nab \cdot \log\Bel + \nab p = \vc{f},\label{e:consermom2} \\
    & \hspace{4cm} \nab \cdot \vc{u} = 0, \label{e:consermass2} \\
    &\f{\de\Bel}{\de t}+(\vc u \cdot \nab)\Bel-\nab \vc{u} \cdot \Bel-\Bel \cdot \nabla \vc{u}^\tsp +\f{1}{\taur}\Bel \cdot \log \Bel = \vt{0}.
\label{e:visfluid1}
\end{align}

\citet{alrashdi2024evolution} have also shown that, in the linearized limit of small elastic stresses and strains, the log-strain model recovers the upper-convected Maxwell model and the first nonlinear correction gives a Giesekus model, as can be readily seen by substituting the approximation $\log\Bel\approx \Bel-\I$ in \cref{e:visfluid1,e:Teldef}.

\subsection{Relevant dimensionless quantities}
The material parameters involved in the log-strain models are the mass density $\rho$, the solvent viscosity $\eta$, the elastic modulus $\kappa$, and the relaxation time $\taur$. By introducing characteristic length $L$ and velocity $U$, one can define several dimensionless numbers.
Besides the classical Reynolds number $\Rey=\rho U L/\eta$ that compares inertial and viscous forces and the Weissenberg number $\Wi=U\taur/L$ that compares the relaxation time $\taur$ with the inverse of the shear rate $\gd=U/L$, we can introduce the ratio
\begin{equation}
\El = \f{\kappa \taur}{\eta},
\end{equation}
that compares the polymeric contribution to the effective viscosity, as given by $\eta_p=\kappa\taur$, with the solvent viscosity $\eta$. 

The three numbers $\Rey$, $\Wi$, and $\El$ constitute a complete set of independent dimensionless quantities.
From these one could also define a ratio $\zeta=\kappa/(\rho U^2)=\El/(\Rey\Wi)$ that compares elastic and inertial effects, essentially taking the ratio between the speed of elastic waves and the characteristic velocity.
From the set of dimensionless quantities just introduced, we see that in a situation where inertia is negligible, we still have two independent effects to be investigated. It is not $\Wi$ alone to determine the properties of the flow, but we need to carefully consider the role of $\El$ as well.

{As a general caveat regarding the comparison of the effects of varying $\Wi$ across different viscoelastic models, we stress that, while the definitions as product of characteristic shear rate and relaxation time seem identical, the difference in constitutive laws may associate to similar values of $\mathit{Wi}$ elastic stresses of different magnitudes.
In particular, this could change significantly various instability thresholds, both physical and numerical.}

\subsection{{Rheometric characterization}}

\begin{figure*}[t]
    \centering
    \begin{subfigure}[b]{0.49\textwidth}
        \caption{\label{f:sstress}}
        \includegraphics[width=\textwidth]{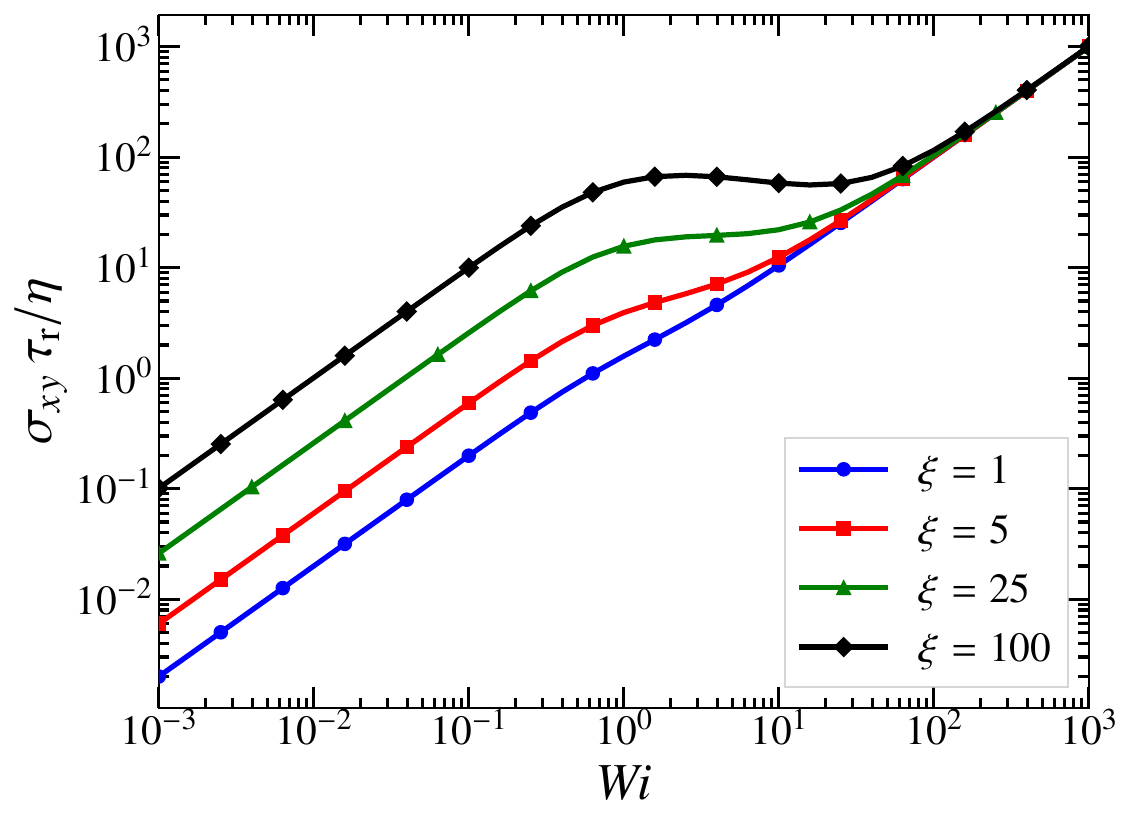}
    \end{subfigure}
    \hfill
    \begin{subfigure}[b]{0.49\textwidth}
        \caption{\label{f:nstressdiff}}
        \includegraphics[width=\textwidth]{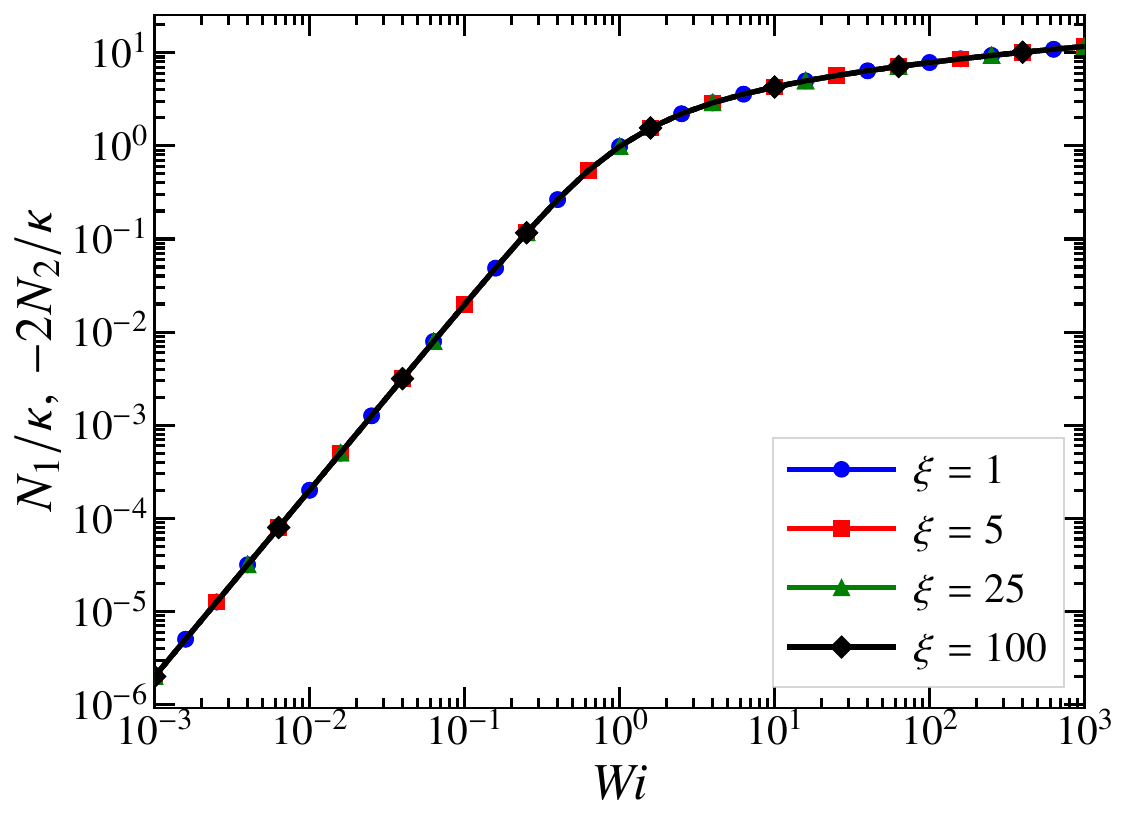}
    \end{subfigure}
    \caption{{Steady state rheological response of the log-strain model: ($a$) rescaled shear stress $\sigma_{xy}\taur/\eta$ as a function of the Weissenberg number $\Wi$, and ($b$) rescaled normal stress differences $N_1/\kappa$ and $-2N_2/\kappa$ as functions of $\Wi$ for different values of the viscosity ratio $\xi$.}}
    \label{f:shear_normal_stress}
\end{figure*}

{As a preliminary characterization of the log-strain model relevant for the subsequent discussion, we present the material functions obtained in steady simple shear by varying the shear rate and the corresponding Weissenberg number $\Wi$ (see \cref{f:shear_normal_stress}).}

{The shear stress increases linearly at small $\Wi$, with an effective viscosity given by $\eta_\mathrm{eff}=\eta+\kappa \tau_\mathrm{r}=\eta(1+\xi)$. Then, we observe a shear-thinning transition followed by the asymptotic growth determined by the solvent viscosity $\eta$. The flow curve remains monotonic up to $\xi\sim 40$. Beyond this value, the appearance of a non-monotonic behavior opens the way to shear-banding phenomena. Our numerical investigation will be limited to regimes such that the flow curve is monotonic and no shear banding occurs.}
{On the other hand, normal stress differences always grow monotonically with $\Wi$. They depend linearly on $\kappa$ and $\xi$, as they are unaffected by the solvent viscosity $\eta$. Moreover, we have $N_2=-N_1/2$ independently of $\xi$.
For small $\Wi$ we find the expected quadratic growth, hindered at larger $\Wi$.}

{An analytical solution \cite{alrashdi2024evolution} shows that the extensional flow behavior is completely characterized by a rate-independent effective viscosity given again by $\etaeff=\eta(1+\xi)$. This entails a flow-type dependence of the rheological response that is more and more significant upon increasing $\xi$.}

\section{Computational method}
\label{sec:computational framework}

The governing equations of the viscoelastic flow problem can be expressed in compact form by calling $\vc{U}=[\vc{u},\,p,\,\Bel]^\tsp$, $\vc{F}(\Belhat)=[\vc{f}+\kappa \nab\cdot\log\Belhat,\,0,\,\vt{0}]^\tsp$ and defining
\allowdisplaybreaks
\begin{align}
    \Adv{t}(\vc{U};\uhat)&:=
    \begin{pmatrix}[1.5]
    \rho \cfrac{\de \vc{u}}{\de t}\\
    0 \\
    \cfrac{\de\Bel}{\de t}
    +\left(\uhat \cdot \nab\right) \Bel - \nab\uhat \cdot \Bel - \Bel \cdot \nab \uhat^\tsp
    \end{pmatrix},\\
    \mathcal{L}_0(\vc{U};\uhat,\Belhat)&:=
    \begin{pmatrix}[1.5]
    \rho\left(\uhat \cdot \nab\right) \vc{u} - 2 \eta \nab \cdot (\nab^S \vc u) + \nab p\\
    \nab \cdot \vc{u}\\
     \cfrac{1}{\taur}\, \Bel \cdot \log\Belhat
    \end{pmatrix}.
\label{e:L0}
\end{align}
We may then write the nonlinear governing equations (\ref{e:consermom2}--\ref{e:visfluid1}) as
\begin{align}
    \Adv{t}(\vc{U};\vc{u}) + \mathcal{L}_0(\vc{U};\vc{u},\Bel) = \vc{F}(\Bel).
\end{align}
In the definition of $\Adv{t}$, $\mathcal{L}_0$, and $\vc{F}$ we introduced $\uhat$ and $\Belhat$ to identify some instances of the velocity and strain fields that will play a specific role in subsequent linearization procedures.

These equations must be complemented by initial and boundary conditions to fully define the problem. On the solid boundaries, the standard no-slip condition $\vc{u}=0$ is imposed on $\del \Omega$ for all times $t$. Along the open boundaries suitable periodic or traction conditions are prescribed depending on the specific flow geometry. 
As will become clear in the discussion of the method, no boundary conditions are prescribed for the tensor $\Bel$. Instead, its evolution is entirely determined by the prescribed initial conditions and the velocity field. The specific boundary conditions adopted for each numerical example are described in the corresponding sections.
The problem is fully specified by the initial conditions for the velocity $\vc{u}$ and the variable $\Bel$, such that at time $t = 0$, $\vc{u} = \vc{u}^0$ and $\Bel = \Bel^0$, where $\vc{u}^0$ and $\Bel^0$ are functions defined over the entire domain $\Omega$.

\subsection{Temporal discretization and generalized Lie derivative formulation}
\label{subsec:temporaldiscretization}

To solve the governing coupled viscoelastic equations while avoiding the steep computational expense of a fully monolithic nonlinear solver at each time step, a semi-implicit first-order Backward Euler scheme is employed. The time interval $[0, t_f]$ is partitioned into a uniform grid of time steps $\delta t$, where $t^n = n\delta t$. For a generic time-dependent variable $f$, the temporal derivative at the current time $t^{n+1}$ is approximated via the backward difference such that
\begin{equation}
    \left.\f{\del f}{\del t}\right|_{t^{n+1}} \approx \f{f^{n+1} - f^n}{\delta t},
\end{equation}
where $f^{n+1}$ represents the unknown field to be solved at the current time, and $f^n$ denotes the known solution from the previous time step. 
The time-marching scheme is adopted here as a numerical tool to obtain the steady-state solution, without explicit consideration of transient dynamics.

%In the present implementation, a semi-implicit Backward Euler scheme is employed for temporal discretization. 
%%% Already said
The velocity-pressure system is solved implicitly at each time step, while nonlinear terms are linearized using known quantities from the previous iteration. To decouple the primary kinematic fields $(\vc{u}^{n+1}, p^{n+1})$ from the explicit nonlinearities of the strain tensor, the convective and constitutive relaxation terms are linearized semi-implicitly. The advecting velocity field $\uhat$ within the momentum and constitutive equations is treated explicitly using the velocity field from the previous time level such that $\uhat = \vc{u}^n$, and therefore 
\begin{equation}
    (\vc{u} \cdot \nabla) \vc{u} \approx (\uhat \cdot \nabla)\vc{u}.
\end{equation}
Similarly, the nonlinear term involving the logarithm of the strain tensor $\Bel$ is linearized by evaluating its logarithmic component using the known state from the preceding step such that 
\begin{equation}
    \log\Bel \approx \log\Belhat,
\end{equation}
where $\Belhat = \Bel^n$.
Consequently, the fully coupled problem is reduced to a staggered sequence of linear algorithmic blocks solved consecutively for the updated velocity, pressure, and strain tensor fields at each increment.

The evolution equation for the strain tensor $\Bel$ is treated using a generalized Lie derivative formulation \citep{medeiros2021second,ashby2025discretisation,lee2011global, lee2006new, lee2011stable}. For a tensor field $\Bel$ that is advected by a velocity field $\vc{u}$, the Lie derivative $\mathcal{L}_{\vc u}(\cdot)$ effectively captures material advection, rotation, and stretching, and is defined as
\begin{align}
\mathcal{L}_{\vc u}\left(\Bel\right) = \left(\vc{u} \cdot \nab\right) \Bel - \nab \vc{u} \cdot \Bel - \Bel \cdot \nab \vc{u}^\tsp.
\end{align}
In the numerical implementation, the advective contribution is not discretized using a standard Eulerian approximation. Instead, a semi-Lagrangian characteristic mapping is employed. This approach ensures frame-invariance and enhances stability. At time $t^n$, the tensor field is updated by first pulling back the configuration along the flow trajectory dictated by the linearized velocity field $\uhat = \vc{u}^{n}$ to a shifted coordinate $\mathbf{x}^* = \mathbf{x} - \delta t\,\hat{\mathbf{u}}$. The purely advected tensor state is subsequently sampled at the shifted coordinate corresponding to the upwind characteristic position, evaluated as $\Bel^* = \Bel^n(\mathbf{x}^*)$.
By substituting this characteristic mapping into the temporal update, the final discrete constitutive update for $\mathbf{B}^{n+1}$ is given by 
\begin{align}\label{e:Lie_der}
\Bel^{n+1} + \f{\delta t}{\taur} \left(\Bel^{n+1} \cdot \log\Bel^n\right) = \Bel^* &+ \delta t \left( \nab \vc{u}^n \cdot \Bel^* + \Bel^* \cdot \nab {\vc{u}^n}^\tsp \right) \nn \\
&+ \delta t^2 \left( \nab \vc{u}^n \cdot \Bel^* \cdot \nab {\vc{u}^n}^\tsp \right).
\end{align}

As we shall see below, the incompressibility constraint $\nabla\cdot\vc u=0$ is weakly enforced in our method. 
Since this is important in guaranteeing that the strain tensor $\Bel$ remains in the physically admissible set of symmetric positive definite tensors, at each increment we apply the projection  
\begin{equation}
\Bel^{n+1} \longmapsfrom \f{\f{1}{2}\left(\Bel^{n+1} + {\Bel^{n+1}}^\tsp\right)}{\left[\det\left(\f{1}{2}\left(\Bel^{n+1} + {\Bel^{n+1}}^\tsp\right)\right)\right]^{1/d}}.
\end{equation} 
{The numerical error on the unit value of the determinant and on the symmetry of $\Bel$ depends naturally on the time step. By performing the projection at each iteration we ensure that the typical size of the correction remains always of the order of $10^{-5}$.}
This strategy leads to a consistent numerical discretization in time of the constrained evolution of $\Bel$.
Now, the time-discretized form of the governing equations can be written compactly as
\begin{align}
    \Adv{\delta t}^*(\vc{U};\vc{u}^n) + \mathcal{L}_0(\vc{U};\vc{u}^n, \Bel^n) = \vc{F}(\Bel^n),
\end{align}
where the time-discretized operator $\Adv{\delta t}^*$ is defined by 
\begin{align}
    \Adv{t}^*(\vc{U}^{n+1};\un)
    ={}&
    \begin{pmatrix}[1.5]
    \rho\, \cfrac{\vc{u}^{n+1}-\vc{u}^n}{\delta t}\\
    0 \\
    \cfrac{\Bel^{n+1}-\Bel^*}{\delta t}
    - \Big[\nab \vc{u}^n \cdot \Bel^*
    + \Bel^* \cdot \nab {\vc{u}^n}^\tsp \\
    \begin{aligned}
    &+\delta t\big(
    \nab \vc{u}^n \cdot \Bel^*
    \cdot \nab {\vc{u}^n}^\tsp
    \big)\Big]
    \end{aligned}
    \end{pmatrix},
\end{align}
and the spatial operator $\mathcal{L}_0$ retains its definition from \cref{e:L0}. 
For ease of notation, the superscript $(n+1)$ is omitted throughout the remainder of this work, with the understanding that unindexed variables refer to the unknown solution at the current time. Quantities evaluated at the previous time step $t^n$ are denoted with a superimposed hat, that is $\hat{(\cdot)} \equiv (\cdot)^n$.

\subsection{Variational formulation and spatial discretization of the time-discretized problem}

Let us introduce the standard functional spaces required to formulate the weak form of our time-discretized problem.
The space of square-integrable functions over a domain $\omega$ is represented by $L^2(\omega)$. The space of functions whose distributional derivatives of the order up to $m\geq 0$ (integer) belong to $L^2(\omega)$ is denoted by $H^{m}(\omega)$.
The space $H_0^1(\omega)$ consists of functions in $H^1(\omega)$ that vanish on the boundary $\partial \omega$. The topological dual of $H_0^1(\omega)$ is denoted by $H^{-1}(\omega)$, with the duality pairing represented by $\langle \cdot,\, \cdot \rangle$. The $L^2$ inner product for scalars, vectors, and tensors within $\omega$ is indicated by $(\cdot,\cdot)_{\omega}$. 
The integral over $\omega$ of the product of two general functions is expressed as $\langle\cdot,\, \cdot \rangle_{\omega}$, with the subscript omitted when $\omega = \Omega$. Norms in a space $X$ are denoted by $\|\cdot\|_X$, except for $X=L^2(\Omega)$, where the subscript is omitted.

With these definitions, the finite element solution spaces for the velocity and pressure fields are defined as $\vc{\mathcal{V}}_0 = H_0^1(\Omega)^d$ and $\mathcal{Q} = L^2(\Omega)/\mathbb{R}$, respectively. The functional space for the symmetric strain tensor $\Bel$ is denoted by $\vc{\mathit{\Upsilon}}$, typically a Sobolev space of tensors satisfying the necessary regularity conditions. The weak form of the problem requires determining 
$\vc{U}=[\vc{u},\,p,\,\Bel]:(0, t_{\rm f})\rightarrow\vc{\mathcal{X}}:= \vc{\mathcal{V}}_0 \times \mathcal{Q} \times \vc{\mathit{\Upsilon}}$.
By testing the time-discretized momentum, mass conservation, and characteristic-mapped constitutive equations against the arbitrary test functions $\vc{V} = [\vc{v},\,q,\,\vc{\chi}] \in \vc{\mathcal{X}}$, the variational problem can be compactly formulated as:
\begin{align}
B(\vc{U}, \vc{V}; \uhat, \Belhat) = L(\vc{V};\uhat,\Belhat),
\label{e:vareq_discrete}
\end{align}
where
\begin{align}
    B(\vc{U}, \vc{V}; \uhat, \Belhat) &= \left( \rho \cfrac{\vc{u}}{\delta t}, \vc{v} \right) +
    \langle \rho \left( \uhat \cdot \nabla \right) \vc{u}, \vc{v} \rangle + \left( \cfrac{\Bel}{\delta t},\vc{\chi}\right) \nn \\
    &\quad + 2 \left( \eta \nabla^{S} \vc{u}, \nabla^{S}\vc{v} \right) - \left( p, \nabla\cdot\vc{v} \right) + \left( q, \nabla\cdot\vc{u} \right) \nn\\
    &\quad + \left( \cfrac{1}{\taur} \Bel \cdot \log\Belhat, \vc{\chi} \right), \label{e:B_discrete_weak}
    \end{align}
    and
    \begin{align}
    L(\vc{V}&;\uhat,\Belhat) = \left( \rho \cfrac{\uhat}{\delta t}, \vc{v} \right) - \kappa \left( \log\Belhat, \nabla^S \vc{v} \right)  + \langle \vc{f}, \vc{v} \rangle \nn\\
    &+ \left( \cfrac{\Bel^*}{\delta t} + \nabla \uhat \cdot \Bel^* + \Bel^* \cdot \nabla \uhat^\tsp + \delta t \left( \nabla \uhat \cdot \Bel^* \cdot \nabla \uhat^\tsp\right), \vc{\chi} \right).
\label{e:L_discrete_weak}
\end{align}

Let us start considering the spatial discretization using the standard Galerkin Finite Element (FE) approximation of the time-discretized variational problem.
Let $\mathcal{T}_h=\{\vc K\}$ represent a FE partition of the domain $\Omega$. The diameter of an element $\vc K\in \mathcal{T}_h$ is denoted by $h_K$ and the diameter of the partition is defined as $h=\text{max} \{h_K\,|\,\vc K\in \mathcal{T}_h\}$. 
From $\mathcal{T}_h$ we may construct the finite element spaces for the velocity $\left(\vc{\mathcal{V}}_h \subset \vc{\mathcal{V}}\right)$, the pressure $\left(\mathcal{Q}_h \subset \mathcal{Q}\right)$, and the strain $\left(\vc{\mathit{\Upsilon}}_h \subset \vc{\mathit{\Upsilon}}\right)$. 
Calling $\vc{\mathcal{X}}_h:=\vc{\mathcal{V}}_h \times \mathcal{Q}_h \times \vc{\mathit{\Upsilon}}_h$, the Galerkin FE approximation of the problem consists in finding $\vc{U}_h:\left(0,t_{\rm f}\right)\rightarrow \vc{\mathcal{X}}_h$, such that 
\begin{align}
    B\left(\vc{U}_h,\vc{V}_h;\,\uhat_h,\Belhath\right)=L\left(\vc{V}_h;\uhat_h,\Belhath\right),
\end{align}
for all $\vc{V}_h=[\vc{v}_h,\,q_h,\,\vc{\chi}_h] \in \vc{\mathcal{X}}_h$, and satisfying the appropriate initial conditions. 

In this study, the discrete spaces are constructed using equal--order continuous linear Lagrange elements ($P_1$--$P_1$) for both the velocity $\vc{\mathcal{V}}_h$ and pressure $\mathcal{Q}_h$ fields, and a piecewise constant discontinuous Galerkin ($DG_0$) approximation is chosen for the strain tensor space $\vc{\mathit{\Upsilon}}_h$. The velocity pressure interpolation indicated is not stable, and this is why we need to use the stabilized formulation described below.

\subsection{Implementation of a residual based stabilized finite element method}

In this work, the formulation associated with the velocity and pressure fields is stabilized using the Variational Multiscale (VMS) framework \citep{hughes1998variational, codina2000stabilization, codina2002stabilized, codina2008analysis}. This treatment eliminates the necessity of satisfying the inf--sup (Ladyzhenskaya--Babuška--Brezzi) condition.
However, the strain tensor field is retained in its standard Galerkin form. We observed that, for the test cases that we considered, VMS stabilization of the tensor field $\Bel$ was not necessary, as the temporal discretization and the treatment of its advective transport provide sufficient stability. Note that we are not interested in considering very small time steps, as we intend to reach the stationary solution through time stepping. Small values of $\delta t$ would possibly require additional stabilization mechanisms.

The VMS method splits the unknown $\vc{U}$ into two components. The first component, $\vc{U}_h$, is resolved within the FE space, while the second component, $\td{\vc{U}}$, represents the sub-grid scale. 
The sub-grid scale must be approximated to accurately capture its effects and ensure a stable formulation. 
The total number of degrees of freedom in this method remains consistent with that of the Galerkin method. Various strategies exist for approximating the sub-scale and defining the space in which it is operated. In this work, we implement the approach proposed by \citet{moreno2019logarithmic}. A brief overview of the stabilization framework is presented below; readers may refer to the cited work for a more comprehensive discussion.

Consider that $\mathcal{L}_0(\uhat,\,\Belhat;\,\cdot)$ is a \textit{linear} operator for a given velocity field $\uhat$. After introducing the sub-scale decomposition and performing integration by parts, the VMS method leads to the problem of finding the FE solution $\vc{U}_h:(0,\,t_{\rm f})\rightarrow \vc{\mathcal{X}}_h$ such that
\begin{align}
    B\left(\vc{U}_h,\vc{V}_h;\,\uhat_h,\Belhath\right)+\sum_K\bigg\langle \td{\vc{U}},\mathcal{L}^{*}(\uhat_h;\vc{V}_h)\bigg\rangle_K=L\left(\vc{V}_h;\uhat_h,\Belhath\right),
\label{e:VMS}
\end{align}
for all $\vc{V}_h \in \vc{\mathcal{X}}_h$, where $\mathcal{L}^{*}(\uhat,\Belhat;\,\cdot)$ is the formal adjoint operator of $\mathcal{L}_0(\uhat,\Belhat;\,\cdot)$, and is calculated as 
\begin{align}
\mathcal{L}^*\left(\uhat,\Belhat;\,\vc{V}\right):=
    \begin{pmatrix}[1.5]
    -\rho (\uhat \cdot \nab) \vc{v} - 2\eta\nab \cdot (\nab^{S}\vc{v}) - \nab q \\   
    -\nab\cdot \vc{v} \\
    \cfrac{1}{\taur} \vc{\chi} \cdot \log\Belhat
    \end{pmatrix}.
\label{e:L0adj}
\end{align}
Here, $\td{\vc{U}}$ is the sub-grid scale, which needs to be approximated without accounting for boundary conditions. The approximation we consider within each element is
\begin{align}
    \td{\vc{U}} = \bs{\alpha}\,\td{P}\, [\bs{F}(\Belhath)- \Adv{t}(\vc{U}_h;\uhat_h)-\mathcal{L}_0(\vc{U}_h;\uhat_h,\Belhath)],
\label{e:Usubgrid}
\end{align}
where $\td{P}$ is the $L^2$ projection onto the space of sub-grid scale and $\bs{\alpha}$ is a diagonal matrix computed within each element. 
The stabilization matrix $\bs{\alpha}$ is defined as
\begin{align}
    \bs{\alpha}=\text{diag}\left(\alpha_1\,\vc{I}_d,\,\alpha_2, \alpha_3 \vc{I}_{d\times d}\right),
\end{align}
with $\vc{I}_d$ the identity on vectors of $\R^d$, $\vc{I}_{d\times d}$ the identity on second order tensor and $\alpha_i$, $i=1,\,2,\, 3$ are stabilizing parameters for the velocity components, the pressure and the stress, respectively. The stabilization parameters $\alpha_1$ and $\alpha_2$ are computed as 
\begin{align}
\begin{split}
    \alpha_1= \left[c_1 \f{\eta}{h_1^2}+c_2 \f{\rho\, |\vc{u}_h|}{h_2}\right]^{-1}, \quad
    \alpha_2= \f{h_1^2}{c_1\alpha_1}.
\end{split}
\label{e:alpha}
\end{align}
As explained above, the treatment of the Lie derivative of tensor $\Belh$ provides enough stability, which is reflected by the fact that we can take $\alpha_3 = 0$.
In these expressions, $h_1$ represents a characteristic length calculated as the square root of the element area in the two-dimensional case and the cubic root of the element volume in the three-dimensional case. On the other hand, $h_2$ corresponds to the characteristic length calculated as the element length in the streamline direction.
The term $|\vc{u}_h|$ is the Euclidean norm of the velocity. The constants $c_i$, $i=1,\,2$, are algorithmic parameters in the formulation, with $c_1 = 4k^4$ and $c_2 = 2k^2$, where $k$ denotes the order of the finite element interpolation.

By substituting the sub--grid scale approximation \cref{e:Usubgrid} into the VMS formulation in \cref{e:VMS}, and using the definitions of  stabilization parameters $\bs{\alpha}$ in \cref{e:alpha} and the adjoint operator in \cref{e:L0adj}, the finite element problem can be expressed as follows: find $\vc{U}_h\in\mathcal{X}_h$ such that
\begin{align}
    B(\vc{U}_h,\vc{V}_h;\uhat_h,\Belhath) &+ S_1(\vc{U}_h,\vc{V}_h; \vc{u}_h) + S_2(\vc{U}_h,\vc{V}_h) \nn \\
    & = L(\vc{V}_h;\uhat_h,\Belhath) + R_1(\vc{u}_h;\vc{V}_h),
\label{e:VMS1}
\end{align}
for all test functions $\vc{V}_h \in \vc{\mathcal{X}}_h$, where
{\allowdisplaybreaks
\begin{align}
    S_1(\vc{U}_h,\vc{V}_h;\uhat_h) &=\sum_K \alpha_1\Big\langle \td{P} \Big[\rho\,\f{\vc{u}_h - \uhat_h}{\delta t}-2 \eta \nab \cdot (\nab^S \vc{u}_h) \nn\\
    &\quad +\rho \left(\uhat_h \cdot \nab\right) \vc u_h + \nab p_h\Big], \, \rho \left(\uhat_h \cdot \nab\right) \vc{v}_h \nn\\
    &\quad + 2 \eta \nab \cdot (\nab^S \vc{v}_h) + \nab q_h\Big\rangle_K,\\
    S_2(\vc{U}_h,\vc{V}_h)&=\sum_K \alpha_2 \Big \langle \td P \left[\nab \cdot \vc u_h\right],\nab\cdot\vc{v}_h \Big\rangle_K,\\
    R_1(\uhat_h;\,\vc{V}_h)&=\sum_K \alpha_1 \Big\langle \td P[\vc{f}+\kappa \nab\cdot \log\Belhath], \,\rho \left(\uhat_h\cdot\nab \right) \vc{v}_h \nn\\
    &\quad + 2 \eta \nab \cdot (\nab^S \vc{v}_h) + \nab q_h \Big\rangle_K.
\end{align}
}In these equations, $\td P$ represents the projection operator, which is restricted to the appropriate space and applied to the components of the finite element residual $\vc{R}_h:=\vc{F}-\mathcal{L}(\vc{U}_h;\Belhath)$. The next step is to define this projection. Specifically, two choices are commonly considered, as outlined by \citet{castillo2014stabilized}. 
The first option is the Algebraic Sub-Grid Scale (ASGS) method, in which $\td P = I$ (the identity operator) when applied to the finite element residuals. The second option is the Orthogonal Sub-Grid Scale (OSGS) method, where $\td P = P_h^\perp = (I - P_h)$, with $P_h$ denoting the $L^2$ projection onto the corresponding finite element space. Regardless of the choice of the projection $\td P$, the method in \cref{e:VMS1} remains consistent, as the terms added to the Galerkin formulation are proportional to the finite element residual $\vc{R}_h$. 
Thus, these terms vanish when the finite element solution is replaced by the solution of the continuous problem $\vc{U}$. For this reason, the method in \cref{e:VMS1} is considered as a residual-based method.

For smooth solutions, the method described in \cref{e:VMS1} is stable and demonstrates optimal convergence for both cases of $\td P = I$ and $\td P =P^\perp_h$. As highlighted by \citet{castillo2014variational} and \citet{moreno2019logarithmic}, the OSGS method generally achieves higher accuracy, while the ASGS method is more cost-effective as projections are not needed and is more robust. 
Building on the approach presented by \citet{castillo2014variational} and \citet{moreno2019logarithmic}, we focus on the case where $\td P =P^\perp_h$ to employ a simplified method based on \cref{e:VMS1}. This simplification involves neglecting the cross local inner--product terms, along with other terms that do not contribute to stability. Following the considerations outlined in \citet{castillo2017finite} for constructing the Split OSGS stabilization in the traditional viscoelastic formulation, we present a method for the present viscoelastic problem: find $\vc{U}_h:(0,\,t_\text{f}) \to \mathcal{X}_h$ satisfying the appropriate initial conditions such that
\begin{align}
    B(\vc{U}_h,\vc{V}_h;\uhat_h,\Belhath) + S_1^\perp(\vc{U}_h,\vc{V}_h; \vc{u}_h) &+ S_2^\perp(\vc{U}_h,\vc{V}_h) \nn\\
    &= L(\vc{V}_h;\uhat_h,\Belhath)
\label{e:VMS2}
\end{align}
for all $\vc{V}_h \in\mathcal{X}_h$, where
\begin{align}
    S_1^\perp(\vc{U}_h,\vc{V}_h;\uhat_h) &= \sum_K \alpha_1\Big\langle P_h^\perp\Big[\nab p_h\Big],\nab q_h\Big\rangle_K \nn\\
    &\quad +\sum_K \alpha_1\Big\langle P_h^\perp\Big[\rho\left(\uhat_h \cdot \nab\right) \vc{u}_h\Big],\rho \left(\uhat_h \cdot \nab\right) \vc{v}_h\Big\rangle_K, \\
    S_2^\perp(\vc{U}_h,\vc{V}_h)&=\sum_K \alpha_2\Big\langle P_h^\perp \Big[\nab \cdot \vc{u}_h\Big],\,\nab \cdot \vc{v}_h\Big\rangle_K.
\end{align}
One may note that both the methods in \cref{e:VMS,e:VMS2} achieve an optimal convergence rate in $h$ for smooth solutions. 
However, previous studies have shown that \cref{e:VMS2} is more robust than \cref{e:VMS} in problems where the solution has strong gradients \citep{moreno2019logarithmic,castillo2017finite}. 
Therefore, \cref{e:VMS2} represents the final stabilized variational formulation used in our numerical solution of the viscoelastic flow problems.

\section{Numerical results and discussion}
\label{sec:numerical results}

In this section, we present our investigation of the dependence of solutions to the log-strain model equations \eqref{e:consermom2}--\eqref{e:visfluid1} on the kinematic and material parameters for three paradigmatic geometries.
All simulations employ as initial conditions $\vc u^0=\vc 0$ and $\Bel^0=\vt I$ and are carried out for sufficiently long times so as to reach a steady state. The reported results concern the velocity field $\vc u$ at steady state.
Flows are {two-dimensional and} driven by pressure gradients, as specified below.
The computational codes for these simulations have been developed using the open-source finite element library \texttt{FEniCSx}~\citep{baratta2023dolfinx}.

The first numerical example presented in \Cref{subsec:channel} studies the flow of a viscoelastic fluid in a straight channel geometry. By varying the average pressure gradient $C$ and the elastic modulus $\kappa$ we explore various combinations of $\Wi$ and $\El$. In \Cref{subsec:fpc}, we present the benchmark problem of a flow past a cylinder with the corresponding drag reduction upon increase of the effective strain rate. 
Next, in \Cref{subsec:contraction}, the 4:1 planar contraction flow is considered. In all these examples, we remain within regimes of $\Rey\ll 1$ (laminar flows) since our aim is to emphasize the interplay between $\El$ and $\Wi$. We note that the implemented stabilized mixed finite element formulation provides stable numerical solutions for the log-strain model even at relatively high values of $\Wi$ {($\sim 35$ in the straight channel, $\sim 7$ in the flow past a cylinder, and $\sim 8$ in the flow through a contraction)}.

To assess convergence in time and ensure that the reported results correspond to the steady-state regime, we monitor the $L^2$ norms of the increments of the field variables $\{\vc{u},\,p,\,\Bel\}$ between successive time steps. The solution is considered to have reached steady-state when these increment norms reach sufficiently low values, indicating  negligible temporal evolution. All simulations reported correspond to this fully converged steady-state regime. In this study, we consider fixed values, normalized to unity, for relaxation time $\taur$, density $\rho$, and solvent viscosity $\eta$ of the viscoelastic fluid and let the elastic modulus $\kappa$ and pressure gradient $C$ change to modify, respectively, $\El$ and $\Wi$.

\subsection{Viscoelastic fluid flow in a straight channel}
\label{subsec:channel}

In this subsection, a pressure-driven straight channel flow is considered as a benchmark to highlight some basic model predictions. The primary motivation for choosing this geometry is that it admits an analytical solution in the Newtonian limit (Poiseuille flow), which provides a reliable reference for comparing the numerical results. 
In particular, we compare the model predictions with the two limits of the Newtonian solvent-only case, with viscosity $\eta$, {and the low-shear-rate limit represented by the solution for an Oldroyd-B model with effective viscosity $\etaeff=\eta+\kappa\taur$ given by the zero-shear viscosity of the log-strain model}. This analysis allows a direct assessment of the consistency of the proposed formulation with known analytical behavior and provides a baseline for evaluating the influence of elasticity in the model.

\begin{figure}[]
    \centering
    \includegraphics[width=0.99
    \linewidth]{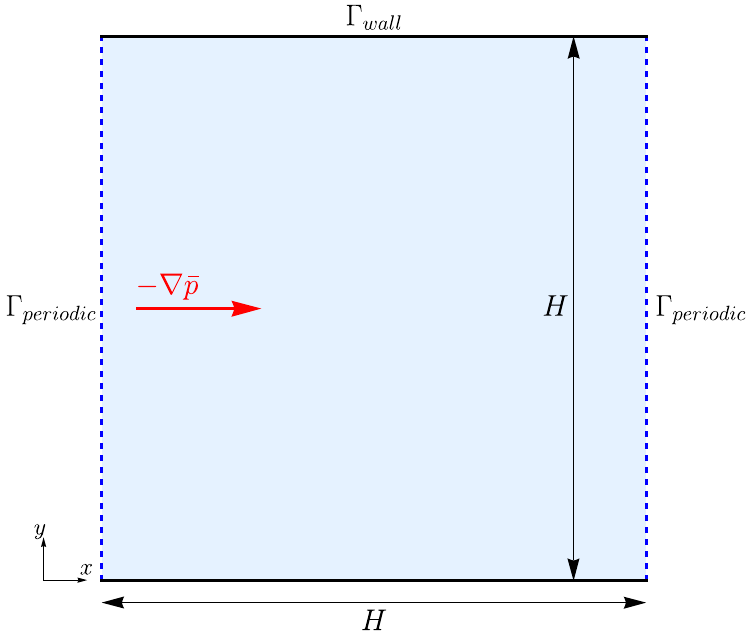}
    \caption{Geometry of the straight channel domain.}
    \label{f:channel}
\end{figure}

\subsubsection{Set up}

We consider a two-dimensional square domain defined by $\Omega = [0,H] \times [0,H]$, where $H$ denotes the width of the channel. The coordinates $x$ and $y$ represent the streamwise and transverse directions, respectively. The flow is confined between two parallel flat plates located at $y=0$ and $y=H$, driven by an externally imposed constant pressure gradient. The geometry of the domain is illustrated in \cref{f:channel}. 

No-slip boundary conditions are imposed on the channel walls $\Gamma_{wall}$, such that the velocity field satisfies $u = (u_x, u_y) = (0, 0)$ at $y = 0$ and $y = H$. In addition, the transverse (vertical) component of the velocity is constrained to vanish throughout the domain, $u_y=0$ in $\Omega$. The streamwise (horizontal) component of the velocity $u_x$ is left free and is determined by the balance of momentum under the imposed pressure gradient. 
Periodic boundary conditions are applied in the streamwise direction. Accordingly, for any field variable $\phi$ belonging to the set $\{\vc{u},\,p,\,\Bel\}$, we impose $\phi(0, y) = \phi(H, y)$ for all $y$ in $[0,H]$.
In our setting, $p$ represents the difference between the physical pressure and a linearly decreasing background pressure $\bar{p}=-C x$. The latter gives rise to the average pressure gradient with intensity $C$, which enters the flow equations as a uniform and constant body force acting in the streamwise direction.

\begin{figure}[t!]
    \centering
    \begin{subfigure}[b]{0.483\textwidth}
        \caption{\label{f:chfig_a}}
        \includegraphics[width=\textwidth]{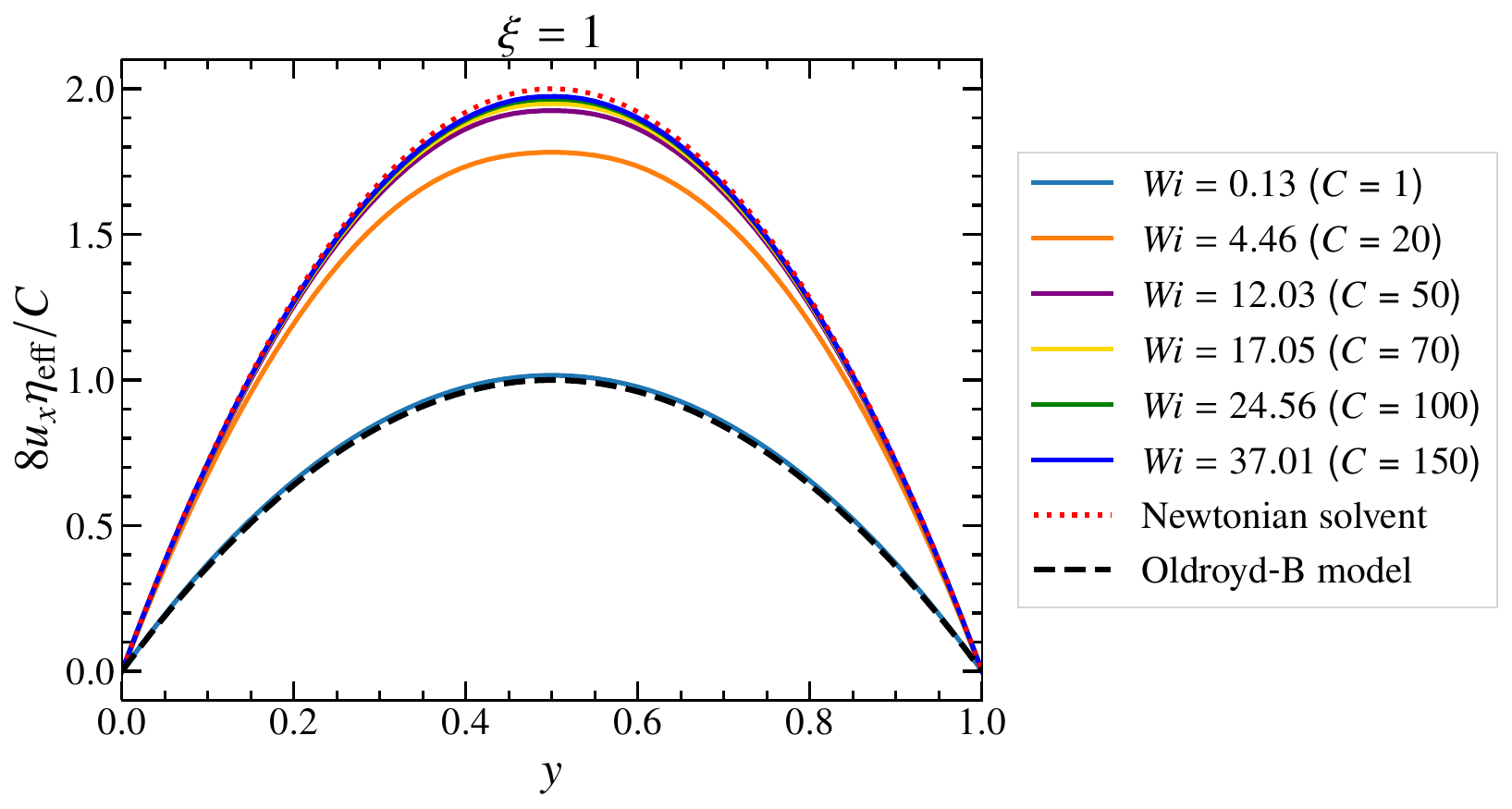}
    \end{subfigure}\\
    \begin{subfigure}[b]{0.483\textwidth}
        \caption{\label{f:chfig_b}}
        \includegraphics[width=\textwidth]{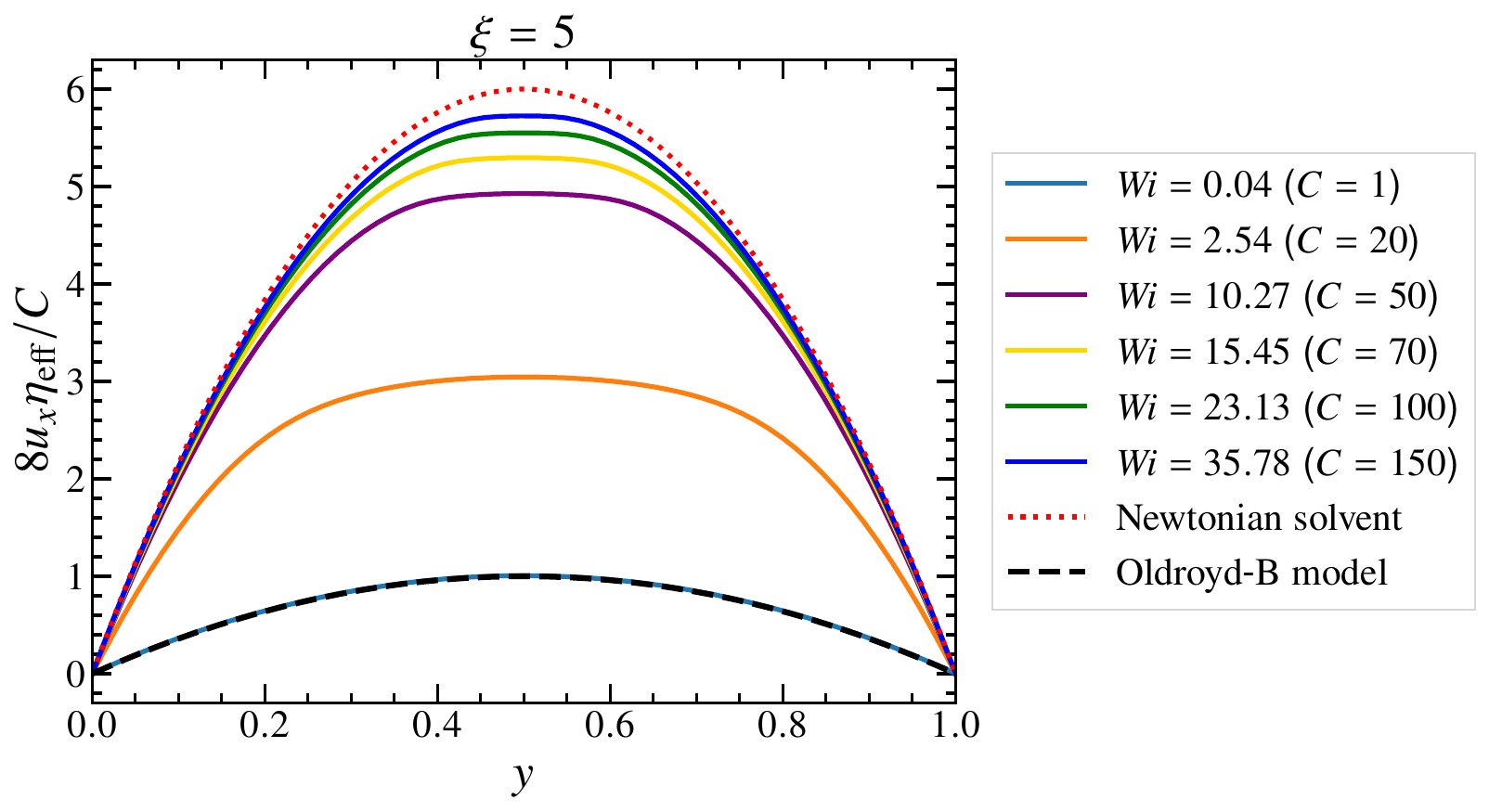}
    \end{subfigure}\\
    \begin{subfigure}[b]{0.483\textwidth}
        \caption{\label{f:chfig_c}}
        \includegraphics[width=\textwidth]{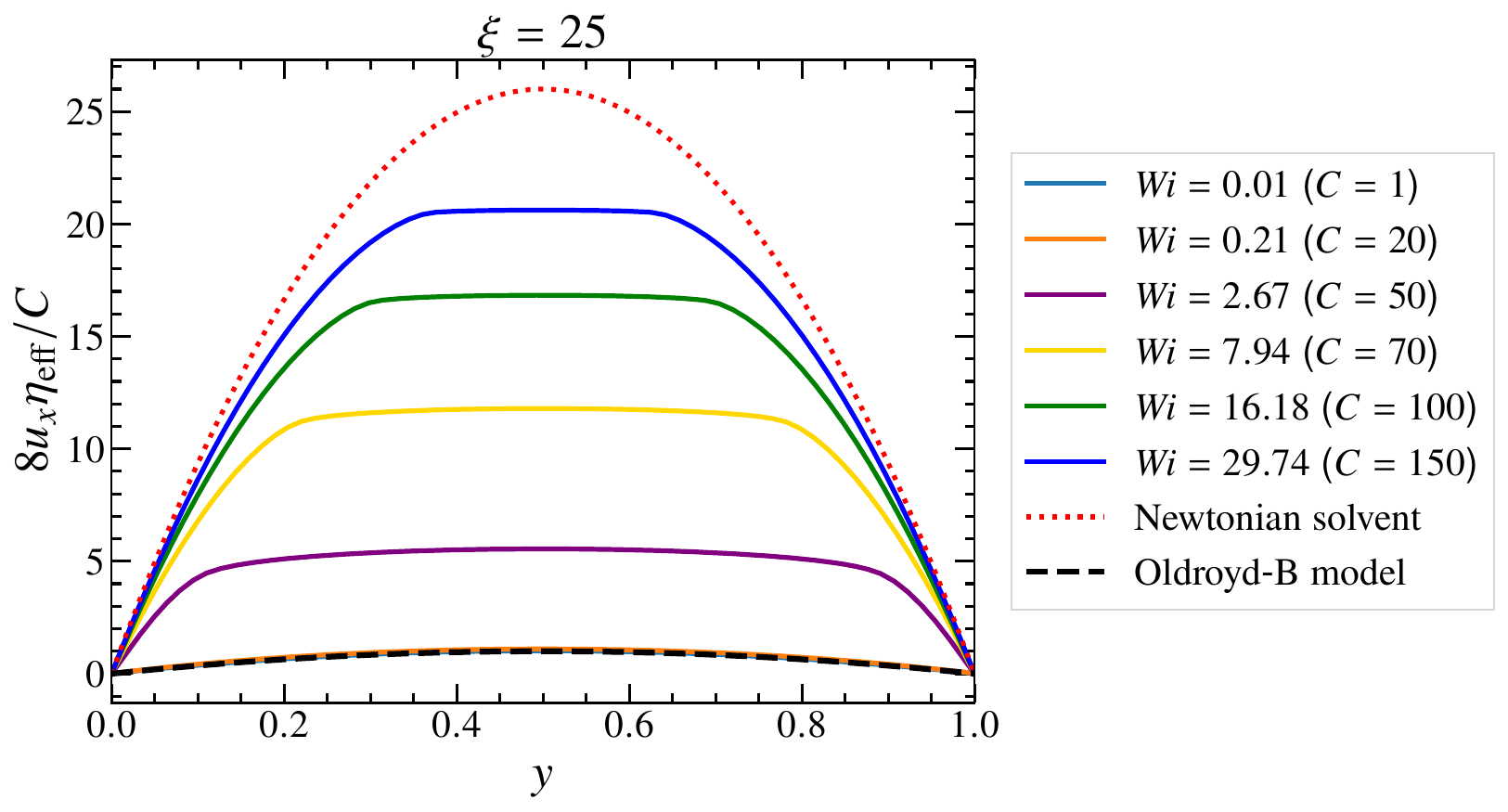}
    \end{subfigure}
    \caption{\emph{The viscosity ratio $\El$ modulates the influence of $\Wi$ on the velocity profile.} Horizontal normalized velocity profiles $8u_x\etaeff/C$, at steady state for different pressure gradients $C$ and corresponding $\Wi$. Results are compared with the solvent-only Newtonian profile with viscosity $\eta$ (red dotted lines) {and the solution of an Oldroyd-B model with solvent viscosity $\eta$ and polymer viscosity given by $\kappa \taur$, so that the total viscosity is $\etaeff=\eta(1+\El)$ (black dashed lines)}. The comparison is shown for three values of the viscosity ratio: (\textit{a}) $\El = 1$, (\textit{b}) $\El = 5$, and (\textit{c}) $\El = 25$.}
    \label{f:chfig}
\end{figure}

The problem is examined across three different elastic moduli, corresponding to three distinct viscosity ratios ($\El=1,5,25$). For each value of $\El$, the localized flow dynamics are subsequently analyzed as a function of the local $\Wi$, that increases upon increasing the pressure gradient $C$. 
In this context, the characteristic velocity $U$ is given by the maximum horizontal velocity in the channel and the characteristic length is $H/2$.

\begin{figure*}[t!]
    \centering
    \includegraphics[width=0.9
    \linewidth]{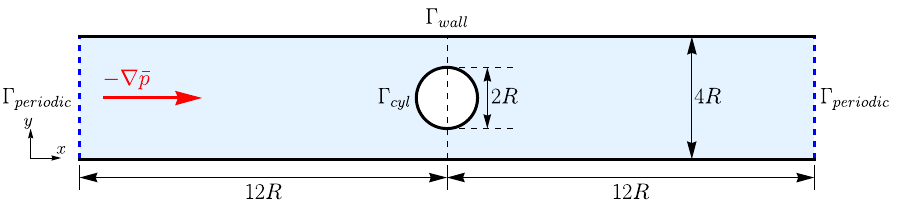}
    \caption{Geometry of the channel with a cylindrical obstacle.}
    \label{f:fpc}
\end{figure*}

The computational domain is discretized using a uniform triangular mesh. A mesh convergence study has been conducted to ensure that the results are independent of the grid (see \ref{app:A}). All simulations for this problem are performed on {a structured triangular mesh with 8,192 elements}.
{The time convergence to the steady state solution was verified by checking the $L^2$-norm of the increments for each unknown field, as reported in Fig.~\ref{f:ch_timeconv} for some illustrative cases.}

\subsubsection{Comparison with the asymptotic limits}

We compare the model predictions with the two limiting cases corresponding to parabolic flows with viscosities $\eta$ and $\etaeff$ as shown in \cref{f:chfig}. For each case, the applied average pressure gradient $C$ is varied to obtain different flow conditions, characterized by the corresponding Weissenberg number $\Wi$.
At a lower value of $\El = 1$ (\cref{f:chfig_a}) the velocity profile remains approximately parabolic along the whole smooth transition between the limiting bounds dictated by the solvent viscosity $\eta$ and the effective zero-shear viscosity $\etaeff$. 
As $\El$ increases, a more and more pronounced departure from the parabolic profile is observed (\cref{f:chfig_b} and \cref{f:chfig_c}). High shear rates near the walls of the channel suppress the polymeric contribution, causing the flow dynamics to be locally governed by the solvent viscosity $\eta$. Near the centerline of the channel, where the shear rate vanishes, the material fully recovers its zero-shear behavior, characterized by $\etaeff$. As a result, we see a plug-like velocity profile typical of strongly shear-thinning media. These spatial variations are consistent with the homogeneous simple shear analysis reported by \citet{alrashdi2024evolution}, which identified a rate-dependent viscosity that interpolates between the limits of $\etaeff$ and $\eta$. 
Larger values of $\El$ are then responsible for a stronger shear thinning behavior.
Nevertheless, we should stress that even in the plug-like region near the center of the channel the flow profile is locally parabolic and not uniform (as it would be in yield-stress fluids). 
Finally, we see that upon increasing $\Wi$ the plug-like region shrinks, since the flow profile must remain within the boundary traced by the solvent-only limit.

\subsection{Viscoelastic fluid flow past a cylinder in a channel}
\label{subsec:fpc}

The second benchmark problem presented in this work is the classical viscoelastic flow past a circular cylinder that represents an obstacle within a straight channel. This geometry introduces further complexity, stagnation points and regions of high strain rates. The primary objective is to evaluate the model's ability to capture the characteristic change in drag coefficient associated with viscoelasticity, that induces a shear thinning behavior. 

\subsubsection{Set up}

We consider a two-dimensional confined flow past a circular obstacle in a channel-like domain defined by $\Omega=[0,L] \times [0,H] \setminus \mathcal{C}$, where the length and height of the domain are taken as $L=24R$ and $H=4R$, respectively, with $R$ denoting the radius of the circular cylinder. The region $\mathcal{C}$ represents the circular domain occupied by the solid cylinder of radius $R$, defined as 
\begin{align}
    \mathcal{C} = \left\{ (x,y) \in \mathbb{R}^2 \;\middle|\; (x - c_x)^2 + (y - c_y)^2 \le R^2 \right\},
\end{align} 
with center located at $(c_x, c_y) = (L/2, H/2)$. The coordinates $x$ and $y$ denote the streamwise (horizontal) and transverse (vertical) directions, respectively. The flow is confined between two parallel walls located at $y=0$ and $y=H$, representing the lower and upper plates of the channel. The geometry considered in this section is depicted in \cref{f:fpc}.

No-slip boundary conditions are imposed on the solid boundaries, namely the channel walls $\Gamma_{wall}$ and the cylinder surface $\Gamma_{cyl}$, such that the velocity field vanishes on these boundaries, $\vc{u} = (u_x, u_y) = (0,0)$ on $y=0$, $y=H$, and on $\Gamma_{cyl}$.
Additionally, the transverse velocity component is constrained to zero at the periodic boundaries $\Gamma_{periodic}$ at $x=0$ and $x=L$, to suppress any spurious cross-stream motion and maintain a predominantly unidirectional flow. 
The streamwise velocity component $u_x$ is left free at the outlet and is obtained as part of the solution, governed by the momentum balance under the imposed pressure gradient. 
Periodic boundary conditions are applied in the streamwise direction. 
Similar to the straight channel case, the flow past the cylinder is driven by an imposed pressure gradient of magnitude $C$, which provides a uniform streamwise driving force in the momentum equation.

This benchmark problem is solved for two values of $\El=1$ and $\El=25$ and for increasing values of the pressure gradient $C$, resulting in different values of $\Wi$. Here, the characteristic velocity $U$ is given by the maximum horizontal velocity, which is achieved at the mid-section of the channel, and the characteristic length is the radius of the cylinder $R$, so that the effective shear rate is $\dot{\gamma}_{\rm{eff}}=U/R$. 

\begin{figure*}[]
    \centering
    \begin{subfigure}[b]{0.48\textwidth}
        \caption{\label{f:dragcoeff_a}}
        \includegraphics[width=\textwidth]{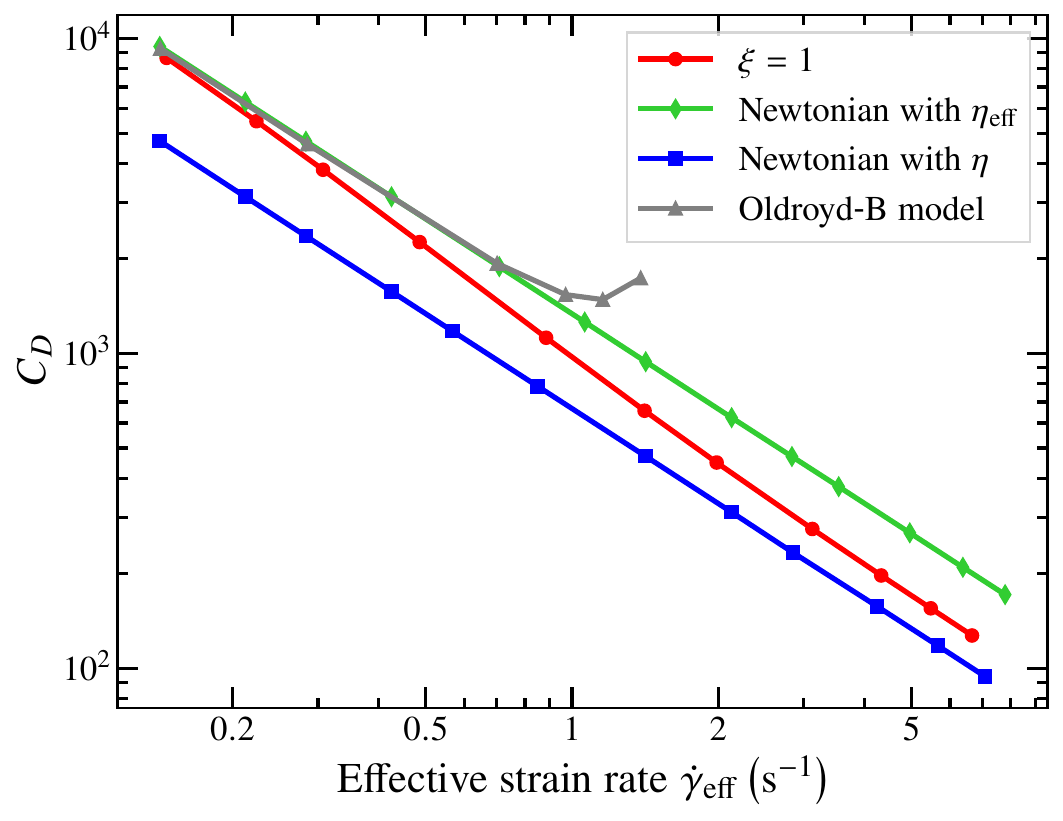}
    \end{subfigure}
    \hfill
    \begin{subfigure}[b]{0.48\textwidth}
        \caption{\label{f:dragcoeff_b}}
        \includegraphics[width=\textwidth]{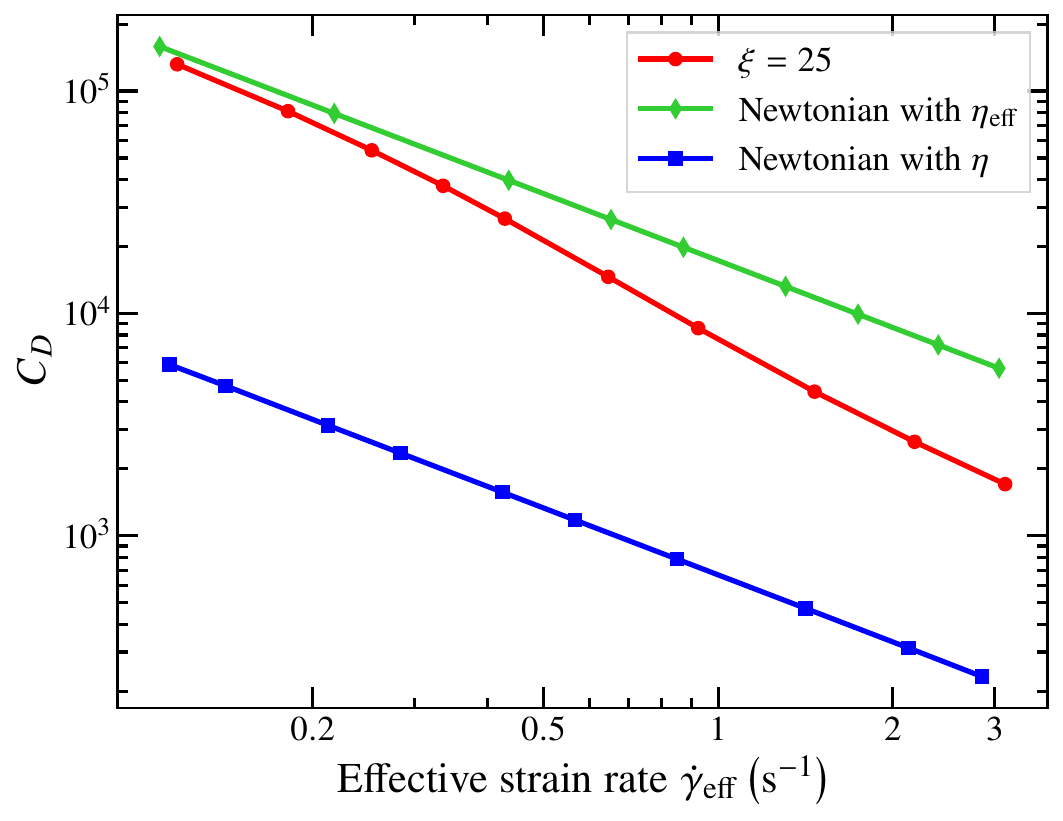}
    \end{subfigure}\\
    \caption{\emph{The log-strain model displays an enhanced drag reduction compared to the Newtonian case.} Variation of drag coefficient $C_D$ with effective strain rate $\dot{\gamma}_{\rm{eff}}$ for (\textit{a}) $\El=1$, and (\textit{b}) $\El=25$ (red circles), compared with Newtonian results with viscosity $\eta = 1$ (blue squares) and $\eta_{\rm{eff}}=\eta(1+\El)$ (green diamonds). {For the case $\El=1$, we include a comparison with the drag coefficient for an Oldroyd-B model with solvent viscosity $\eta$ and polymer viscosity given by $\kappa \taur$, so that the total viscosity is $\etaeff=\eta(1+\El)$ (gray triangles). For low shear rates this follows the Newtonian profile up until $\gd_\mathrm{eff}\approx 1$, where a minimum is reached followed by an increase in the drag, in agreement with the low-$\Rey$ results previously reported in the literature \citep{claus2013viscoelastic}.}}
    \label{f:dragcoeff}
\end{figure*}

The mesh for the computational domain is generated using \texttt{Gmsh} by discretizing a rectangular channel domain of size $(L\times H)$ from which a circular disk representing the obstacle is subtracted. 
{A triangular mesh that provides sufficient resolution near the cylinder is used to capture the strong gradients in the vicinity of the obstacle.
The mesh, with 31,332 elements,} is then imported into \texttt{FEniCSx} with physical groups marking the inlet, outlet, walls, and obstacle boundary for applying the boundary conditions. Mesh convergence is verified through successive refinements to ensure that the solution is independent of the discretization.
{The convergence to the steady state solution was verified by checking the $L^2$-norm of the increments for each unknown field (see Fig.~\ref{f:fpc_timeconv}).}

\subsubsection{Simulation results}
\label{subsec:Drag}

For the different computed flows of the log-strain model, we consider {the drag force per unit length} $\bar{F}_x$ exerted on the cylinder by the fluid and the {dimensionless} drag coefficient is computed using the relation {$C_D=\bar{F}_x/(\rho U^2 R)$}.
The same procedure is applied to two Newtonian limiting cases {and, for $\El=1$, to the results of an Oldroyd-B model.} The resulting drag coefficient values are compared to quantify the drag reduction achieved by the present model as a function of the effective strain rate (\cref{f:dragcoeff}).
{In the Newtonian cases, both the effective strain rate and the flow rate are proportional to the maximum velocity $U$, so that a comparison with matched $\dot{\gamma}_\mathrm{eff}$ gives the same information as a comparison with matched flow rate.
For the non-Newtonian flows, the changing flow profiles disrupt a precise proportionality, but the difference is minor, so we only present the drag coefficient as a function of $\dot{\gamma}_\mathrm{eff}$.}

At sufficiently low effective strain rates, the macroscopic material response converges toward that of a Newtonian fluid governed by the zero-shear effective viscosity $\etaeff${, that also coincides with the prediction of the Oldroyd-B model}. As the effective strain rate increases, a distinct transition occurs, marked by a shift of the drag coefficient from this Newtonian limit with $\etaeff$. 
This phenomenon is qualitatively independent of the viscosity ratio $\El$, but this parameter affects the overall magnitude of the effect.

\begin{figure*}[]
    \centering
    \begin{subfigure}[b]{0.49\textwidth}
        \caption{\label{f:SL_fpc_a}}
        \includegraphics[width=\textwidth]{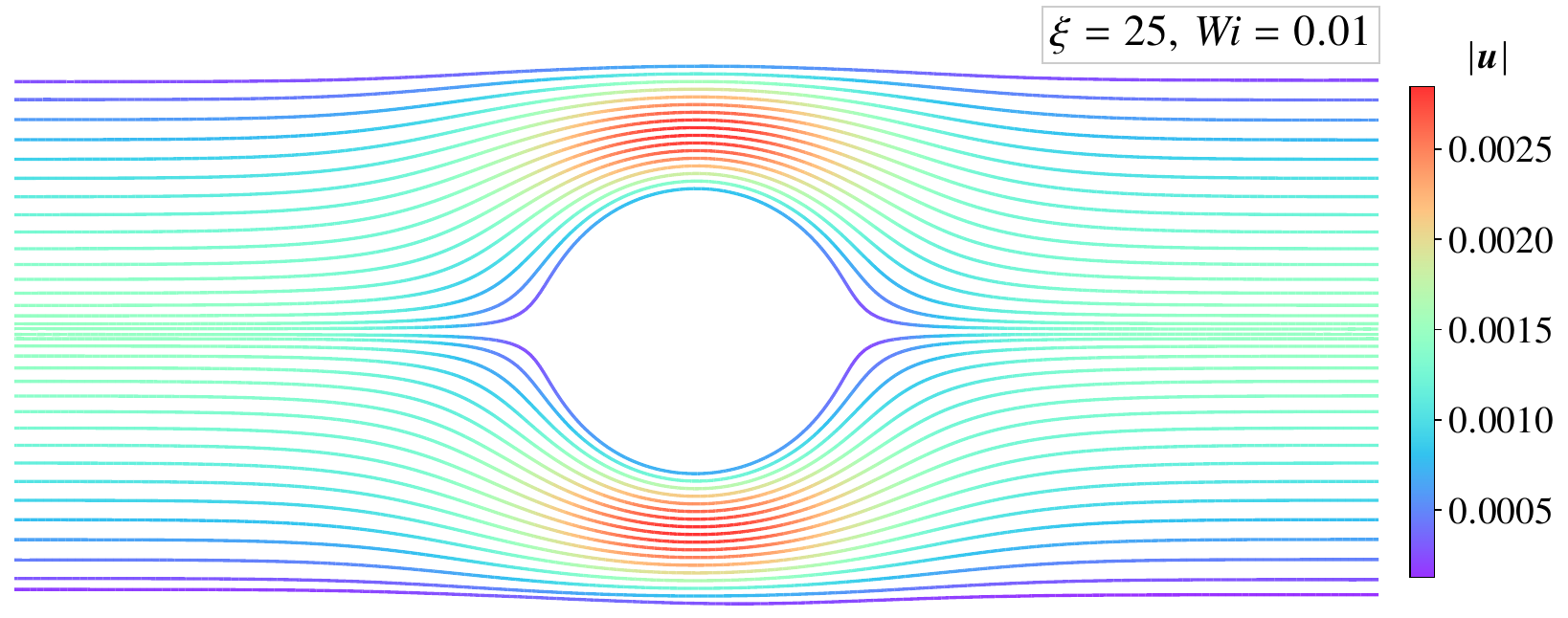}
    \end{subfigure}
    \hfill
    \begin{subfigure}[b]{0.49\textwidth}
        \caption{\label{f:SL_fpc_b}}
        \includegraphics[width=\textwidth]{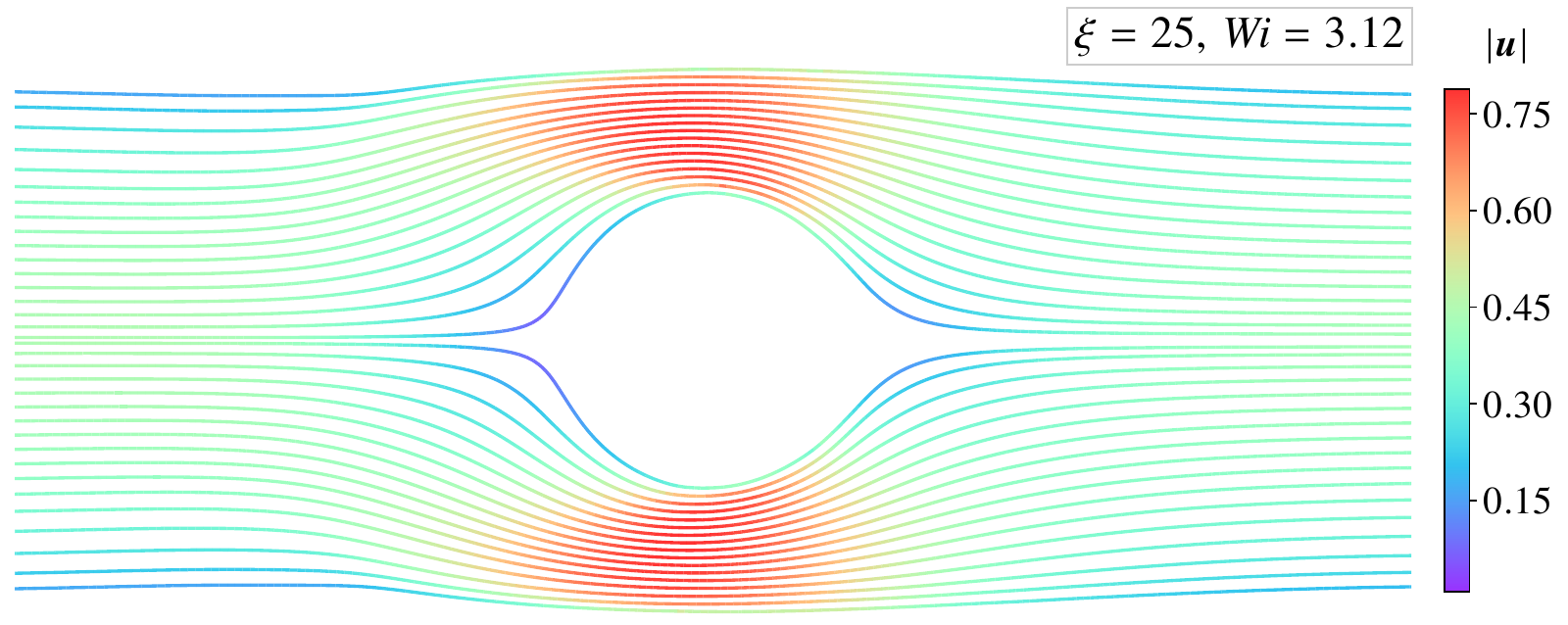}
    \end{subfigure}\\
    \begin{subfigure}[b]{0.49\textwidth}
        \caption{\label{f:SL_fpc_c}}
        \includegraphics[width=\textwidth]{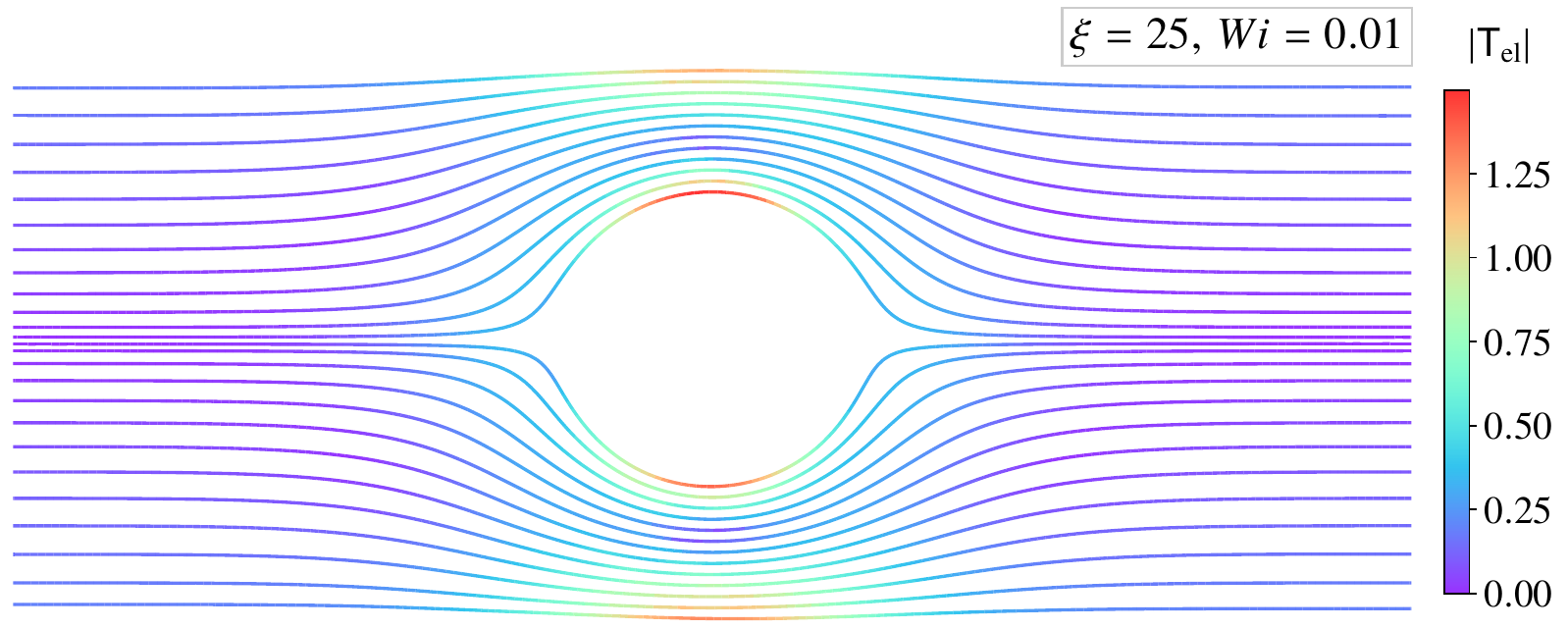}
    \end{subfigure}
    \hfill
    \begin{subfigure}[b]{0.49\textwidth}
        \caption{\label{f:SL_fpc_d}}
        \includegraphics[width=\textwidth]{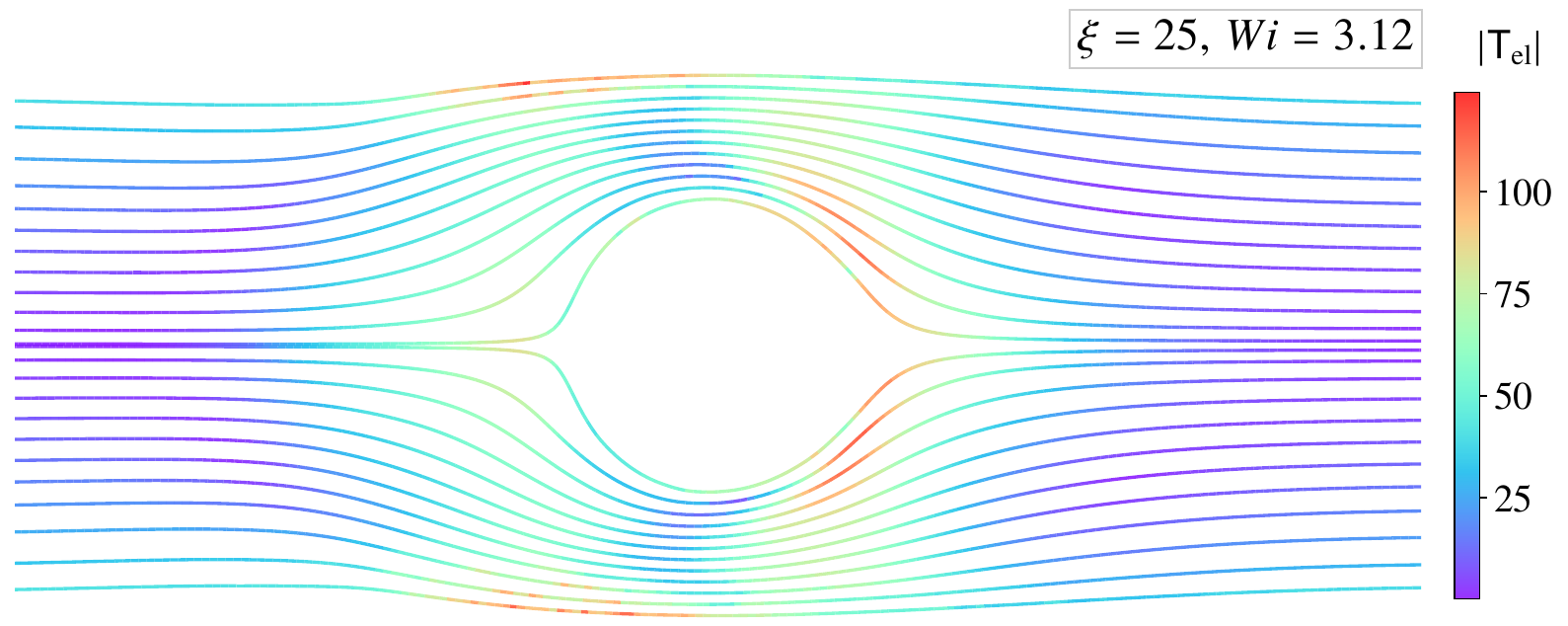}
    \end{subfigure}
    \caption{{Steady streamline patterns in the flow past a cylinder for different values of $\Wi$ at fixed $\xi = 25$. Panels ($a,b$) show the streamlines colored by the velocity magnitude $|\vc{u}|$, and ($c,d$) show the same streamlines colored by the elastic stress intensity $|\Tel|$.}}
    \label{f:SL_fpc}
\end{figure*}

{In the flow past a cylinder, when the viscosity ratio is sufficiently large, we can see a dramatic difference between the solution at small and high values of $\Wi$ (\cref{f:SL_fpc}). At low $\Wi$ we observe an upstream-downstream symmetry of all quantities, including flow-type, velocity magnitude, and elastic stress intensity (\cref{f:SL_fpc_a,f:SL_fpc_c}), with a behavior very close to that of a Newtonian fluid. 
Upon increasing $\Wi$, the increasing influence of elastic memory and the associated growth of normal stress differences break the symmetry (\cref{f:SL_fpc_b,f:SL_fpc_d}).
Most notably, the regions with the highest elastic stress along the channel walls are upstream of the cylinder, while in the vicinity of the obstacle there is a large high-stress region only downstream. This is clearly at variance with respect to the low-$\Wi$ regime, where the highest elastic stress at the boundaries appears near the central cross section.}

\subsection{Viscoelastic fluid flow in a 4:1 planar contraction}
\label{subsec:contraction}

\begin{figure*}[t!]
    \centering
    \includegraphics[width=0.9\linewidth]{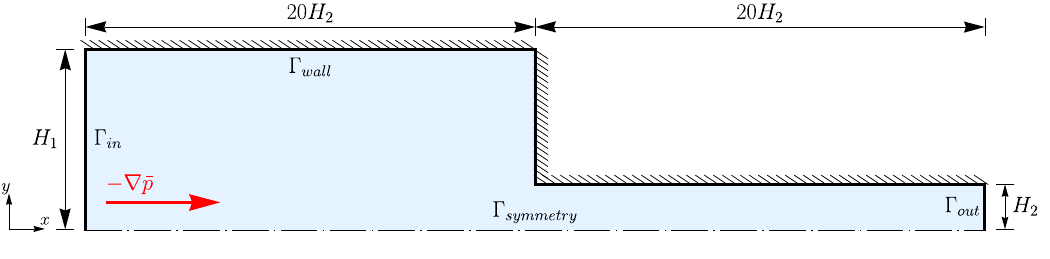}
    \caption{Geometry of the planar 4:1 contraction.}
    \label{f:contraction}
\end{figure*}

\begin{figure*}[t!]
    \centering
    \includegraphics[width=0.9\textwidth]{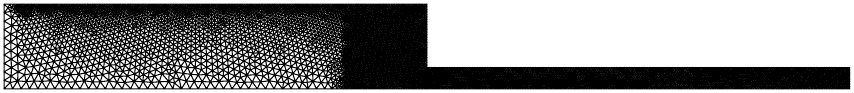}
    \caption{Computational mesh used in the 4:1 contraction problem.}
    \label{f:mesh_contraction}
\end{figure*}

The next benchmark considered here is the 4:1 planar contraction problem. This configuration is widely used as a test case for viscoelastic flow simulations. Compared to other standard benchmarks, it becomes particularly challenging in the high-elasticity regime due to the presence of a sharp re-entrant corner at the contraction. This geometric singularity can lead to significantly strong stress concentrations and pronounced gradients in the flow field. Moreover, the flow necessarily features different regions dominated by the different local flow types of simple shear, extension, and rotation. As a result, this problem offers a broad assessment of the log-strain model predictions.

\subsubsection{Set up}

The key characteristics of this configuration are as follows. Owing to symmetry, only half of the domain is considered in the computation, as illustrated in \cref{f:contraction}. The inlet boundary $\Gamma_{in}$ has a height $H_1=4H_2$, with $H_2$ the height of the outlet boundary $\Gamma_{out}$. The total length of the domain is $L = 40H_2$, with the contraction located at $L/2$.
No-slip boundary conditions are imposed on the solid walls $\Gamma_{wall}$ such that the velocity field satisfies $u=(u_x,u_y) = (0,0)$. Symmetry is enforced along the centerline. At the inlet $x=0$ and outlet $x=L$, the transverse velocity component $u_y$ is constrained to zero to suppress cross-stream motion, while the streamwise component $u_x$ is left free and determined as part of the solution. 
Similar to the earlier examples, the flow is driven by an imposed average pressure gradient along the streamwise direction.

The computational mesh is created using \texttt{Gmsh} and consists of uniform triangular discretization of the domain. Local mesh refinement is applied in the upstream region before the contraction, throughout the narrow contraction zone, and along the walls to accurately resolve steep velocity gradients and potential vortex structures. Coarser elements are employed in regions away from these flow features to reduce computational cost. After generating the mesh, it is imported into \texttt{FEniCSx}, where different physical groups are defined to facilitate the application of boundary conditions. Mesh convergence is verified through successive mesh refinements to ensure that the solution is independent of the spatial discretization. \Cref{f:mesh_contraction} shows the computational mesh {with 9,454 elements} used for the present analysis. {The convergence to the steady state solution was again verified by checking the $L^2$-norm of the increments for each unknown field (see Fig.~\ref{f:cont_timeconv}).}

\begin{figure*}[t!]
    \centering
    \begin{subfigure}[b]{0.325\textwidth}
        \caption{\label{f:SL_a}}
        \includegraphics[width=\textwidth]{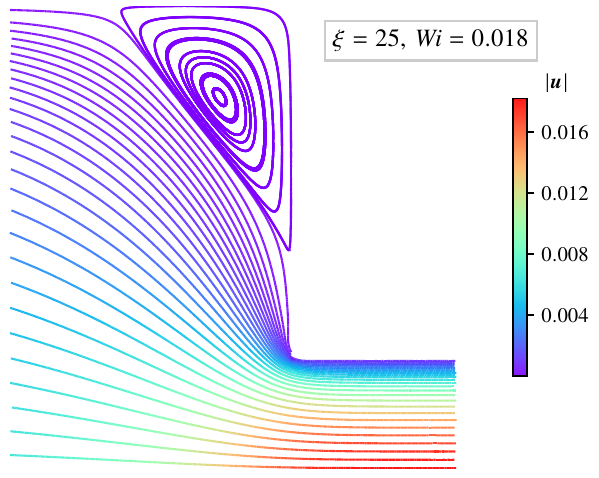}
    \end{subfigure}
    \hfill
    \begin{subfigure}[b]{0.325\textwidth}
        \caption{\label{f:SL_b}}
        \includegraphics[width=\textwidth]{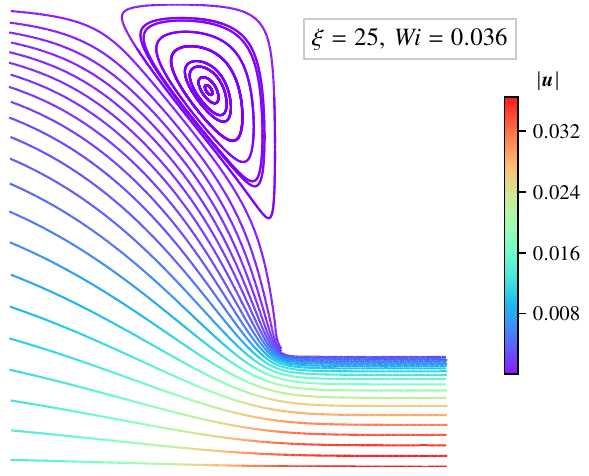}
    \end{subfigure}
    \begin{subfigure}[b]{0.325\textwidth}
        \caption{\label{f:SL_c}}
        \includegraphics[width=\textwidth]{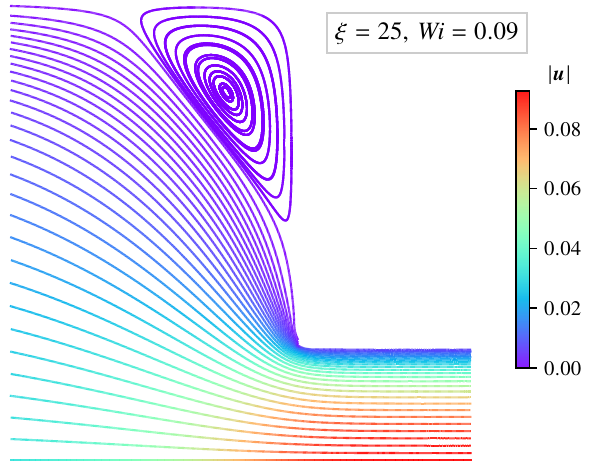}
    \end{subfigure}\\
    \begin{subfigure}[b]{0.325\textwidth}
        \caption{\label{f:SL_d}}
        \includegraphics[width=\textwidth]{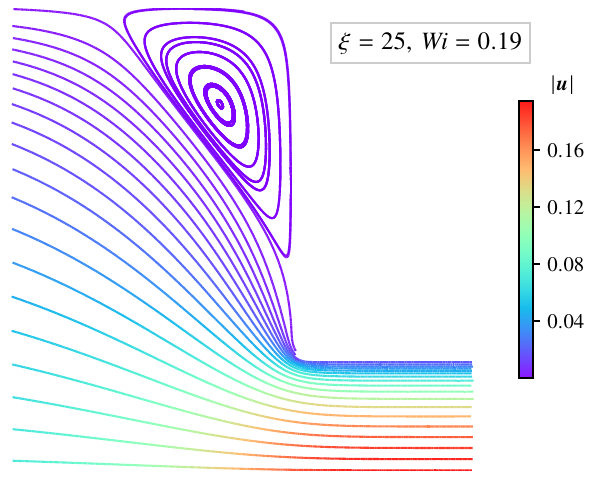}
    \end{subfigure}
    \hfill
    \begin{subfigure}[b]{0.32\textwidth}
        \caption{\label{f:SL_e}}
        \includegraphics[width=\textwidth]{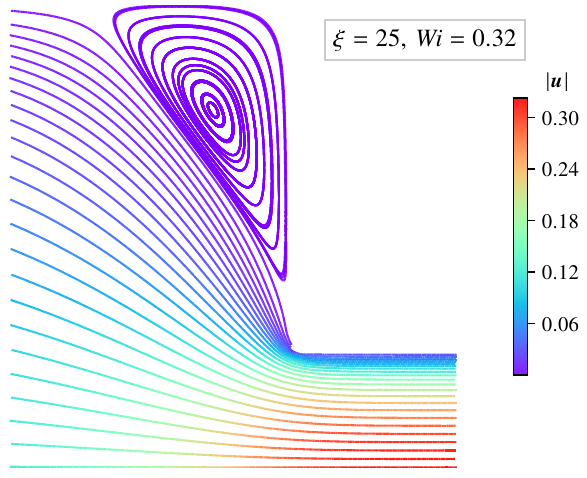}
    \end{subfigure}
    \hfill
    \begin{subfigure}[b]{0.32\textwidth}
        \caption{\label{f:SL_f}}
        \includegraphics[width=\textwidth]{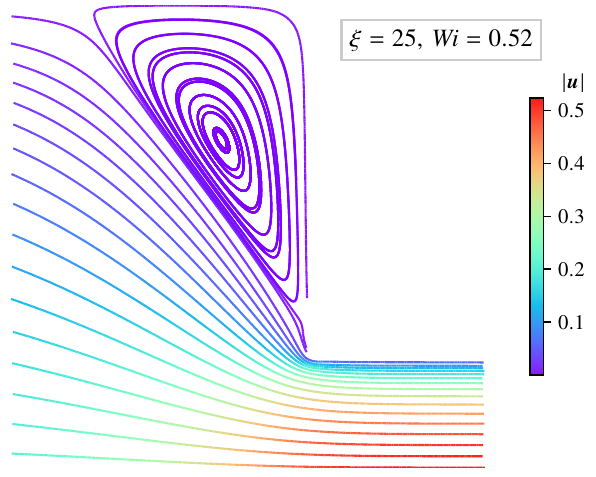}
    \end{subfigure}
    \caption{Steady streamline patterns in the 4:1 planar contraction colored by the velocity magnitude $|\vc{u}|$ for different values of $\Wi$ and fixed $\El=25$.}
    \label{f:SL}
\end{figure*}

\begin{figure*}[t!]
    \centering
    \begin{subfigure}[b]{0.325\textwidth}
        \caption{\label{f:SL_SI_a}}
        \includegraphics[width=\textwidth]{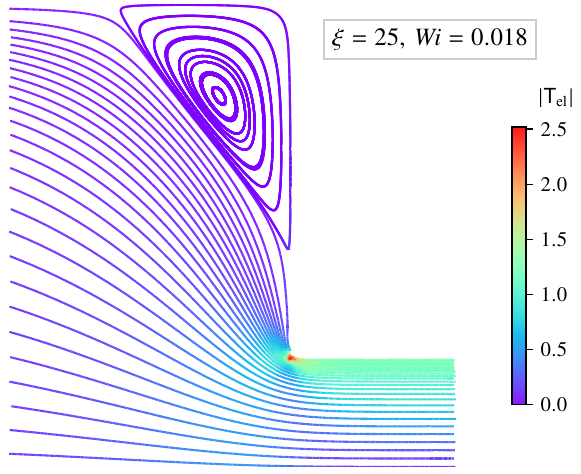}
    \end{subfigure}
    \hfill
    \begin{subfigure}[b]{0.325\textwidth}
        \caption{\label{f:SL_SI_b}}
        \includegraphics[width=\textwidth]{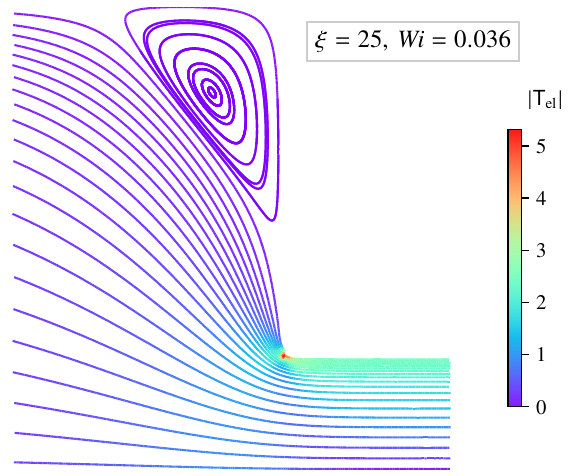}
    \end{subfigure}
    \begin{subfigure}[b]{0.325\textwidth}
        \caption{\label{f:SL_SI_c}}
        \includegraphics[width=\textwidth]{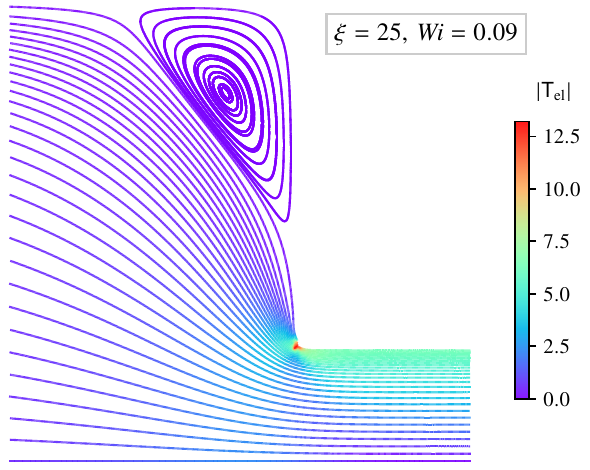}
    \end{subfigure}\\
    \begin{subfigure}[b]{0.325\textwidth}
        \caption{\label{f:SL_SI_d}}
        \includegraphics[width=\textwidth]{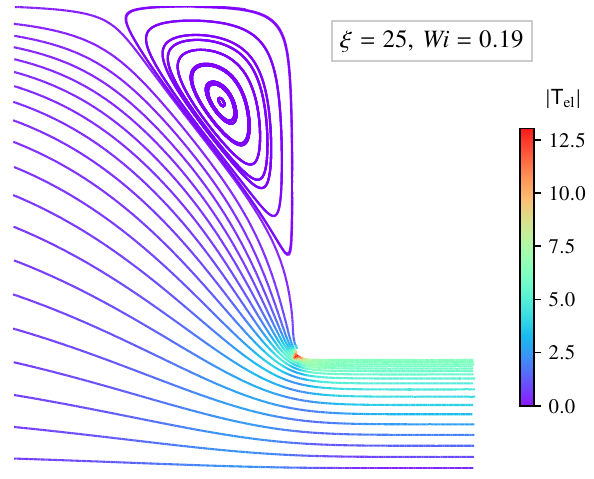}
    \end{subfigure}
    \hfill
    \begin{subfigure}[b]{0.32\textwidth}
        \caption{\label{f:SL_SI_e}}
        \includegraphics[width=\textwidth]{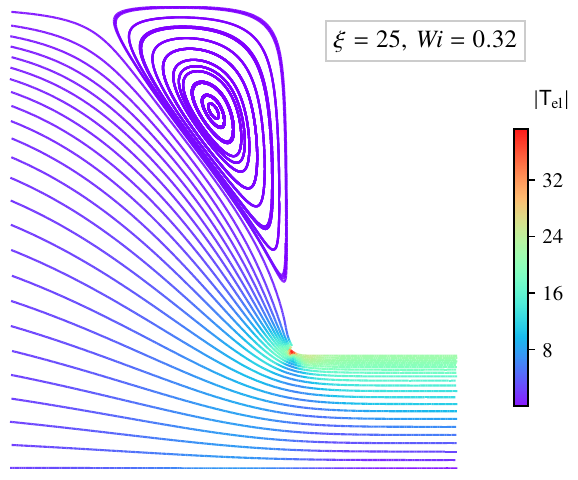}
    \end{subfigure}
    \hfill
    \begin{subfigure}[b]{0.32\textwidth}
        \caption{\label{f:SL_SI_f}}
        \includegraphics[width=\textwidth]{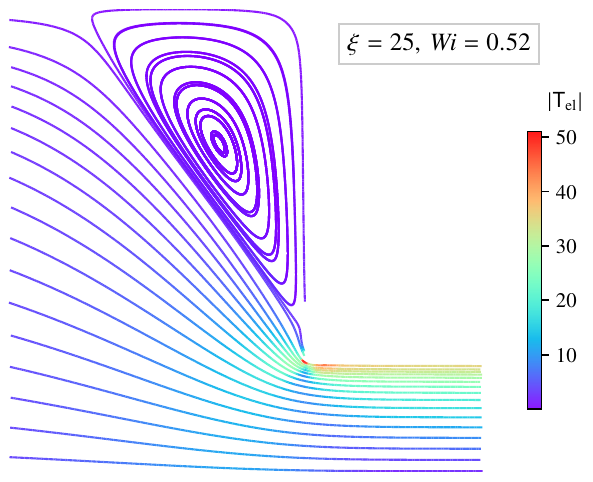}
    \end{subfigure}
    \caption{{Streamlines in the 4:1 planar contraction colored by the elastic stress intensity $|\Tel|$ for different values of $\Wi$ at fixed $\El=25$.}}
    \label{f:SL_SI}
\end{figure*}

\subsubsection{Analysis of the flow behavior in the 4:1 planar contraction}

\begin{figure*}[t!]
    \centering
    \begin{subfigure}[b]{0.48\textwidth}
        \caption{\label{f:contr_sec_a}}
        \includegraphics[width=\textwidth]{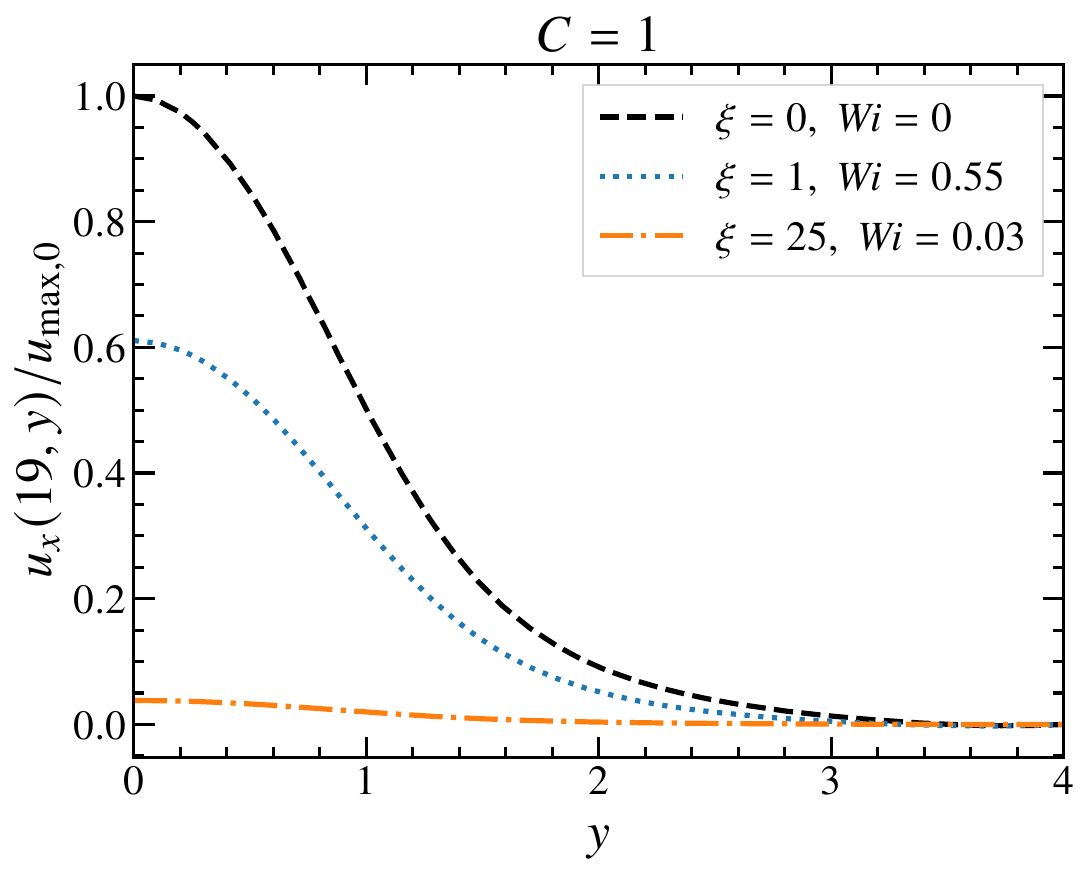}
    \end{subfigure}
    \hfill
    \begin{subfigure}[b]{0.48\textwidth}
        \caption{\label{f:contr_sec_b}}
        \includegraphics[width=\textwidth]{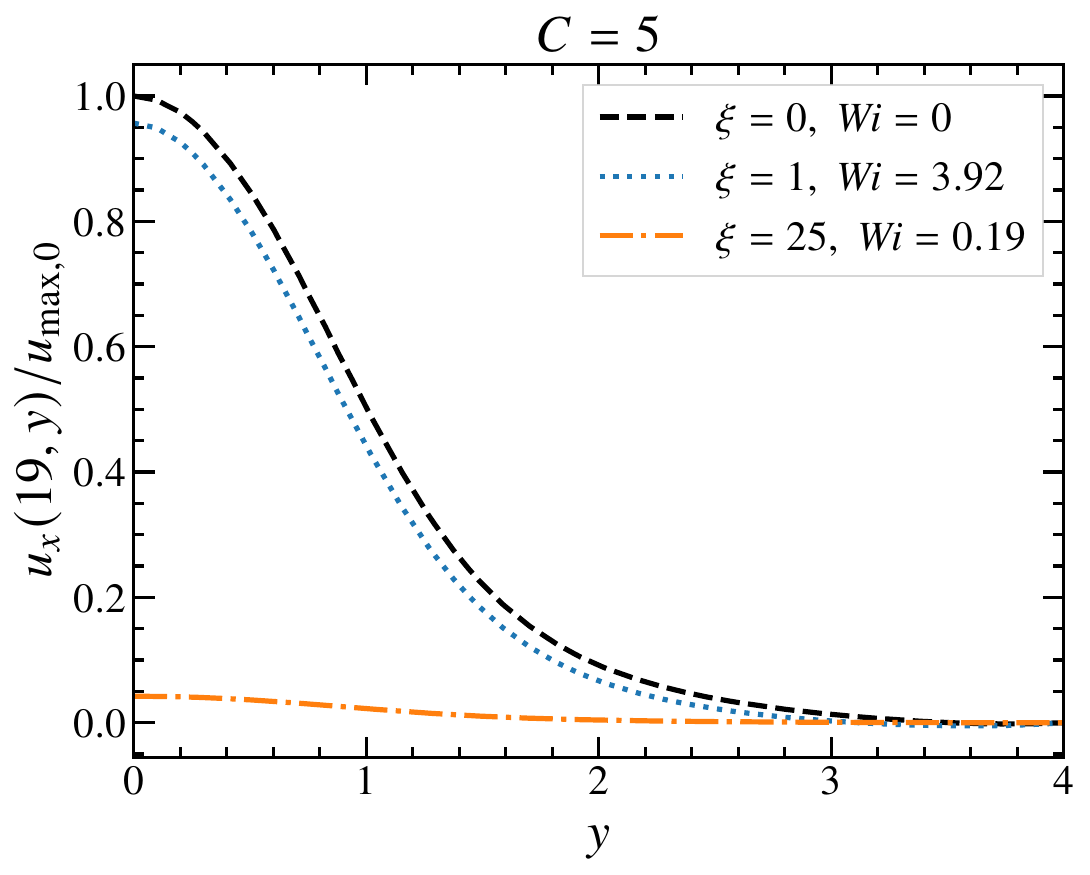}
    \end{subfigure}\\
    \begin{subfigure}[b]{0.48\textwidth}
        \caption{\label{f:contr_sec_c}}
        \includegraphics[width=\textwidth]{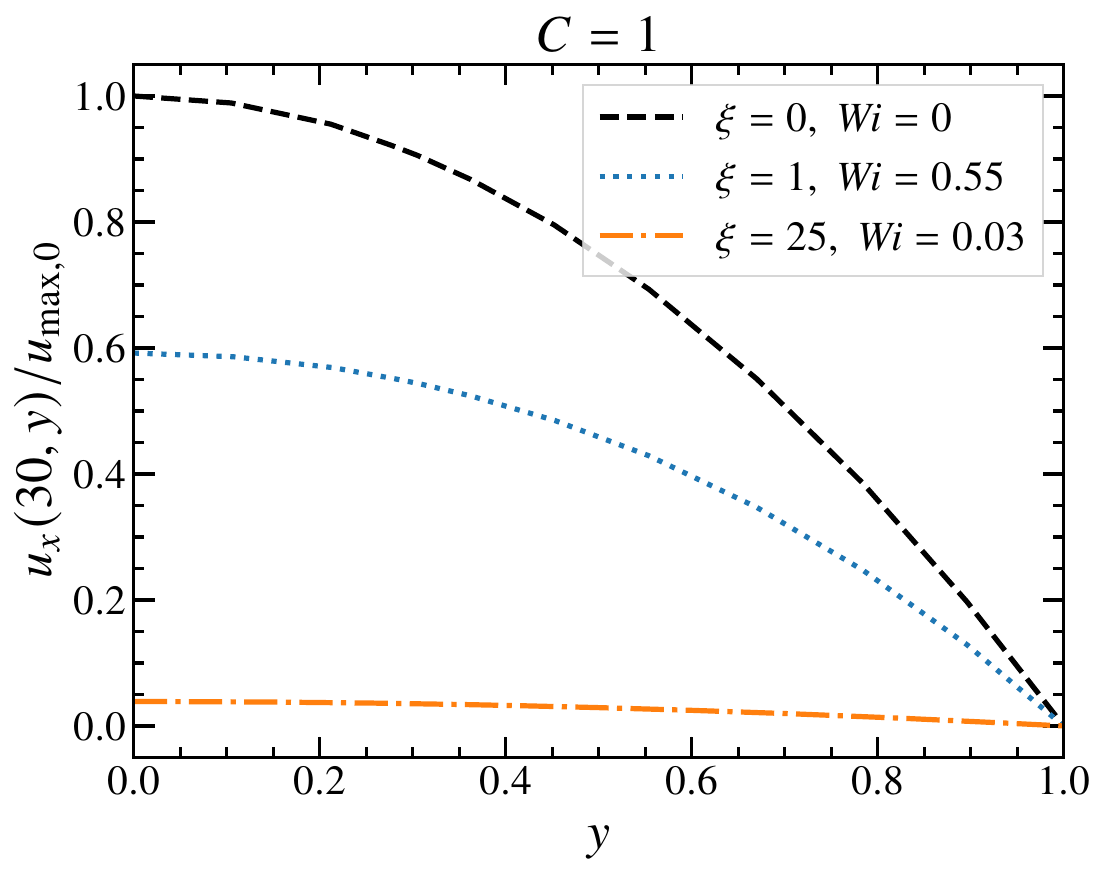}
    \end{subfigure}
    \hfill
    \begin{subfigure}[b]{0.48\textwidth}
        \caption{\label{f:contr_sec_d}}
        \includegraphics[width=\textwidth]{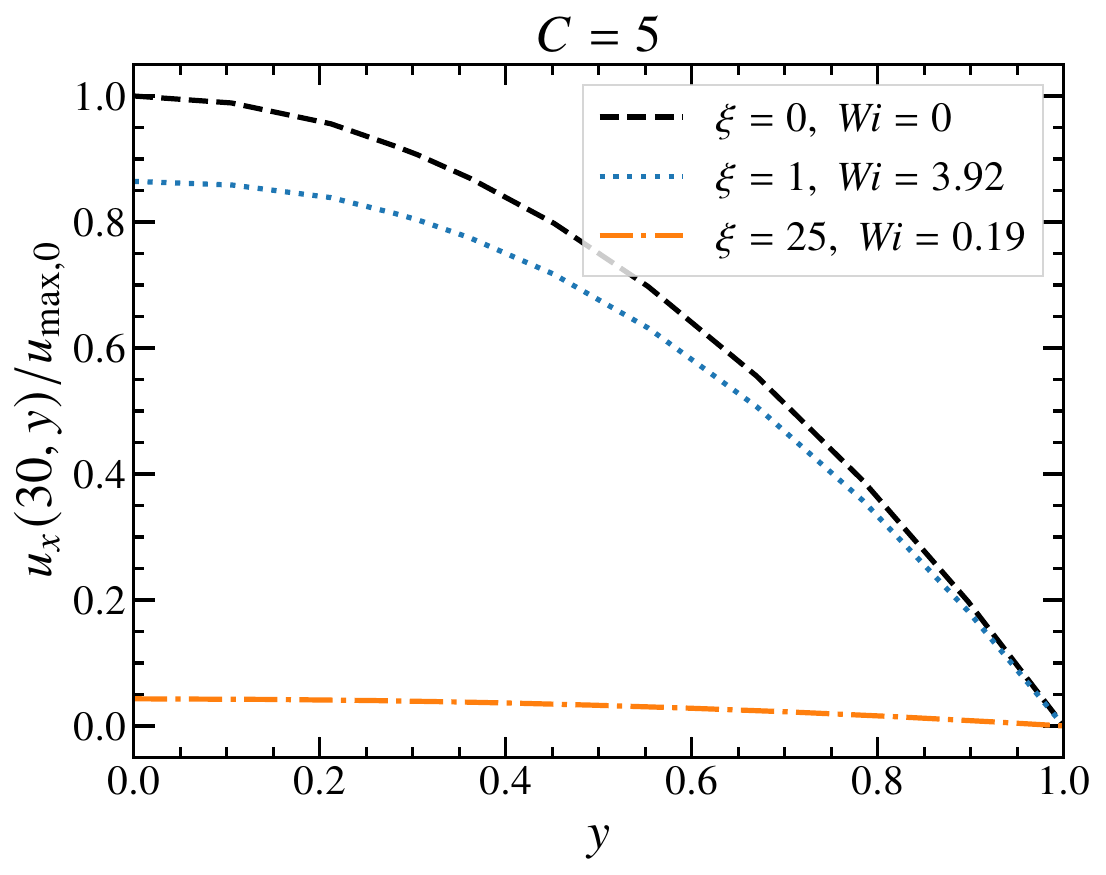}
    \end{subfigure}
    \caption{Horizontal velocity $u_x$ at steady state for pressure gradient $C = 1.0$ and $C=5.0$ along the lines at ($a,\,b$) $x = 19$, and ($c,\,d$) $x = 30$. The values are normalized with respect to the maximum velocity $u_{\max,0}$ for $\El = 0$.}
    \label{f:contr_sec}
\end{figure*}

Several key observations can be deduced from the streamline patterns shown in \cref{f:SL} for $\El = 25$. A non-monotonic variation in the corner vortex size is observed as $\Wi$ increases. Within a very narrow regime of $\Wi$, the vortex size initially undergoes a slight contraction before subsequently expanding at higher $\Wi$. Notably, the nucleation of secondary lip vortices is entirely absent across the investigated parametric range. This is consistent with the shear-thinning behavior of the log-strain model. {The same streamlines, coloured by the elastic stress intensity $|\Tel|$, are shown in \cref{f:SL_SI} for different values of $\Wi$. The maximum stress intensity remains concentrated at the contraction corner and increases with increasing $\Wi$.}

The absence of a secondary lip vortex in the presence of strong shear-thinning is well-documented in entry geometries. For instance, in the study conducted by \citet{alves2003benchmark}, a clear distinction was observed between constant-viscosity and shear-thinning viscoelastic models. While constant-viscosity models, like the Oldroyd-B model, yield severe elastic stress concentrations that promote secondary vortex nucleation near the re-entrant corner, the shear-thinning Phan-Thien-Tanner (PTT) framework completely suppresses this localized phenomenon.
This observation is also consistent with the numerical simulations of a 4:1 planar contraction flow for Giesekus fluids reported by \citet{choi1988numerical}, where no secondary or lip vortices were observed.

In \cref{f:contr_sec}, the horizontal velocity profiles of the viscoelastic flow at $x = 19$ (\cref{f:contr_sec_a,f:contr_sec_b}) and $x = 30$ (\cref{f:contr_sec_c,f:contr_sec_d}) are illustrated. The results are shown for the present model with $\El = 1$ and $\El = 25$, along with the Newtonian limit using the solvent viscosity of $\eta = 1$ and corresponding to $\El=0$ and $\Wi=0$.
The simulations were performed for two distinct values of the imposed pressure gradient, $C = 1$ and $C = 5$, with the velocity fields normalized against the Newtonian reference case. 
For an equal gradient $C$, the results indicate that increasing $\El$ leads to a marked reduction in the maximum velocity $u_{x,\max}$. Consequently, the achieved $\Wi$ is significantly lower in our computation at high $\El$. This behavior is governed by the increased elastic resistance of the fluid which naturally leads to a larger energy dissipation.

\begin{figure*}[]
    \centering
    \begin{subfigure}[b]{0.495\textwidth}
        \caption{\label{f:contr_sec_y_a}}
        \includegraphics[width=\textwidth]{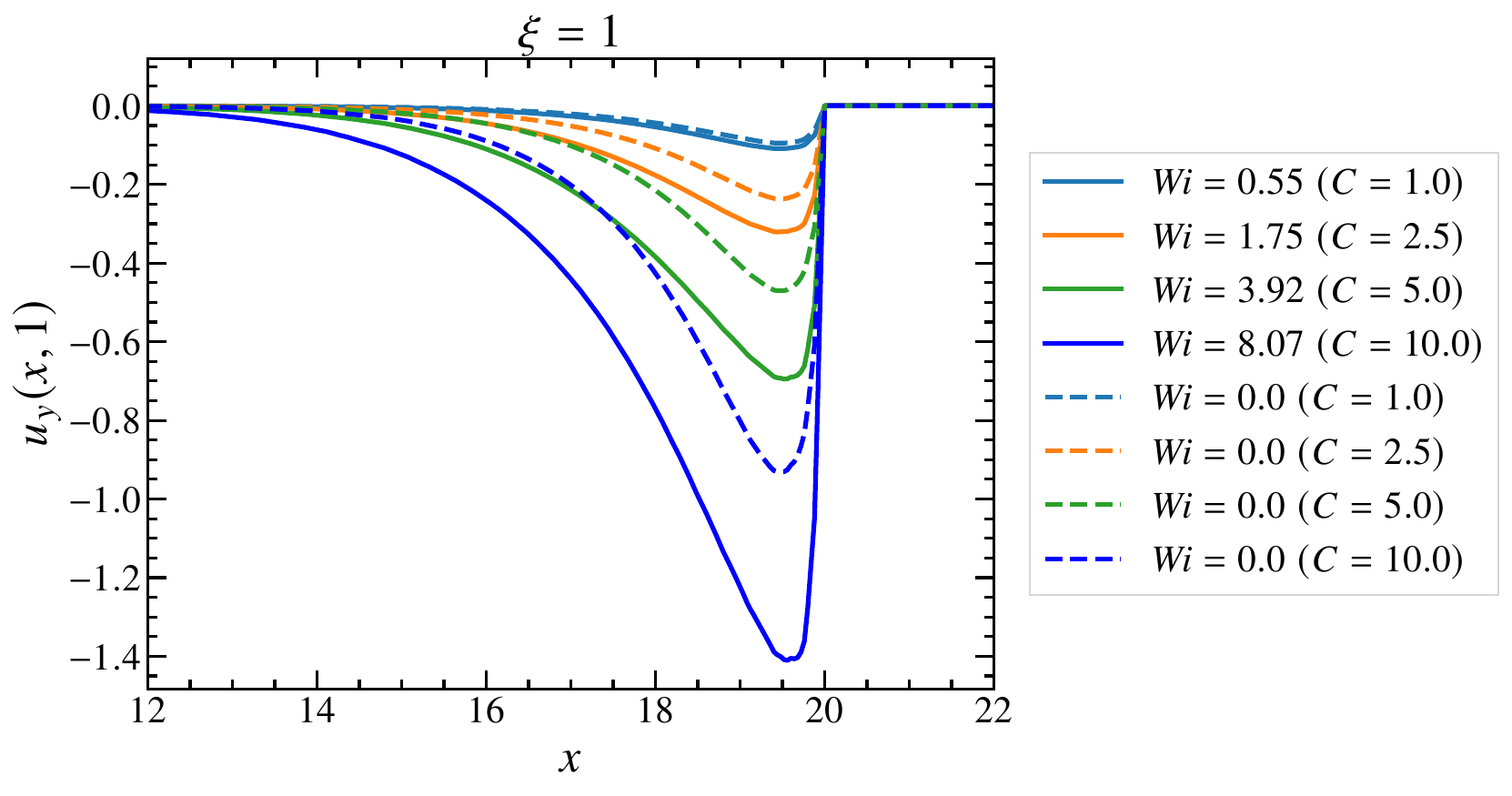}
    \end{subfigure}
    % \hfill
    \begin{subfigure}[b]{0.495\textwidth}
        \caption{\label{f:contr_sec_y_b}}
        \includegraphics[width=\textwidth]{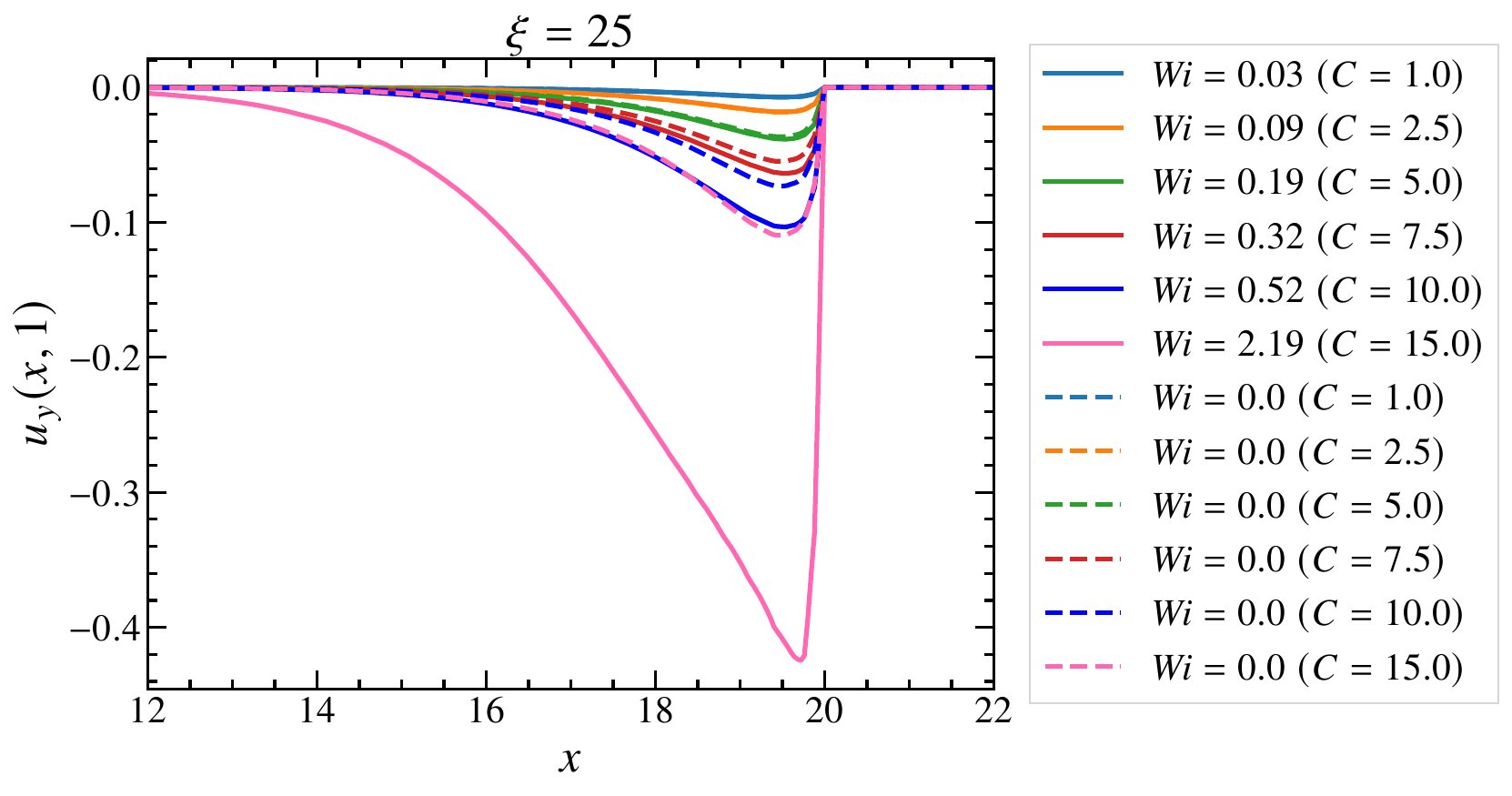}
    \end{subfigure}
    \caption{Vertical velocity $u_y$ at steady state along the line at $y = 1$ (top wall of the narrow region) for two different values of viscosity ratio $(a)$ $\El = 1$, and $(b)$ $\El = 25$, including a comparison with a Newtonian case $(\Wi=0$, dashed lines) with viscosity $\etaeff$.}
    \label{f:contr_sec_y}
\end{figure*}

The profile of the vertical velocity along the line $y = 1$ (top wall of the narrow region) for the present model and the Newtonian limit with viscosity $\eta_{\mathrm{eff}}$ is shown in \cref{f:contr_sec_y}. A comparison is provided for $\El = 1$ (\cref{f:contr_sec_y_a}) and $\El = 25$ (\cref{f:contr_sec_y_b}).
We observe that the deviation between the model and the Newtonian limit increases with increasing $\Wi$. 
This elastic contribution significantly modifies the pressure-velocity coupling compared to the purely dissipative Newtonian case.
For larger $\El$, the difference between the two profiles is already apparent even at relatively moderate values of $\Wi$, as seen in \cref{f:contr_sec_y_b}. 
This again highlights a key feature of the current model, which is the independent role of the viscosity ratio that governs the extent to which variations in $\Wi$ influence the flow behavior.

\subsection{Comparison of the log-strain model with a Generalized Newtonian Fluid model}
\label{subsec:compare_GNF}

In this section, we compare the solutions of the log-strain model against a reduced GNF model for the flows in a straight channel and past a cylinder.
Our aim is to assess the importance of those rheological features that go beyond a rate-dependent viscosity in determining the flow profile.
To this end, we consider a GNF model that features a rate-dependent viscosity matching that predicted by the log-strain model in steady simple shear flows.

To construct the framework for the GNF model, we consider a steady simple shear flow. The Cartesian coordinate system is oriented such that the $x$-axis aligns with the primary flow direction, while the $y$-axis represents the velocity gradient direction normal to the solid boundaries. Given these kinematic constraints, the velocity vector field simplifies to $\vc{u} = \{u(y), 0\}$.
The rate of strain tensor $\DD$, defined as the symmetric component of the velocity gradient tensor, is given by $\DD = \f{1}{2}[\nab \vc{u} + (\nab \vc{u})^\tsp]$. Evaluating this expression for our one-dimensional velocity field yields
\begin{align}
\DD = \f{1}{2} 
\begin{pmatrix} 0 & \del u/\del y \\ 
\del u/\del y & 0  
\end{pmatrix}.
\end{align}
The shear-rate derived from the second invariant $\gd(y) = \sqrt{2(\DD : \DD)}$ maps directly to the localized velocity gradient as $\gd(y) = \vert\del u / \del y\vert$. 
The total Cauchy stress tensor $\vc{\sigma}$ for an incompressible GNF accounts for the isotropic pressure field $p$ and a shear-rate-dependent  viscosity function $\eta_0(\gd)$, satisfying the constitutive expression
\begin{align}
    \vc{\sigma} = -p \I + 2 \eta_0(\gd) \DD.
\end{align}
The primary shear stress component operating in the streamwise direction is $\sigma_{xy} = \eta_0(\gd)\left(\del u / \del y\right)$.
In the absence of gravitational forces and accelerating inertial terms, Cauchy’s momentum equation simplifies to the static balance $\nab \cdot \vc{\sigma} = \vc{0}$. Evaluating the streamwise ($x$-direction) component of this vector equation yields 
\begin{align}
    -\f{\del p}{\del x} + \f{\del}{\del y}\left( \eta_0(\gd) \f{\del u}{\del y} \right) = 0.
\end{align}
Because the flow is fully developed, the pressure gradient is decoupled from the velocity field and reduces to a spatial constant, $-\del p / \del x = C$. This simplifies the governing momentum equation to a direct balance between the divergence of the internal shear stress and the constant pressure drive
\begin{align}
    \f{\del}{\del y} \left[ \eta_0(\gd) \f{\del u}{\del y} \right] = -C.
\end{align}
Applying the chain rule to the spatial derivative of the effective viscosity yields the final non-linear governing differential equation for the velocity profile
\begin{align}
    \left. \left[ \eta_0(\gd) + \f{\del u}{\del y} \f{\del \eta_0(\gd)}{\del \gd} \right]\f{\del^2 u}{\del y^2} \,\right|_{\gd = \f{\del u}{\del y}} =- C.
\end{align}
This macroscopic field relation highlights the strong non-linear coupling between the local velocity gradient and the spatial distribution of fluid resistance. To solve this equation, an analytical closure relation for $\eta_0$ must be chosen.

\subsubsection{Viscosity mapping and model parameterization}
 
To compute the effective shear viscosity predicted by the log-strain model, we need to follow the evolution of the strain tensor $\Bel$, as expressed in \Cref{e:visfluid1},
under a homogeneous simple shear flow.
The shear stress $\sigma_{xy}$ combines the Newtonian solvent resistance and the extra polymeric stress scaled by the elastic modulus $\kappa$, leading to $\sigma_{xy} = \eta \dot{\gamma} + \kappa (\log {\Bel})_{12}$.
We computed such an evolution for the case $\El=25$ exploring the range $\gd \in [10^{-3}, 10^3]$. 

\begin{figure}[t!]
    \centering
    \includegraphics[width=0.9\linewidth]{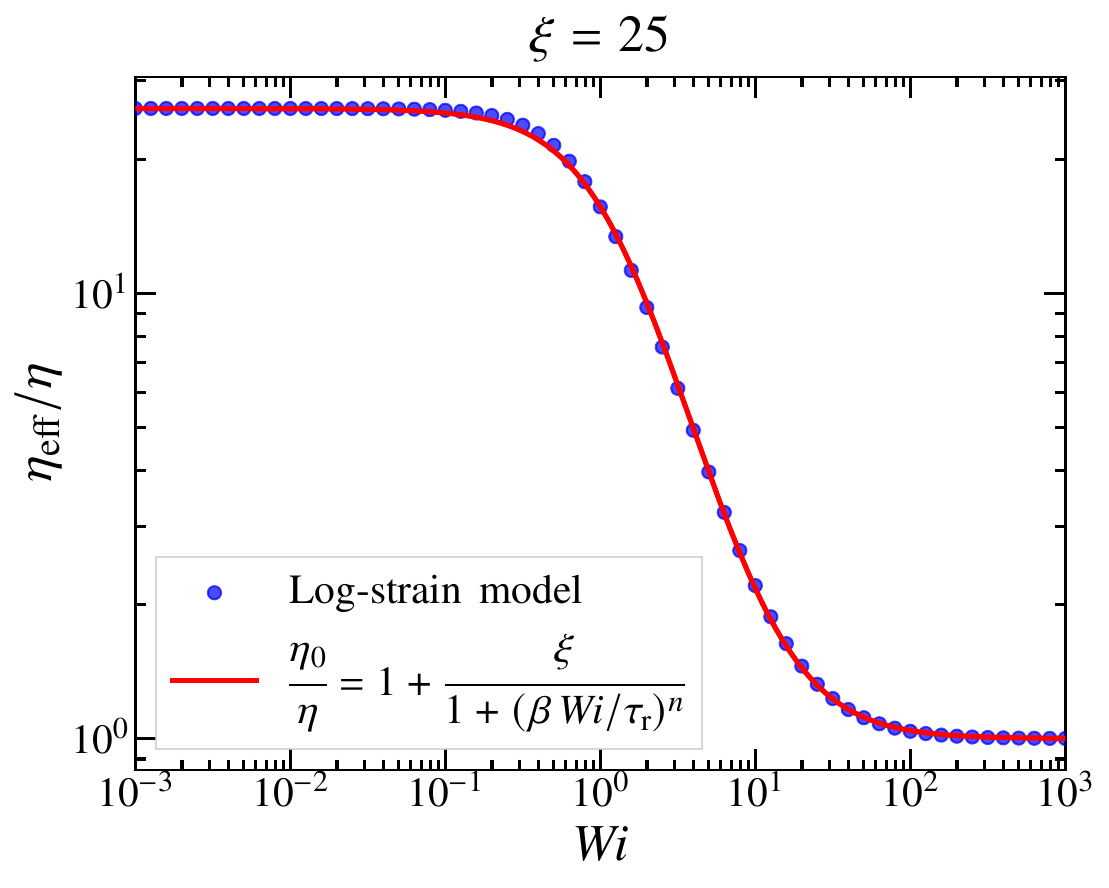}
    \caption{Normalized effective viscosity $\etaeff/\eta$ as a function of $\Wi=\gd\taur$. Circles report the data from the log-strain model. The red solid line indicates the fitted Cross law for the effective viscosity used in the GNF model. The fitting parameters are $\beta = 0.788$, and $n = 1.458$.}
    \label{f:GNF_Fitting}
\end{figure}

The resulting discrete effective viscosity profile was computed using the stress-to-shear-rate ratio at steady state as
\begin{align}
    \etaeff = \eta + \kappa \cfrac{(\log {\Bel})_{12}}{\gd}=\eta\left(1+\frac{\El}{\Wi}(\log {\Bel})_{12}\right).
\end{align}
The numerical data points revealed a characteristic shear-thinning response, transitioning from a low-shear Newtonian plateau to the high-shear solvent baseline as shown in \cref{f:GNF_Fitting}. To embed these data into the GNF model, the curve was fitted using the Cross-type empirical expression
\begin{align}
    \eta_0 = \eta \left(1+ \f{\El}{1 + (\beta \Wi/\taur)^n}\right),
\end{align}
where the quantity $\kappa \taur=\eta\El$ denotes the asymptotic polymeric viscosity increment under zero-shear conditions. A non-linear regression analysis of the data generated by the model was executed to extract the optimal material time constant $\beta$ and the power-law exponent $n$.

\Cref{f:GNF_Fitting} illustrates the non-linear regression fit. Overall, the analytical Cross-model provides a good quantitative fit across the majority of the shear-rate domain. A minor mismatch occurs within the intermediate shear-rate region ($\Wi \sim 0.3$), where the analytical curve exhibits a slightly broader transition into the shear-thinning regime than the full model. The optimized parameters ($\beta$ and $n$) are subsequently used in the numerical solver implemented in \texttt{FEniCSx} to simulate the flow behavior of the reduced GNF framework. 

\subsubsection{Comparison of flow profiles}

\begin{figure}[b!]
    \centering
    \includegraphics[width=0.95\linewidth]{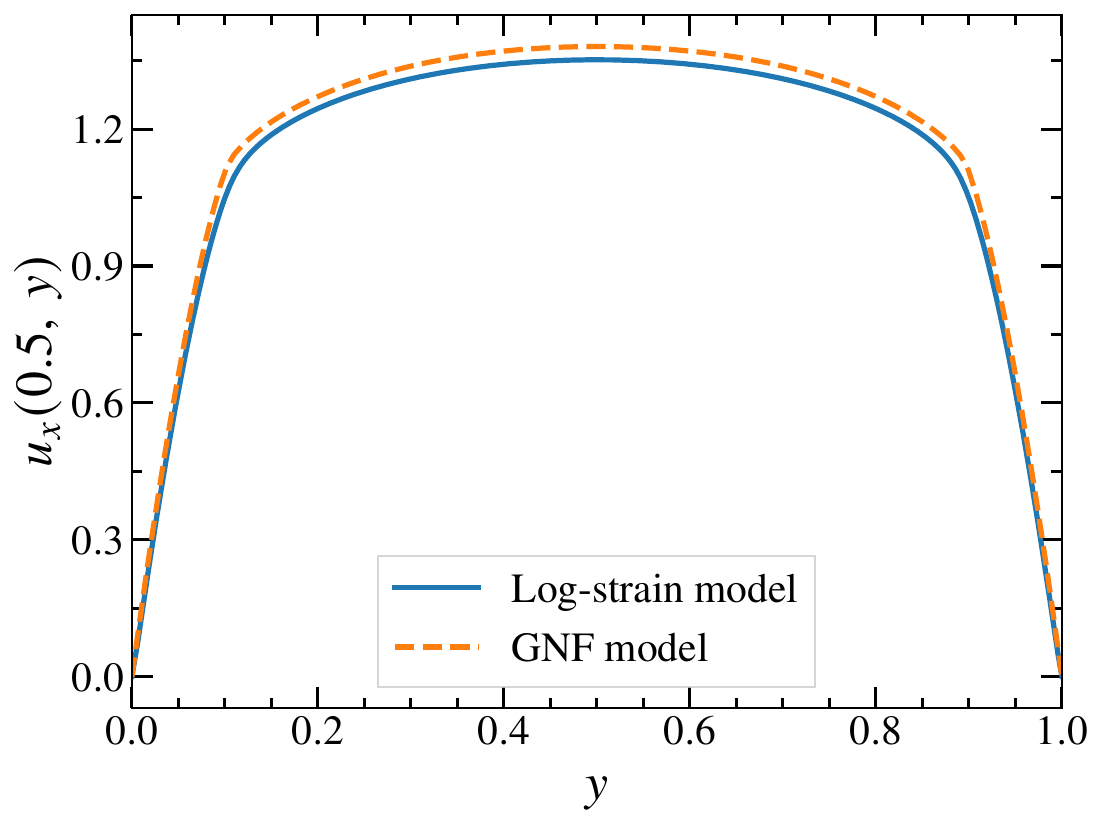}
    \caption{Horizontal velocity $u_x$ at steady state in a straight channel for the log-strain model (blue solid line) and the GNF model (orange dashed line). The viscosity ratio is $\El=25$ and the driving pressure gradient is given by $C=50$.}
    \label{f:compare_channel}
\end{figure}
 
We first examine the flow behavior in a straight channel. The channel geometry and boundary conditions are the same as those described in \cref{subsec:channel}.
Since in the laminar channel flow normal stress differences are balanced out and the local flow type is everywhere that of a simple shear, we expect to observe almost identical profiles for the log-strain model and the GNF model.
Indeed, the steady-state horizontal velocity profiles, presented in \cref{f:compare_channel}, display only a slight discrepancy. To rigorously quantify this variation, a spatial integration of the absolute velocity difference $\left[\int |u_{x,\text{Model}} - u_{x,\text{GNF}}| \, dy\right]$ was performed across the channel height, yielding a global relative error of $2.63\%$.
We ascribe this minor difference to the marginal mismatch in the analytical Cross-model fit observed within the intermediate shear-rate regime ($\Wi\sim 0.3$) where the GNF curve underestimates the numerical data. As the local shear rate drops from its maximum at the walls toward zero at the centerline, it passes through this shear-thinning transition window. Because the lower predicted viscosity of the GNF model yields less local resistance to flow in this zone, it naturally results in a slight over-estimation of the fluid velocity, placing the GNF profile marginally above the log-strain model near the center of the channel.

\begin{figure}[t!]
    \centering
    \begin{subfigure}[b]{0.48\textwidth}
        \caption{\label{f:compare_fpc_a}}
        \includegraphics[width=\textwidth]{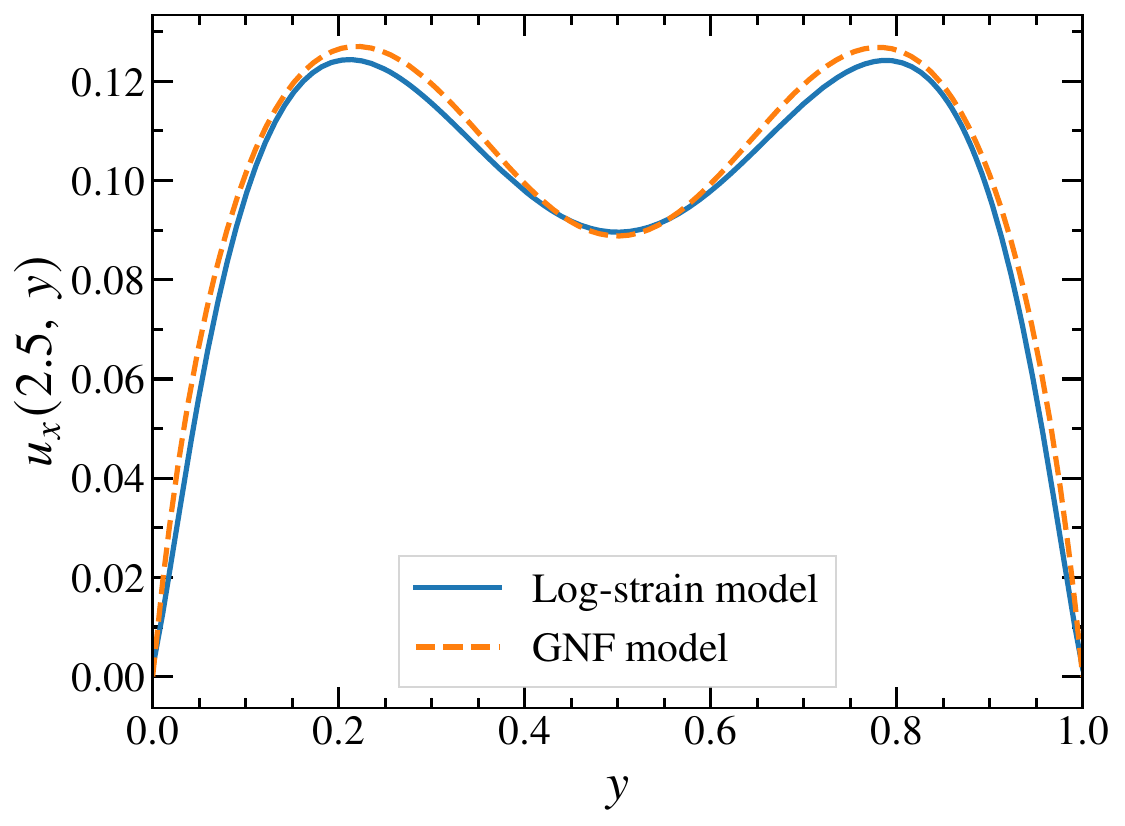}
    \end{subfigure}
    \hfill
    \begin{subfigure}[b]{0.48\textwidth}
        \caption{\label{f:compare_fpc_b}}
        \includegraphics[width=\textwidth]{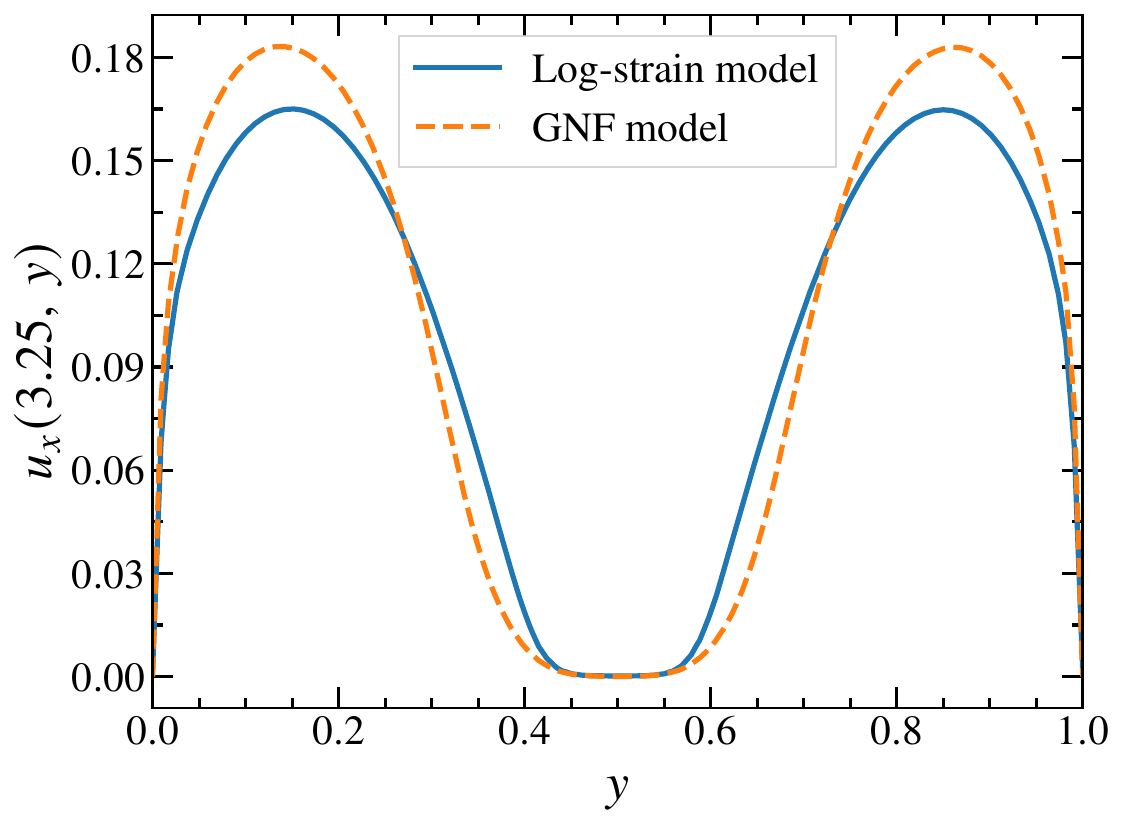}
    \end{subfigure}
    \caption{Horizontal velocity $u_x$ profiles at steady state in a flow past a cylinder of radius $R$ and centered at $(c_x,c_y)$ for the log-strain model (blue solid lines) and the GNF model (orange dashed lines). The viscosity ratio is $\El=25$ and the driving pressure gradient is given by $C=50$. The data are taken for two cross sections at ($a$) $x=c_x-2R$, and ($b$) $x=c_x+R$.}
    \label{f:compare_fpc}
\end{figure}

\begin{figure}[]
    \centering
    \begin{subfigure}[b]{0.48\textwidth}
        \caption{\label{f:SL_fpc_model_a}}
        \includegraphics[width=\textwidth]{SLines_ux_FPC_k25_pg95.pdf}
    \end{subfigure}\\
    \begin{subfigure}[b]{0.48\textwidth}
        \caption{\label{f:SL_fpc_GNF_b}}
        \includegraphics[width=\textwidth]{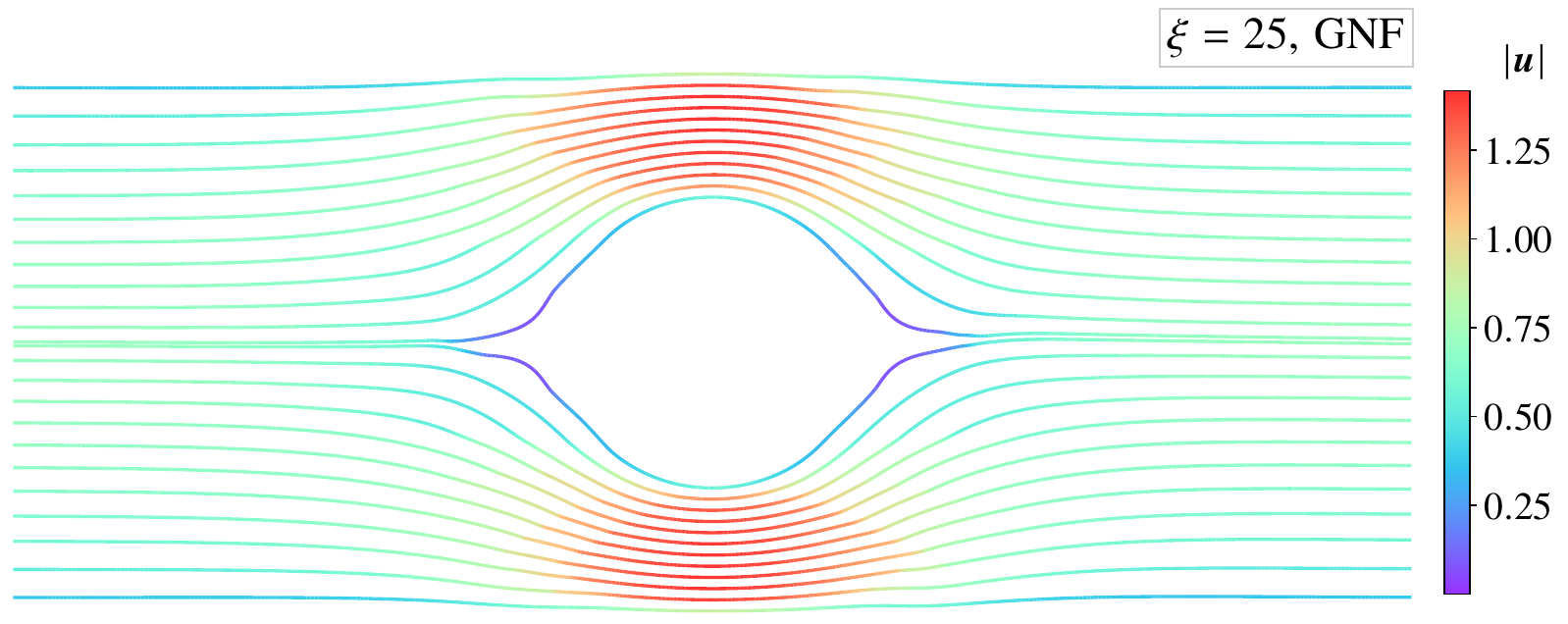}
    \end{subfigure}
    \caption{{Steady streamline patterns for flow past a cylinder at different values of $\Wi$, with fixed $\xi=25$ and driving pressure gradient $C=95$ for the ($a$) log-strain model and ($b$) GNF model. \textit{The log-strain model exhibits a loss of flow symmetry, whereas the GNF model retains symmetry about the cylinder centreline.}}}
    \label{f:SL_fpc_Model_GNF}
\end{figure}

Next, we compare the steady-state horizontal velocity $u_x$ profiles for the flow past a cylinder of radius $R$. The geometry and boundary conditions are the same as those described in \cref{subsec:fpc}.
The velocity field for the full model and the GNF model is evaluated at two distinct cross-sections, one upstream of the obstacle, at $x=c_x-2R$, and the other downstream, at $x=c_x+R$ tangent to the cylinder, as shown in \cref{f:compare_fpc}. Noticeable differences between the log-strain model and the GNF model appear in these profiles.
Importantly, both panels of \Cref{f:compare_fpc} show that the velocity profiles for the two models do not differ merely by a scaling factor (as was the case in the channel flow). This indicates that the mismatch in the fitting of the viscosity is not likely to be responsible for the discrepancy.

{The observed behavior of the log-strain model is partly due to the presence of a significant region dominated by extensional flows, both upstream and downstream of the obstacle. In such conditions the GNF model features a rate dependence that lowers the viscosity, while the log-strain model maintains a more significant dissipation.
Nevertheless, the comparison that we make in  \Cref{f:SL_fpc_Model_GNF} shows that flow-type dependence is not the only source of differences. Indeed, since we are at low Reynolds numbers, the flow-type distribution for a Newtonian fluid is symmetric with respect to the upstream-downstream reflection and so is the GNF solution, even in the presence of higher shear rates (\cref{f:SL_fpc_GNF_b}). On the other hand, the log-strain model presents the upstream-downstream symmetry at low $\Wi$, but that symmetry is broken at higher $\Wi$ (\cref{f:SL_fpc_model_a}), where the effect of elastic memory, with the consequent growth of normal stress differences, becomes relevant in determining the flow.}

Our observations highlight the fact that, while the GNF model remains suitable for simple shear-dominated steady flows, the log-strain model presents a much richer rheology, that is necessary to discuss viscoelastic flows in complex geometries.

\section{Conclusions}
\label{sec:conclusions}

With this study we investigated the interplay of the Weissenberg number $\Wi$ and of the ratio $\El$ between the polymeric and solvent viscosity contributions in a recently proposed viscoelastic fluid model. 
This tensorial model is characterized by an elastic stress tensor that depends logarithmically on a tensorial measure of elastic (or recoverable) strain, hence the name of \emph{log-strain model}. 
We explored the model behavior in three paradigmatic planar flows: the flow in a straight channel, the flow past a cylinder, and the 4:1 contraction problem.
Moreover, a comparison with a suitably defined Generalized Newtonian Fluid model has been carried out in two examples to highlight the importance of effects that cannot be captured simply by a rate-dependent viscosity.

We find that the value of the viscosity ratio is crucial in determining to which extent non-Newtonian flow profiles are observed upon increasing $\Wi$. Indeed, for finite values of $\El$, the channel flow profiles interpolate between a quasi-Newtonian behavior at low $\Wi$ and another parabolic profile to which the velocity tends asymptotically for larger and larger values of $\Wi$. In the intermediate regime, the flow profile presents a plug-like region around the center of the channel, which shrinks upon increasing $\Wi$.
The latter phenomenology can be observed only if $\El$ is not too small (in which case the profile is parabolic for any $\Wi$), showing that the viscosity ratio $\El$ plays a major role in generating flows that are manifestly non-Newtonian. 
This is mainly due to the fact that $\El$ dictates to what extent the log-strain model features a shear-thinning rheology.

The flow past a cylinder offers an opportunity to further highlight the properties of the log-strain model. 
The computation of the drag exerted on the cylinder under steady laminar flow for different values of $\Wi$ and $\El$ shows a drag reduction effect typical of shear-thinning fluids.
Moreover, for $\El\gg 1$ the model displays a strong dependence on the local flow type of its rheological curves. Indeed, while shear thinning appears in simple shear, there is no rate-dependence of the viscosity in extensional flows. The consequences of this and of the presence of elastic memory on the predicted flows are shown by means of the comparison between the log-strain model solutions and those of a Generalized Newtonian Fluid model.

The flow profiles obtained in the benchmark 4:1 contraction problem display, for sufficiently large values of $\El$, a non-monotonic dependence of the size of the recirculating region at the contraction corner with increasing $\Wi$. The appearance of a secondary lip vortex was not observed, in keeping with the typical behavior of shear-thinning fluid models for which the high-strain-rate region generated by the contraction leads to a lowering of the viscous resistance, that in this way cannot generate the secondary recirculation.

To perform our simulations we employed a finite-difference discretization in time combined with a stabilized mixed finite element formulation based on the Variational Multiscale method for the spatial discretization and with a discretization of the Lie derivative approach in the constitutive equation.
In the log-strain model, the elastic strain plays the role of a conformation tensor and its evolution equation inherently preserves its determinant and positive definiteness. 
These properties also need to be enforced in the computational method and their loss may be critical in determining the simulation break-down. 
We observed that the intrinsic mathematical structure of the log-strain model, combined with the implemented stabilization scheme, enables the computation of flows at relatively high values of $\Wi$ {($\sim 35$ in the straight channel, $\sim 7$ in the flow past a cylinder, and $\sim 8$ in the flow through a contraction)}.

The present analysis motivates multiple directions for future study. 
Of particular relevance will be the identification of computational methods that can allow to efficiently simulate three-dimensional flows as well as two-dimensional flows with more challenging geometries or at higher values of the Reynolds number $\Rey$. Indeed, investigating the role of inertia and its interplay with elasticity and relaxation mechanisms will be important for many applications. Even though our computational method performed well, we anticipate the need of more sophisticated approaches to deal with such a task. 
From the point of view of the constitutive model, it will be advantageous to consider the presence of multiple relaxation modes, either by the superposition of different elastic stress contributions (with the evolution of the corresponding elastic strains) or by considering a single mode with a relaxation time that can vary depending on the local state of the material.

\section*{Declaration of competing interest}
The authors declare that they have no known competing financial interests or personal relationships that could have appeared
to influence the work reported in this paper.

\section*{Data availability}
Data will be made available on request.

\section*{Acknowledgements}
N. Dash and G. G. Giusteri would like to thank Hirofumi Notsu (Faculty of Mathematics and Physics, Kanazawa University, Japan) for several useful discussions. R. Codina gratefully acknowledges the support received through the ICREA Aca\-dèmia Research Program of the Catalan Government.

\appendix

\section{Numerical convergence tests}
\label{app:A}

\begin{figure*}[t!]
    \centering
    \begin{subfigure}[b]{0.35\textwidth}
        \caption{\label{f:ch_timeconv_a}}
        \includegraphics[width=\textwidth]{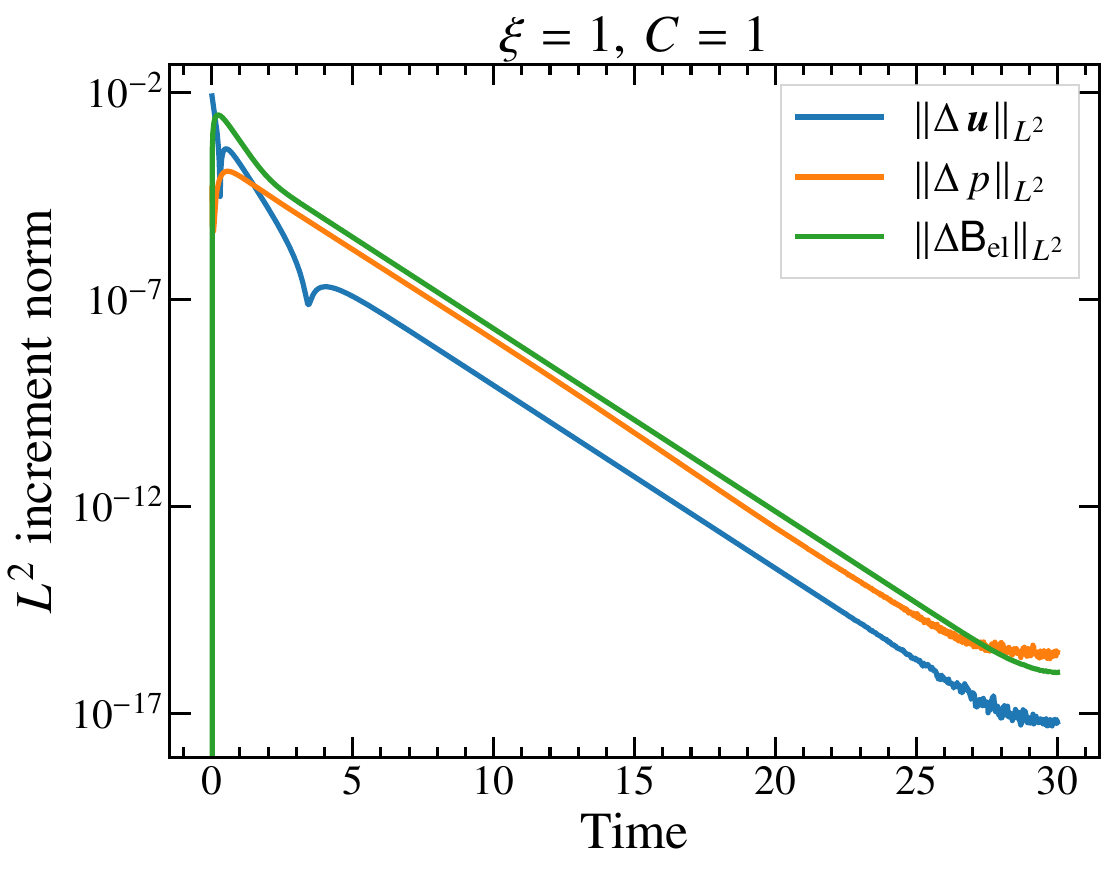}
    \end{subfigure}
    % \hfill
    \hspace{1cm}
    \begin{subfigure}[b]{0.35\textwidth}
        \caption{\label{f:ch_timeconv_b}}
        \includegraphics[width=\textwidth]{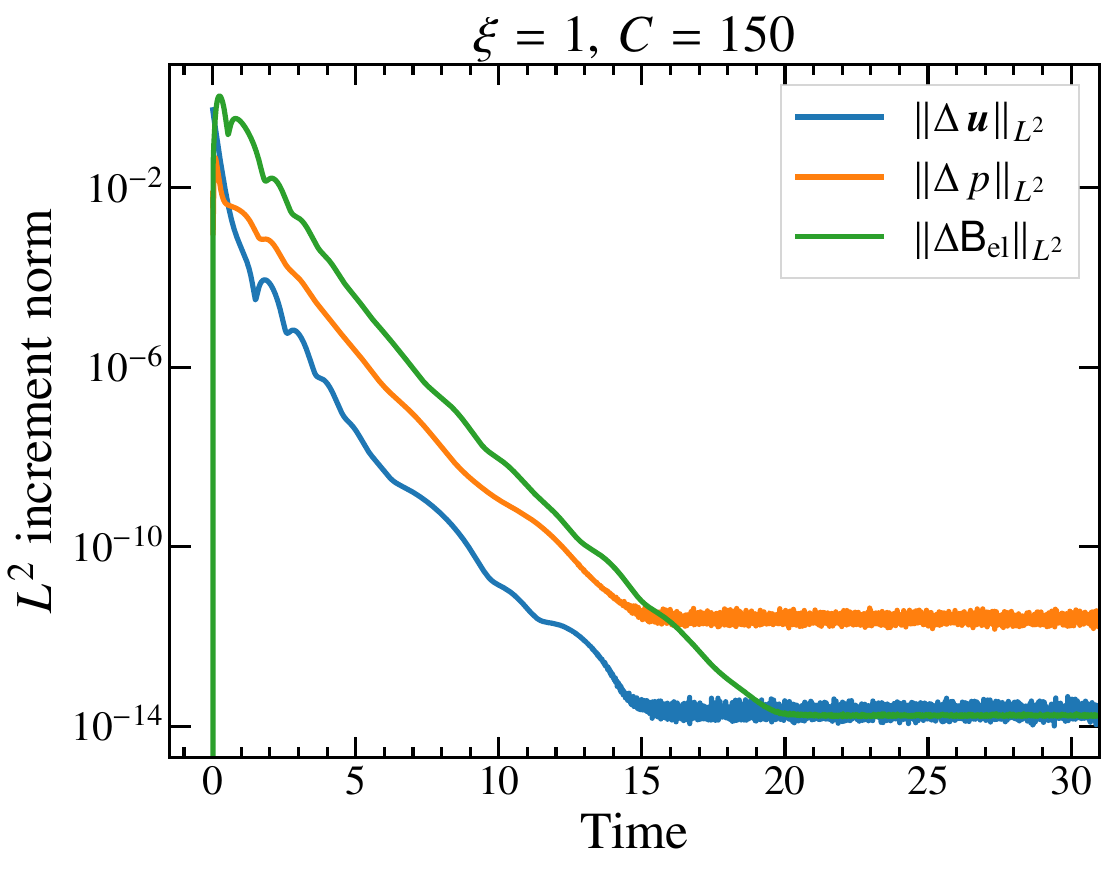}
    \end{subfigure}\\
    \begin{subfigure}[b]{0.35\textwidth}
        \caption{\label{f:ch_timeconv_c}}
        \includegraphics[width=\textwidth]{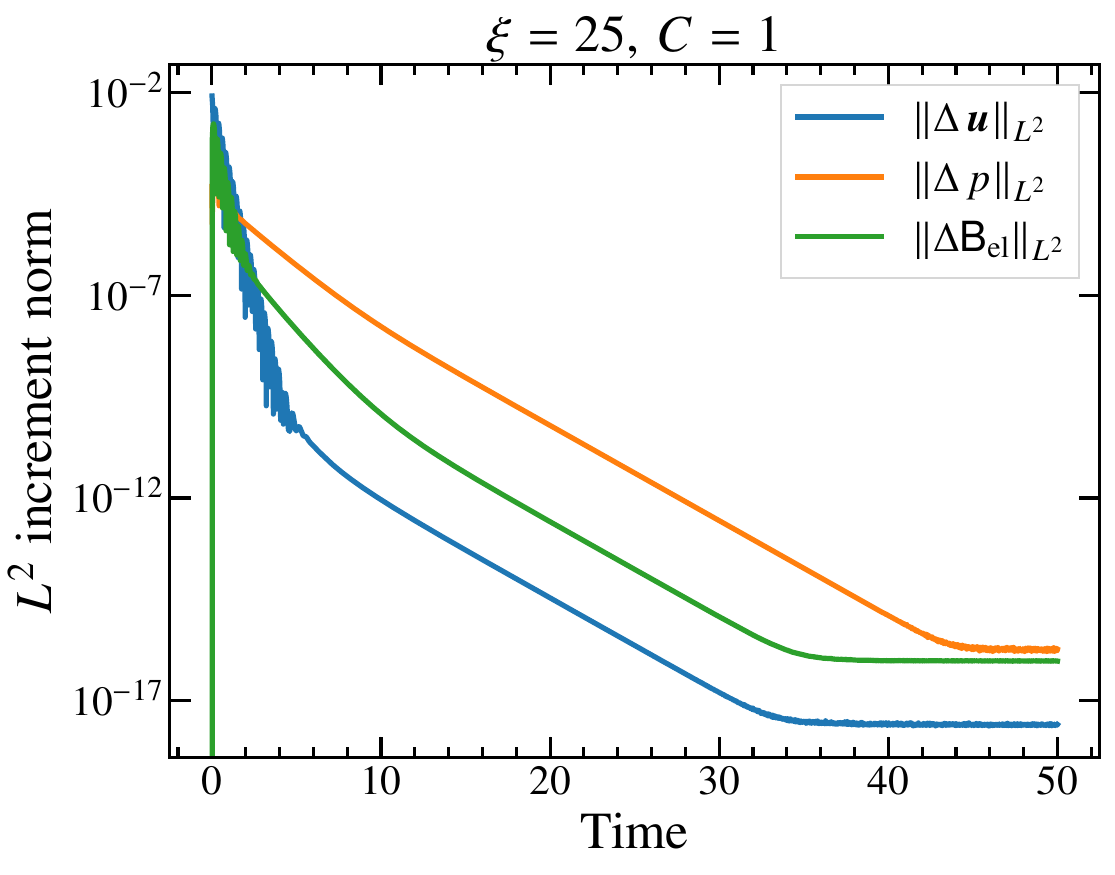}
    \end{subfigure}
    % \hfill
    \hspace{1cm}
    \begin{subfigure}[b]{0.35\textwidth}
        \caption{\label{f:ch_timeconv_d}}
        \includegraphics[width=\textwidth]{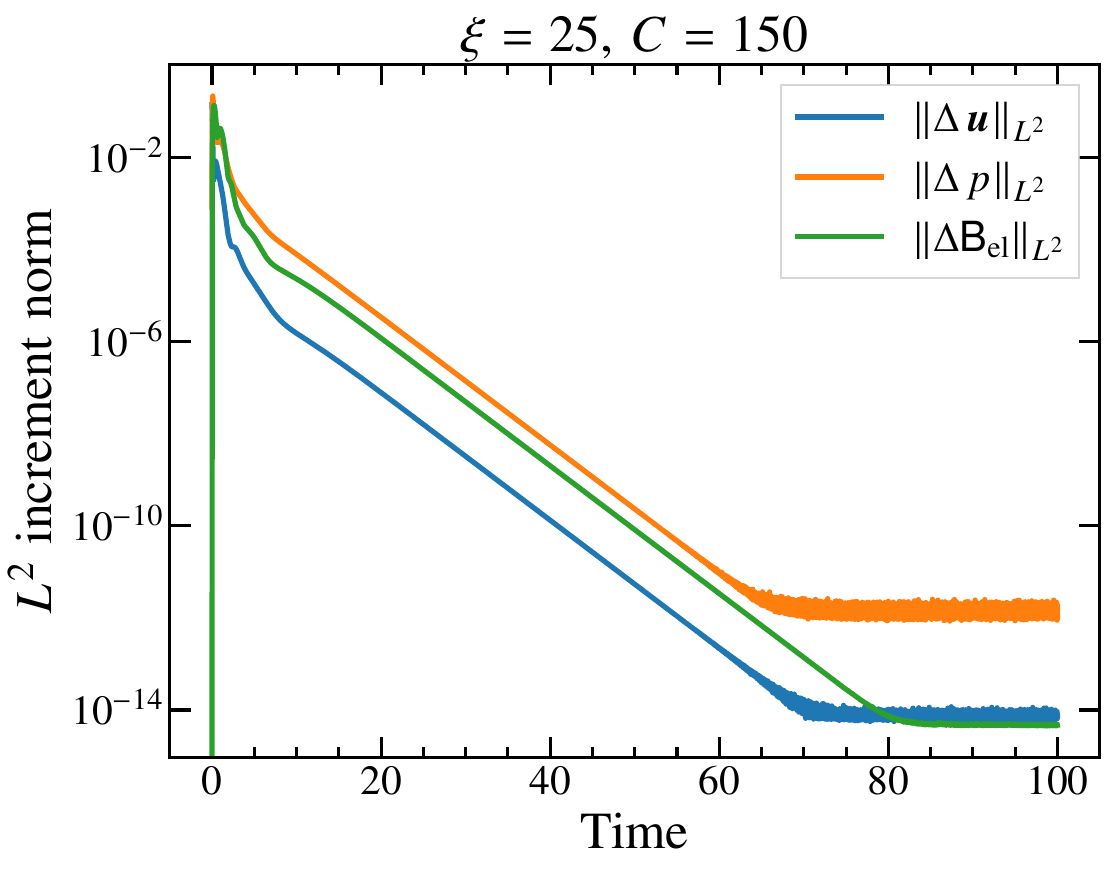}
    \end{subfigure}
    \caption{{Time convergence to the steady state of the channel flow simulations for ($a,b$) $\xi=1$, and ($c,d$) $\xi=25$, with different pressure gradients $C$ as indicated.}}
    \label{f:ch_timeconv}
\end{figure*}

\begin{figure*}[t!]
    \centering
    \begin{subfigure}[b]{0.35\textwidth}
        \caption{\label{f:fpc_timeconv_a}}
        \includegraphics[width=\textwidth]{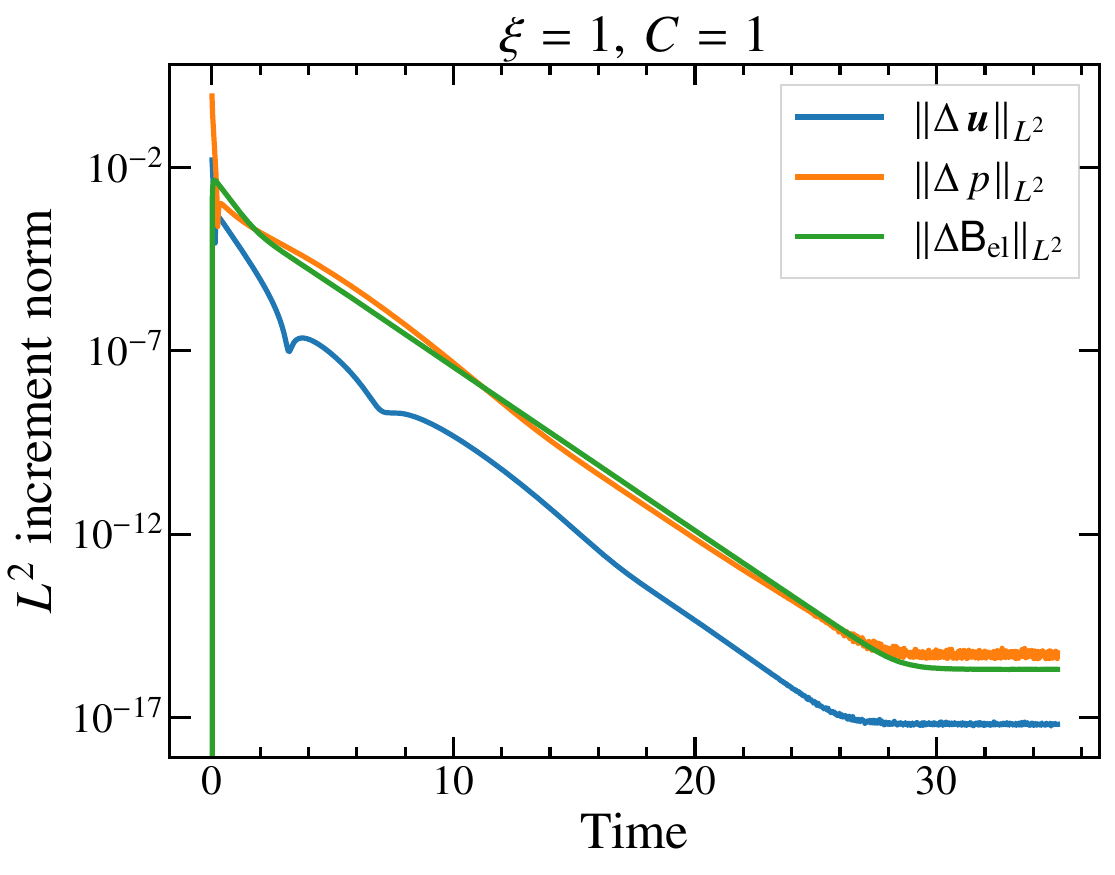}
    \end{subfigure}
    % \hfill
    \hspace{1cm}
    \begin{subfigure}[b]{0.35\textwidth}
        \caption{\label{f:fpc_timeconv_b}}
        \includegraphics[width=\textwidth]{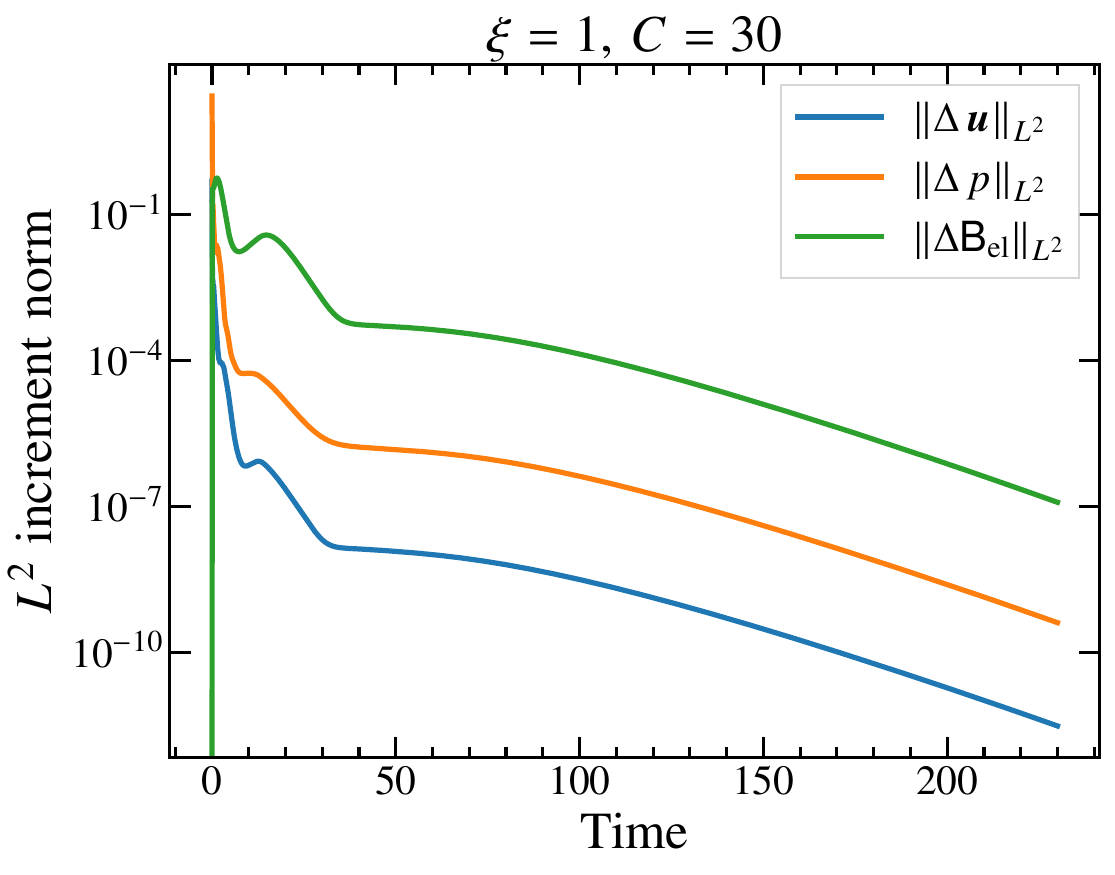}
    \end{subfigure}\\
    \begin{subfigure}[b]{0.35\textwidth}
        \caption{\label{f:fpc_timeconv_c}}
        \includegraphics[width=\textwidth]{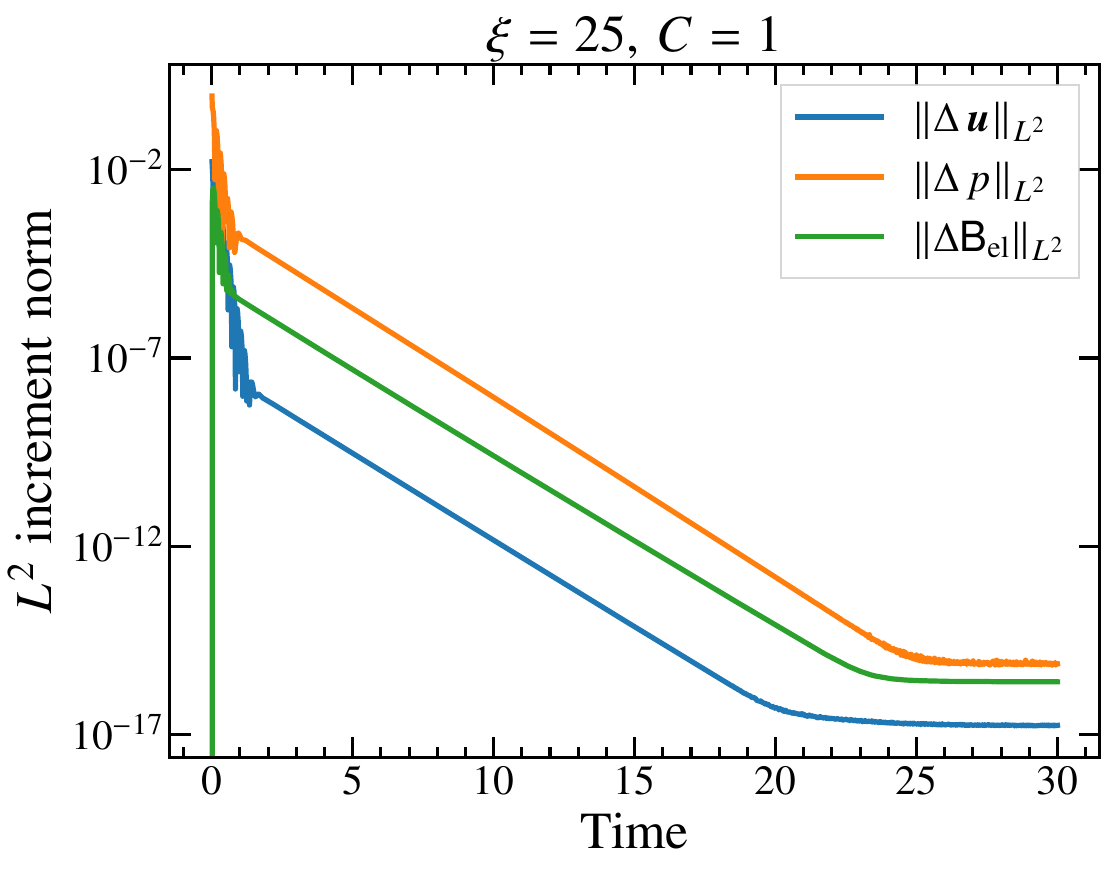}
    \end{subfigure}
    % \hfill
    \hspace{1cm}
    \begin{subfigure}[b]{0.35\textwidth}
        \caption{\label{f:fpc_timeconv_d}}
        \includegraphics[width=\textwidth]{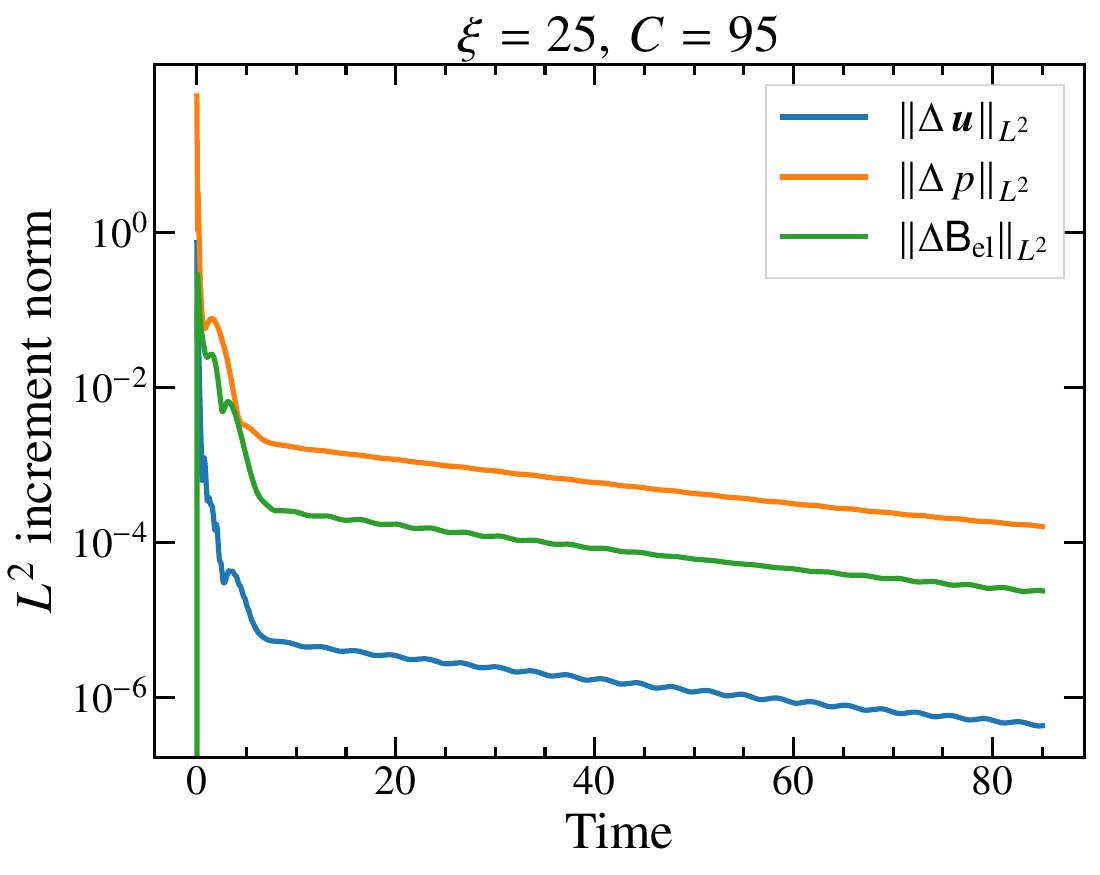}
    \end{subfigure}
    \caption{{Time convergence to the steady state of the flow past a cylinder simulations for ($a,b$) $\xi=1$, and ($c,d$) $\xi=25$, with the different pressure gradients $C$ as indicated.}}
    \label{f:fpc_timeconv}
\end{figure*}

\begin{figure*}[t!]
    \centering
    \begin{subfigure}[b]{0.35\textwidth}
        \caption{\label{f:cont_timeconv_a}}
        \includegraphics[width=\textwidth]{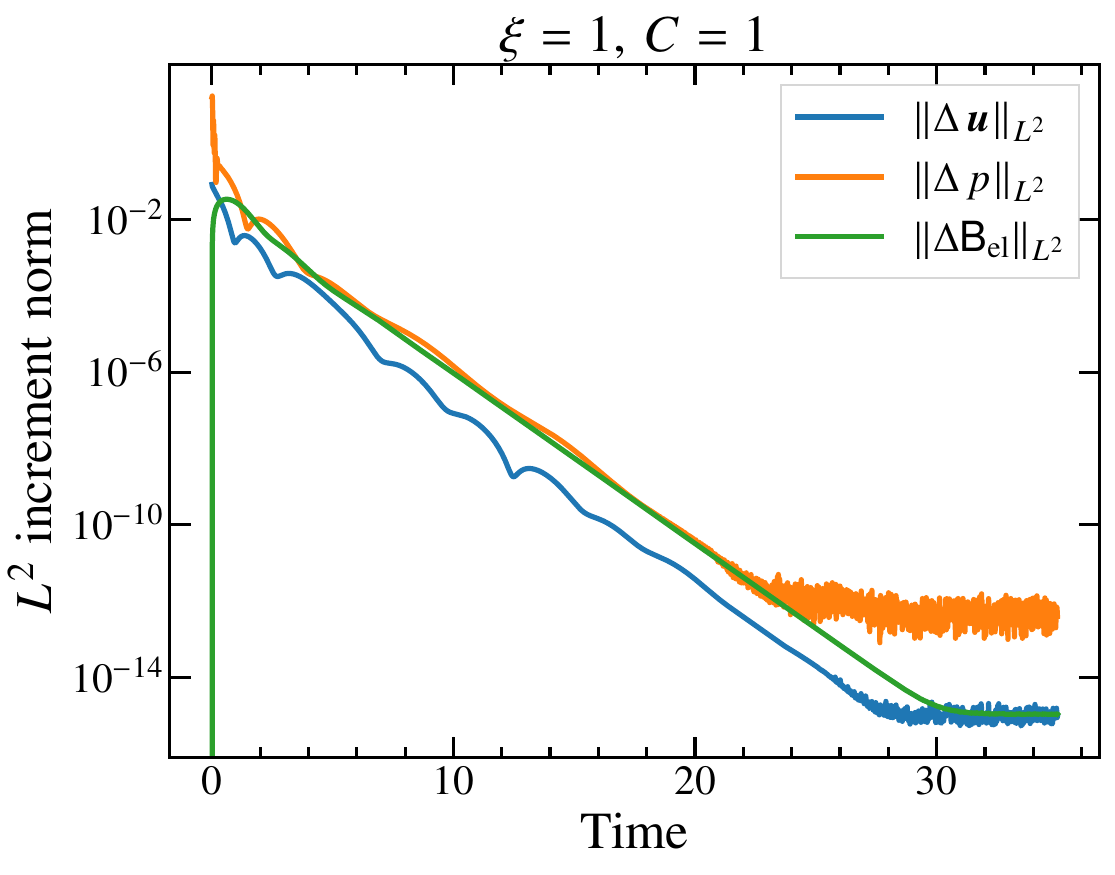}
    \end{subfigure}
    \hspace{1cm}
    % \hfill
    \begin{subfigure}[b]{0.35\textwidth}
        \caption{\label{f:cont_timeconv_b}}
        \includegraphics[width=\textwidth]{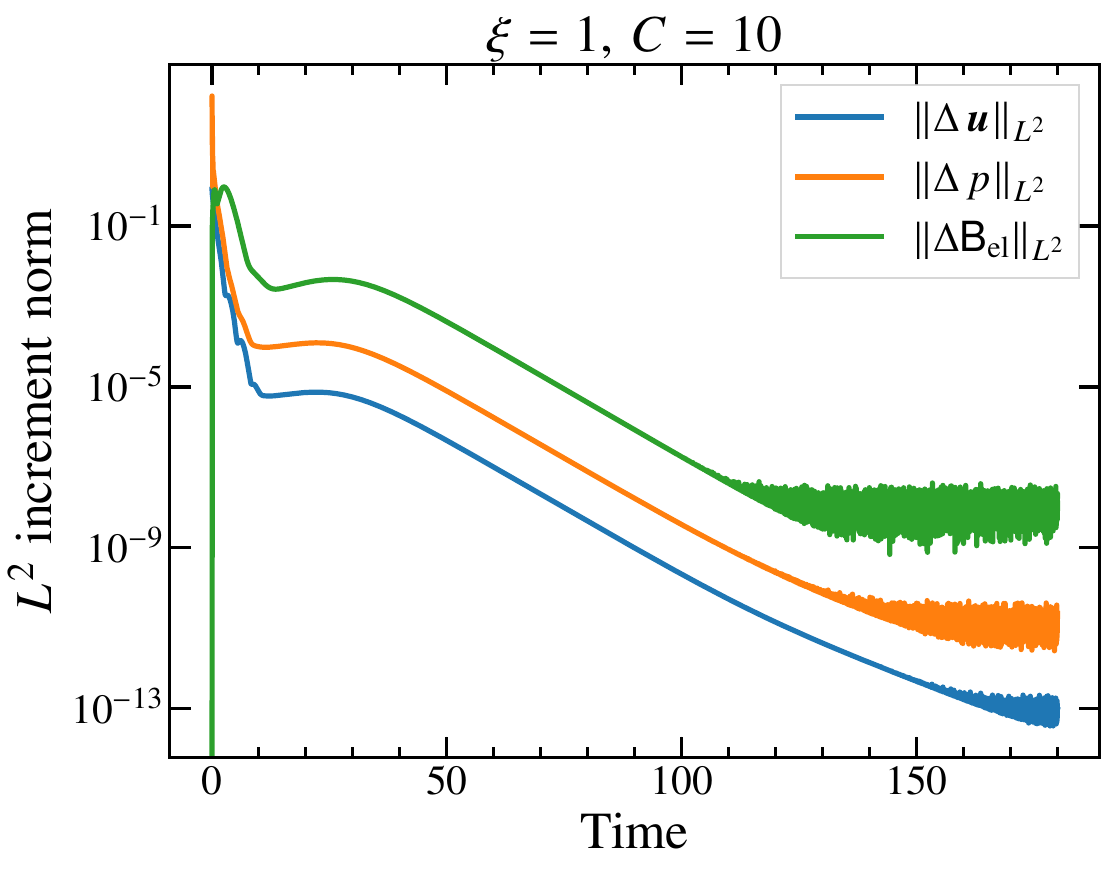}
    \end{subfigure}\\
    \begin{subfigure}[b]{0.35\textwidth}
        \caption{\label{f:cont_timeconv_c}}
        \includegraphics[width=\textwidth]{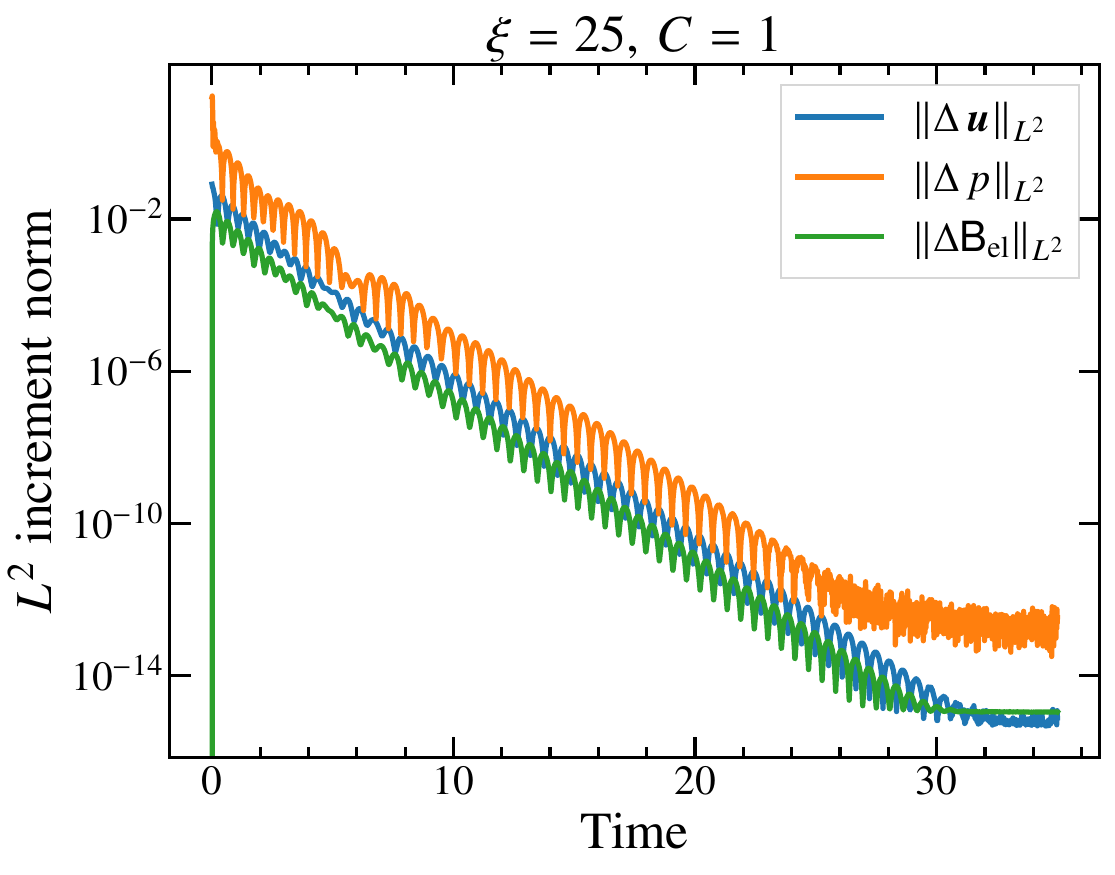}
    \end{subfigure}
    % \hfill
    \hspace{1cm}
    \begin{subfigure}[b]{0.35\textwidth}
        \caption{\label{f:cont_timeconv_d}}
        \includegraphics[width=\textwidth]{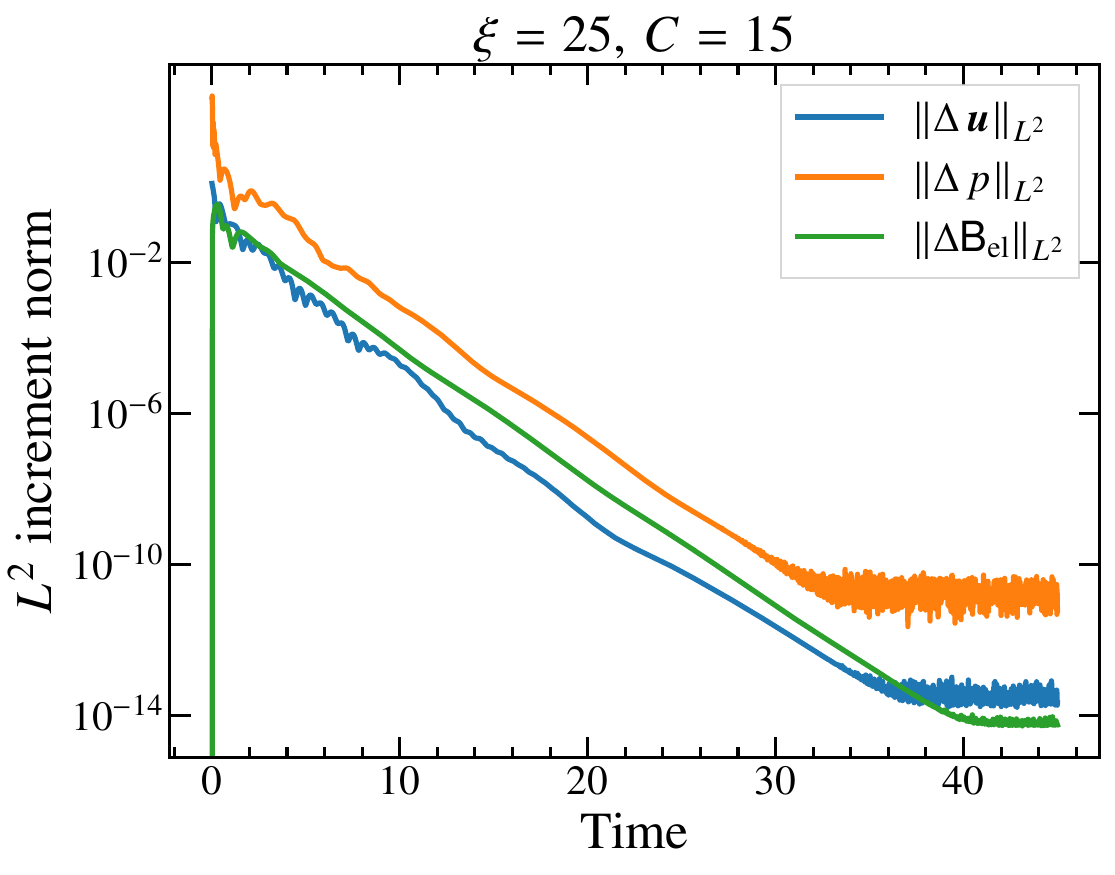}
    \end{subfigure}
    \caption{{Time convergence to the steady state of the flow through a 4:1 contraction for ($a,b$) $\xi=1$, and ($c,d$) $\xi=25$, with the different pressure gradients $C$ as indicated.}}
    \label{f:cont_timeconv}
\end{figure*}

{Time convergence to the steady state was monitored in all simulations by tracking the $L^2$-norms of the increments in velocity $\vc{u}$, pressure $p$, and strain tensor $\Bel$ between successive time steps. These increments were computed as the differences between the current and previous solutions, and their norms were evaluated over the computational domain. The resulting histories were recorded throughout each simulation and plotted on a logarithmic scale. 
For the channel flow simulations, the time step was set to $\Delta t=0.01$ for pressure gradients $C\leq 50$ and to $\Delta t=0.004$ for $C>50$. A time step of $\Delta t=0.01$ was used for both flow past a cylinder and flow through a 4:1 contraction. 
The time convergence results for representative simulations of channel flow, flow past a cylinder, and flow through a 4:1 contraction are shown in \cref{f:ch_timeconv,f:fpc_timeconv,f:cont_timeconv}, respectively.
The decrease in the increment norms for all unknown fields confirms that steady state solutions were attained in all simulations to within a sufficiently small tolerance.}

\begin{table}[b!]
\centering
\setlength{\tabcolsep}{4pt}
\caption{{Mesh convergence analysis for channel flow with $\xi=25$.}}
\label{t:mesh_ch}
\begin{tabular}{@{}cccc@{}}
\toprule
$C\,$(Pa/m) & Mesh elements & Relative error \\
\midrule
1  & 2,048  & 3.73\% \\
   & 8,192  & 1.26\% \\
   & 32,768 & Reference \\
\midrule
70 & 2,048 & 0.51\% \\
   & 8,192 & 0.13\% \\
   & 32,768 & Reference \\
\bottomrule
\end{tabular}
\end{table}

\begin{table}[b!]
\centering
\setlength{\tabcolsep}{4pt}
\caption{{Mesh convergence analysis for flow past a cylinder with $\xi=25$.}}
\label{t:mesh_fpc}
\begin{tabular}{@{}cccc@{}}
\toprule
$C\,$(Pa/m) & Mesh elements & Relative error \\
\midrule
1  & 7,900   & 3.88\% \\
   & 31,332  & 0.84\% \\
   & 124,247 & Reference \\
\midrule
50 & 7,900 & 11.21\% \\
   & 31,332 & 3.41\% \\
   & 124,247 & Reference \\
\bottomrule
\end{tabular}
\end{table}

\begin{table}[t!]
\centering
\setlength{\tabcolsep}{4pt}
\caption{{Mesh convergence analysis for planar 4:1 contraction flow simulations with $\xi=25$.}}
\label{t:mesh_cont}
\begin{tabular}{@{}cccc@{}}
\toprule
$C\,$(Pa/m) & Mesh elements & Relative error \\
\midrule
1  & 4,330 & 0.67\% \\
   & 6,134 & 0.42\% \\
   & 9,454 & Reference \\
\midrule
15 & 4,330 & 3.59\% \\
   & 6,134 & 1.94\% \\
   & 9,454 & Reference \\
\bottomrule
\end{tabular}
\end{table}

{As far as spatial discretization is concerned, 
mesh convergence analyses for the computationally more demanding cases of the three viscoelastic flow problems are considered. For each problem, three meshes of increasing refinement were used. The corresponding numbers of elements and relative $L^2$-errors with respect to the solution obtained on the finest mesh are reported in \Cref{t:mesh_ch,t:mesh_fpc,t:mesh_cont}.
The results indicate that the numerical solutions are sufficiently insensitive to further mesh refinement, supporting the adequacy of the selected meshes for the simulations presented in this work.}

{The flow in a straight channel offers the opportunity of comparing the finite-element approximation with a semi-analytical profile obtained through interpolation and quadrature of the flow curve. Indeed, the relevant component of the flow equation in the lower part of the channel for $y\in [0,H/2]$ becomes
\[
\frac{\de }{\de y}\sigma_{xy}(\dot{\gamma}(y))=-C,
\]
that entails
\[
\dot{\gamma}(y)=\sigma_{xy}^{-1}(-C(y-H/2)),
\]
whence the quadrature expression
\begin{equation}\label{e:semianalytical}
u(y)=\int_{0}^y\sigma_{xy}^{-1}(-C(\hat{y}-H/2))\,d\hat{y}.
\end{equation}}

\begin{figure}[]
    \centering
    \includegraphics[width=0.9\linewidth]{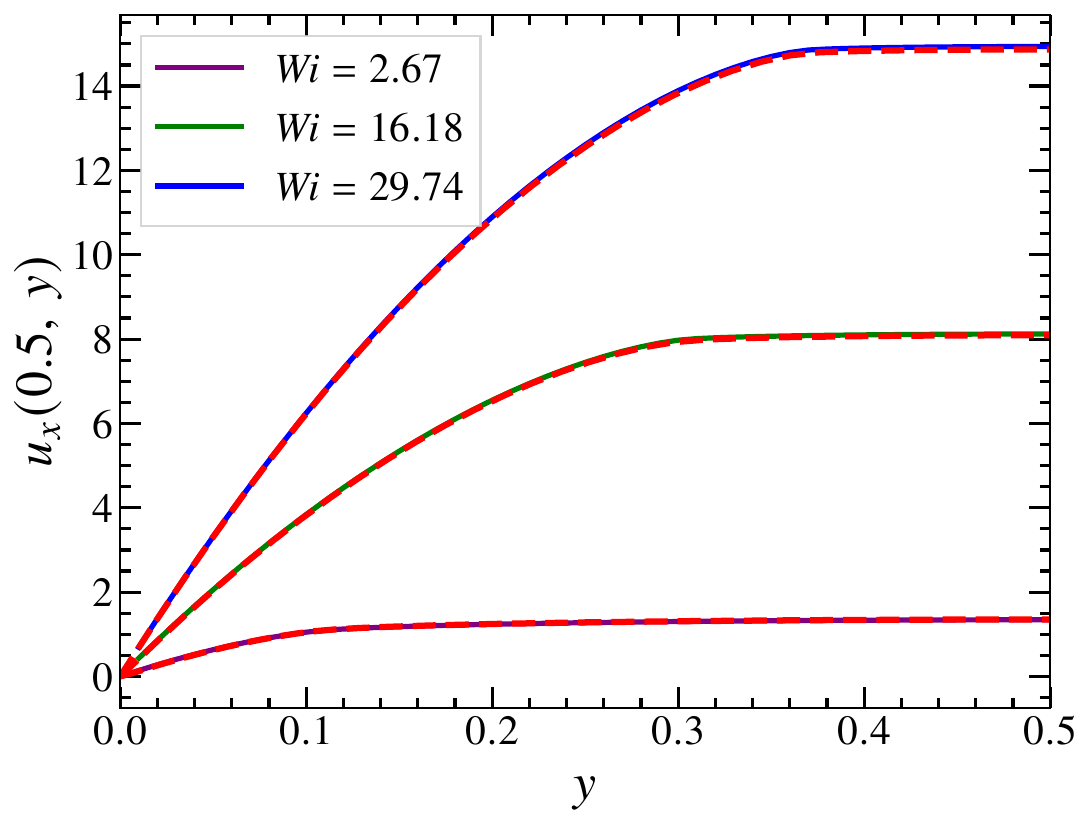}
    \caption{Comparison between the finite-element solution (solid lines) for the channel flow and the semi-analytical solution (dashed lines) given by \cref{e:semianalytical} for the case with $\xi=25$.}
    \label{f:analytical_compare_channel}
\end{figure}

{While the analytical inversion of the flow curve $\sigma_{xy}(\dot{\gamma})$ is rarely possible, we can numerically sample such an inverse, interpolate, and integrate to obtain an approximation of the velocity profile. The result is compared with the finite-element approximation for the log-strain model with $\xi=25$ at three values of $\Wi$ in \cref{f:analytical_compare_channel}, showing an excellent agreement between the two computations.
This is an additional verification of the convergence towards the exact solution.}

\bibliographystyle{elsarticle-num-names}

\bibliography{references}

\end{document}